\documentclass[twocolumn]{jpsj2} 
\def\eg{e.g.}

\def\etal{{\it et\ al.}}

\newcommand{\lsim}
 {\ \raise.35ex\hbox{$<$}\kern-0.75em\lower.5ex\hbox{$\sim$}\ }
\newcommand{\gsim}
 {\ \raise.35ex\hbox{$>$}\kern-0.75em\lower.5ex\hbox{$\sim$}\ }
\def\journal #1#2#3#4{#1 {\bf #2} (#4) #3}
\def\PR{Phys.\ Rev.}
\def\PRB{Phys.\ Rev.\ B}
\def\PRL{Phys.\ Rev.\ Lett.}
\def\PRS{Proc.\ R.\ Soc.}
\def\APNY{Ann.\ Phys.\ (New York)}

\def\SSC{Solid State Commun.}

\def\JMMM{J.~Magn.~Magn.~Mater.}

\def\JPParis{J.~Phys.~(Paris)}

\def\JPSJ{J.\ Phys.\ Soc.\ Jpn.}

\def\RMP{Rev.\ Mod.\ Phys.}
\def\PTP{Prog.\ Theor.\ Phys.}

\title{Mott Transitions and $d$-Wave Superconductivity in Half-Filled-Band 
Hubbard Model on Square Lattice with Geometric Frustration}

\author{Hisatoshi \textsc{Yokoyama},\thanks{E-mail address: yoko@cmpt.phys.tohoku.ac.jp} Masao \textsc{Ogata}$^{1}$ and Yukio \textsc{Tanaka}$^{2}$}

\inst{Department of Physics, Tohoku University, Sendai 980-8578 \\
$^{1}$Department of Physics, University of Tokyo, Bunkyo-ku, Tokyo 113-0033 \\
$^{2}$Department of Applied Physics, Nagoya University,  Nagoya 464-8603
}

\abst{
Mechanisms of Mott transitions and $d_{x^2-y^2}$-wave superconductivity 
(SC) are studied in the half-filled-band Hubbard model on square lattices 
with a diagonal hopping term ($t'$), using an optimization (or correlated) 
variational Monte Carlo method. 
In the trial wave functions, a doublon-holon binding effect is introduced 
in addition to the onsite Gutzwiller projection. 
We mainly treat a $d$-wave singlet state and a projected Fermi sea. 
In both wave functions, first-order Mott transitions without direct 
relevance to magnetic orders occur at $U=U_{\rm c}$, which is 
approximately the bandwidth, for arbitrary $t'/t$. 
These transitions originate in the binding or unbinding of a doublon to 
a holon. 
$d$-wave SC appears in a narrow range immediately below $U_{\rm c}$. 
The robust $d$-wave superconducting correlation is necessarily 
accompanied by enhanced antiferromagnetic correlation; the strength of 
SC decreases, as $t'/t$ increases.
}

\kword{Mott transition, superconductivity, condensation energy, 
antiferromagnetic correlation, spin gap, high-$T_{\rm c}$ cuprate, 
Hubbard model, variational Monte Carlo method, frustration}

\begin{document}
\maketitle

\section{\label{sec:Intro}Introduction}

\begin{figure}[hob]
\begin{center}
\includegraphics[width=5.6cm,clip]{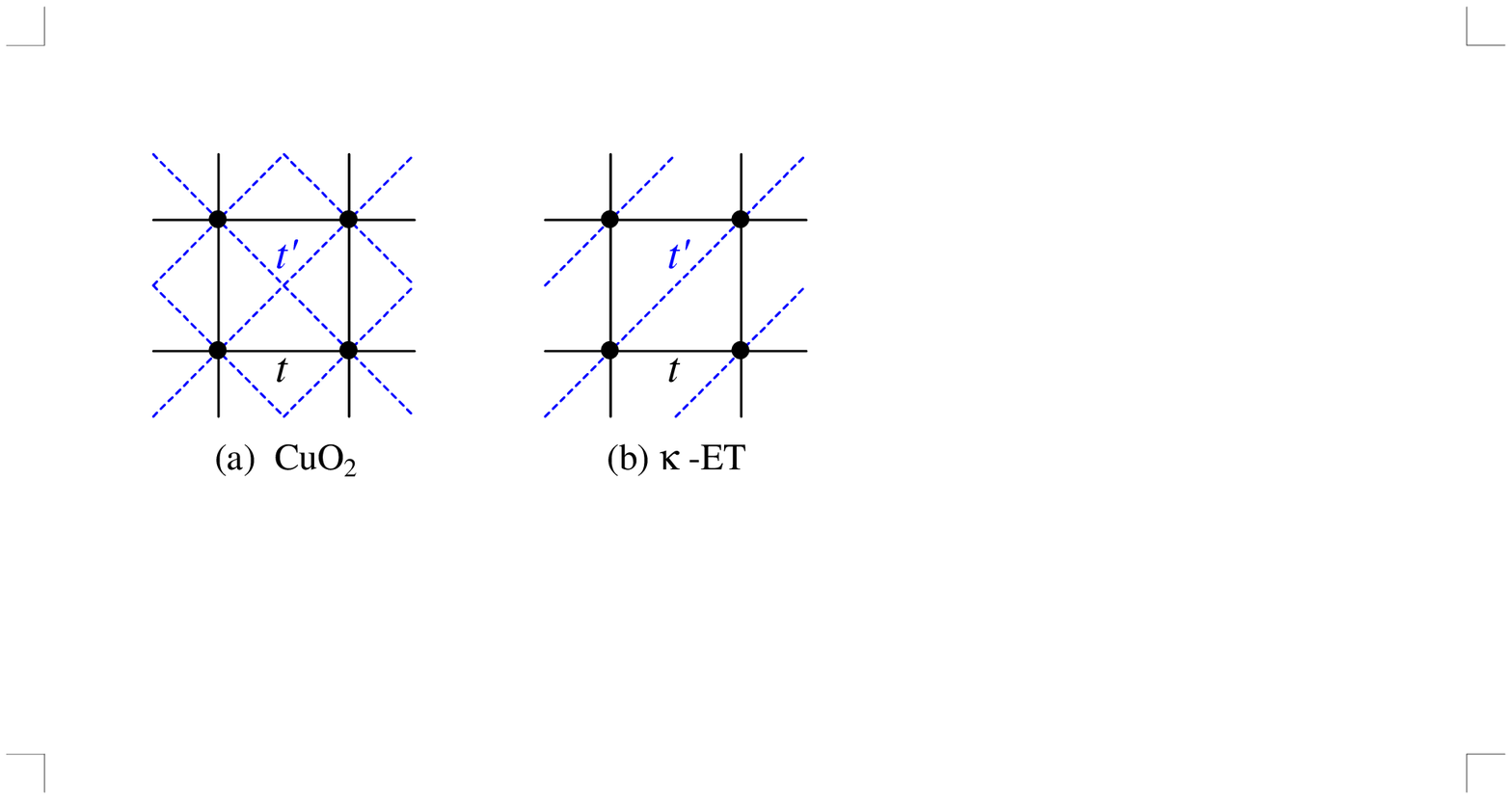}
\end{center}
\vskip -5mm
\caption{(Color online)
Lattice structure and hopping integrals $t$ and $t'$, 
(a) studied in this work, and 
(b) often used for $\kappa$-BEDT-TTF salts. 
Lattice sites are denoted by dots. 
}
\label{fig:model}
\end{figure}

In connection with the superconductor-insulator transitions in organic 
compounds $\kappa$-(BEDT-TTF)X,\cite{ET} the half-filled-band Hubbard 
model on anisotropic triangular lattices [Fig.~\ref{fig:model}(b)] 
\cite{Kino} has been intensively studied. 
In addition, a recent experimental study \cite{Naito} has reported 
that a series of film samples of nondoped high-$T_{\rm c}$ cuprates 
(parent materials of electron-doped systems) do not become 
antiferromagnetic (AF) insulators but exhibit metallic properties 
including superconductivity (SC) at 21 K. 
Thus, it is important to grasp the mechanisms of Mott transitions 
and SC, if any, in half-filled-band Hubbard models with frustration. 
Because such phenomena arise at intermediate correlation strength 
($U\sim W$; $W$ being the bandwidth), one has to use a method that 
can reliably treat both strongly correlated and weakly correlated 
regimes. 
As one of such methods, the optimization (or correlated) variational 
Monte Carlo (VMC) method \cite{Umrigar} has rapidly progressed in 
recent years for the study of ground-state properties. \cite{OVMC05}
\par 

Using this method, the present authors have recently studied the Hubbard 
model on the anisotropic triangular lattice [Fig.~\ref{fig:model}(b)]. 
\cite{Wataorg} 
With a projected $d$-wave singlet state, it is found that 
(1) a conductive-to-nonmagnetic-insulator (Mott) transition occurs 
at $U=U_{\rm c}$, which is somewhat smaller than $W$, and is caused 
by the binding (and unbinding) of a doublon to a holon; 
(2) $d_{x^2-y^2}$-wave SC appears in the vicinity of both the Mott 
transition and the AF phase, and develops together with the short-range 
AF correlation (or fluctuation). 
\par

In weakly frustrated cases, the two lattices in Figs.~\ref{fig:model}(a) 
and \ref{fig:model}(b) have a common characteristic wave number 
${\bf G}=(\pi,\pi)$ in spin correlation. 
In strongly frustrated cases, however, the characteristic wave numbers 
become different; for instance, 120-degree-structure spin correlation 
is probably dominant in (b) for $t'\sim t$, \cite{tri} 
whereas the collinear structure is favored in (a). 
In this paper, we carry out similar detailed calculations for the lattice, 
often treated in the context of high-$T_{\rm c}$ cuprates ($t$-$t'$-$U$ 
model) [Fig.~\ref{fig:model}(a)], and reveal whether or not the above 
mechanisms of the Mott transition and of SC also work here. 
In addition to the $d$-wave singlet state, we study Mott transitions in 
the projected Fermi sea. 
\par 
 
\vspace{-0.5cm}

The organization of this paper is as follows: 
In \S\ref{sec:model}, we introduce the model and method we use. 
In \S\ref{sec:MottSC} and \S\ref{sec:MottFS}, we discuss Mott transitions 
in the $d$-wave singlet state and in the projected Fermi sea, respectively. 
Section \ref{sec:SC} is assigned to the stability and properties of 
the $d$-wave superconducting (SC) state. 
In \S\ref{sec:AF}, we construct a ground-state phase diagram based on 
the present VMC calculations, and address important problems with 
respect to antiferromagnetism (AF). 
In \S\ref{sec:summary}, we briefly summarize the main results and discuss 
the subject further. 
\par

Part of the present results have been reported in a previous letter. 
\cite{YTOT} 

\section{\label{sec:model}Model and Method}
%
In \S\ref{sec:Hub}, the model we study is introduced, and related 
studies are summarized. 
In \S\ref{sec:WF}, we briefly review the background of variational 
wave functions for the Hubbard model in the research on the Mott 
transition. 
In \S\ref{sec:WFnow}, we give an account of the wave functions used 
in this paper and the conditions of VMC calculations. 
\par

\subsection{\label{sec:Hub}Hubbard model on frustrated square lattices}
In this paper, we study the Hubbard model \cite{Hubbard,Kanamori,Gutz} 
on a square lattice with diagonal transfer $t'$ [Fig.~\ref{fig:model}(a)], 
\begin{equation}
{\cal H}={\cal H}_{\rm kin}+{\cal H}_U
        =\sum_{{\bf k}\sigma} \varepsilon({\bf k})
        c^\dagger_{{\bf k}\sigma}c_{{\bf k}\sigma} 
         +U\sum_jn_{j\uparrow}n_{j\downarrow},
\label{eq:model}
\end{equation}
\begin{equation}
\varepsilon({\bf k})=-2t(\cos k_x+\cos k_y)-4t'\cos k_x\cos k_y, 
\label{eq:bareband} 
\end{equation}
with $U, t>0$. 
Equation (\ref{eq:model}) has been often used as a simple model 
that captures the essence of cuprates. \cite{Anderson} 
Here, we concentrate on the half-filled band 
($n=N_{\rm e}/N_{\rm s}=1$; $N_{\rm e}$: electron number and 
$N_{\rm s}$: site number) and consider doped cases in a forthcoming 
publication. 
We exclusively treat the case of $t'\le 0$, because the behavior 
for $t'(>0)$ is identical to that for $-t'$, owing to the particle-hole 
symmetry at $n=1$. 
Note that the negative sign of $t'/t$ is in agreement with the 
hole-doped case of high-$T_{\rm c}$ cuprates. 
When $|t'/t|$ is varied, the features of the bare band, 
eq.~(\ref{eq:bareband}), abruptly change at $|t'/t|=0.5$, at which 
low-lying energy levels become strongly degenerate; 
the van Hove singularity points depart from $(\pi,0)$ and $(0,\pi)$ 
for $|t'/t|>0.5$. 
Furthermore, the plausible values of high-$T_{\rm c}$ cuprates are 
considered to be $|t'/t|=0.1$-0.3. \cite{hightc}
Thus, we restrict the range of frustration strength to 
$0\le|t'/t|\le 0.5$ in this paper. 
\par

Despite the importance of the model, reliable knowledge is limited, 
particularly in the intermediate and strong coupling regimes. 
For the pure square lattice ($t'=0$), it is believed that the ground 
state is insulating with a long-range AF order for $U>0$, owing to 
the complete nesting condition. \cite{Hirsch}
For frustrated cases ($t'\ne 0$), however, it is urgently required 
to clarify the properties of the conductor-insulator 
transition. \cite{Moriya,DMFT,Kashima,Mizusaki} 
For SC at half filling, although the anisotropic triangular lattices 
have often been studied, \cite{Tremblay} studies on the present 
lattice are rare to our knowledge. 
\par

\subsection{\label{sec:WF}Historical background of wave functions}

As a many-body trial wave function, the Jastrow type \cite{Jastrow}
is useful and has been often applied: $\Psi={\cal P}\Phi$, where ${\cal P}$ 
denotes a many-body correlation (Jastrow) factor composed of projection 
operators, and $\Phi$ is a one-body wave function usually given by a Slater 
determinant. 
For the Hubbard model, more than four decades ago, Gutzwiller introduced 
the celebrated onsite projection, \cite{Gutz}
\begin{equation}
{\cal P}_{\rm G}=\prod_j\left[1-(1-g)n_{j\uparrow}n_{j\downarrow}\right], 
\end{equation}
which has primary importance for arbitrary parameters in the Hubbard model. 
Although the Gutzwiller wave function (GWF), 
$
\Psi_{\rm G}^{\rm FS}={\cal P}_{\rm G}\Phi_{\rm F}
$
($\Phi_{\rm F}$: Fermi sea), looks simple, it is generally difficult 
to accurately calculate expectation values using it. 
Hence, a mean-field-type approximation [now called a Gutzwiller approximation 
(GA)] was introduced by Gutzwiller himself, \cite{GutzA} and was used 
and extended by many researchers for the following two decades. \cite{Voll} 
However, variation theory loses its various advantages when additional 
approximations such as GA are applied, and consequently it becomes difficult 
to improve the wave function. 
To break this deadlock, VMC methods \cite{VMC} have been applied to this 
problem; \cite{YS1,GJR} thereby and by subsequent exact analytic 
treatment in one dimension, \cite{Metzner} the precise behavior 
of $\Psi_{\rm G}^{\rm FS}$ was clarified for the first time. 
Although ${\cal P}_{\rm G}$ is indispensable for treating the Hubbard model, 
its independent use leads to the following physically unsatisfactory 
results: 
(1) The momentum distribution function $n({\bf k})$ tends to be an 
increasing function of $|{\bf k}|$, 
(2) $2k_{\rm F}$ anomalies in the spin [charge density] structure 
factor $S({\bf q})$ [$N({\bf q})$] cannot be properly represented, and
(3) a Mott transition cannot be described, in addition to the problem 
of a considerably high variational energy. 
\par

\begin{figure}[hob]
\begin{center}
\includegraphics[width=8.7cm,clip]{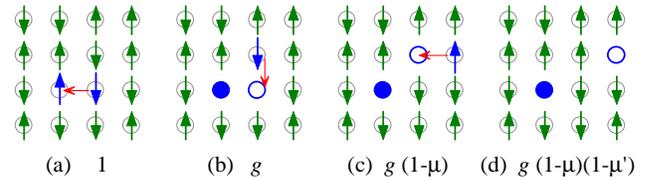} 
\end{center}
\vskip -5mm
\caption{(Color online) 
Weight assignment of Jastrow factor 
${\cal P}={\cal P}_Q{\cal P}_{\rm G}$ depending on local electron 
configuration. 
Each circle indicates a site. 
A solid (open) circle indicates a doublon (holon). 
Thin arrows denote virtual hopping processes in the strong-coupling 
expansion. 
(a) Configuration with no doublon; 
a basis for $U/t\rightarrow\infty$. 
(b) A doublon sits at a nearest neighbor of a holon;  
a virtual state in the second order of strong-coupling expansion in $t/U$.  
(c) A doublon sits at a diagonal neighbor of a holon; 
a virtual state in the second (fourth) order in $t'/U$ ($t/U$).  
(d) A doublon sits at a farther site from a holon; 
a higher-order virtual state. 
For the case of eq.~(\ref{eq:PQ1}), we set $\mu'=0$. 
}
\label{fig:weight}
\end{figure}

Although the electron-electron interaction in the Hubbard model is 
limited to within a single site, its effect reaches distant sites. 
Therefore, to overcome the above shortcomings of ${\cal P}_{\rm G}$, 
one needs to add intersite correlation (long-range Jastrow) factors. 
In cases of low electron density, distance-dependent long-range 
Jastrow factors are useful. \cite{YS3} 
On the other hand, at half filling, the short-range part of the Jastrow 
factor is predominant owing to the screening effect. 
Castellani \etal\ \cite{Castellani} derived an effective Hamiltonian 
of the Hubbard model, taking account of both spin and charge degrees 
of freedom. 
In their effective Hamiltonian, an exchange term between a doubly 
occupied site (doublon) and an empty site (holon) appears, indicating 
that a doublon-holon correlation is inherent in the Hubbard model. 
Kaplan \etal\ \cite{Kaplan} actually showed by studying one-dimensional 
(1D) small clusters that the binding of a doublon to a holon is 
important for large $U/t$ to reduce the energy at half filling. 
Using exact diagonalization, Yokoyama and Shiba \cite{YS3} studied 
the ground-state wave function of Hubbard rings at half filling, at 
which the ground state is known to be insulating for $U/t>0$. \cite{LiebWu} 
They found that, for large $U/t$, the magnitudes of the coefficients 
of bases with one doublon (and one holon) decrease exponentially 
as a function of the distance between doublon and holon. 
This means that a doublon is bound to a holon within the decay 
distance in an insulating state. 
\par

A simplified wave function that reflects the above arguments is 
written as $\Psi_Q={\cal P}_Q{\cal P}_{\rm G}\Phi$. \cite{Kaplan,YS3} 
Here, the doublon-holon binding factor ${\cal P}_Q$ is limited to 
the nearest-neighbor part:
\begin{eqnarray}
{\cal P}_Q=
\prod_i{\bigl(1-\mu Q_i^{\tau} \bigr)}, 
\label{eq:PQ1}
\end{eqnarray}
\begin{eqnarray}
Q_i^{\tau}=\prod_{\tau} 
\left[d_i(1-e_{i+\tau})+e_i(1-d_{i+\tau})\right], 
\label{eq:Qi}
\end{eqnarray} 
where $d_i=n_{i\uparrow}n_{i\downarrow}$, 
$e_i=(1-n_{i\uparrow})(1-n_{i\downarrow})$, and $\tau$ varies over all the 
nearest neighbors. 
A variational parameter $\mu$ ($0\le\mu\le 1$) controls the strength of 
doublon-holon binding in the nearest-neighbor sites; 
as $\mu$ increases, a doublon tends to adhere to a holon, and for 
$\mu\rightarrow 1$, a doublon cannot leave a holon, as shown 
in Fig.~\ref{fig:weight}. 
\par

This Jastrow factor, ${\cal P}_Q{\cal P}_{\rm G}$, can also be derived 
naturally from the strong-coupling expansion. 
It is known that the GWF with $g=0$ is an extremely good trial state for 
the 1D Heisenberg model, \cite{GJR,YO1D} and that the Gutzwiller-type 
functions, ${\cal P}_{\rm G}\Phi_{\rm AF}$ and 
${\cal P}_{\rm G}\Phi_{\rm BCS}$ 
($\Phi_{\rm AF}$: Hartree-Fock-type AF state; $\Phi_{\rm BCS}$: 
$d_{x^2-y^2}$-wave BCS state), yield quantitatively reasonable results 
for the 2D $t$-$J$ model. \cite{YSSC,Gros,YO2D} 
These favorable properties of the GWF for strong-coupling models 
can be applied to the Hubbard model by considering a canonical 
transformation, 
${\cal H}_{t-J}\sim e^{iS}{\cal H}_{\rm Hub} e^{-iS}$ 
($t/U\rightarrow 0$), \cite{Harris} and similarly, 
\begin{equation}
\frac{\langle\Psi_{\rm G}|{\cal H}_{t-J}|\Psi_{\rm G}\rangle}
{\langle\Psi_{\rm G}|\Psi_{\rm G}\rangle}
\sim
\frac{\langle\Psi_{\rm G}e^{iS}|{\cal H}_{\rm Hub}|e^{-iS}\Psi_{\rm G}\rangle}
{\langle\Psi_{\rm G}e^{iS}|e^{-iS}\Psi_{\rm G}\rangle}. 
\end{equation}
Thus, an improved wave function for the Hubbard model is given by 
applying the strong-coupling expansion, $e^{-iS}$, to a Gutzwiller-type 
function $\Psi_{\rm G}$ ($={\cal P}_{\rm G}\Phi$). 
Considering virtual hopping processes in this expansion, as shown 
in Fig.~\ref{fig:weight}, one easily notices that the first-order terms 
of this expansion roughly correspond to ${\cal P}_Q\Psi_{\rm G}$. 
In addition, it was found that a more direct form of $e^{-iS}\Psi_{\rm G}$ 
yields improved results \cite{Otsuka} similar to those of 
${\cal P}_Q\Psi_{\rm G}$, mentioned in the following sections. 
\par 

In the early VMC study of the projected Fermi sea, 
$
\Psi_Q^{\rm FS}={\cal P}_Q{\cal P}_{\rm G}\Phi_{\rm F}, 
$ 
for the 1D and 2D square lattices, Yokoyama and Shiba \cite{YS3} 
concluded that $\Psi_Q^{\rm FS}$ corrects the shortcomings (1) and (2), 
mentioned above, of the GWF, but a Mott transition does not arise, even 
in $\Psi_Q^{\rm FS}$. 
Subsequently, Millis and Coppersmith \cite{Millis} also concluded, 
by calculating the zero-frequency part of the optical conductivity 
using a VMC technique, that a Mott transition cannot be described in 
terms of this type of wave function. 
However, as we have repeatedly explained in previous papers, 
\cite{Yoko,Wataorg} 
these early studies were not sufficiently thorough or careful 
to arrive at the correct conclusion that the doublon-holon binding 
factor ${\cal P}_Q$ is essential for describing a Mott transition. 
Actually, the existence of Mott transitions has been confirmed 
using ${\cal P}_Q$ for various systems, \cite{note1D} such as the 
square lattice, \cite{Yoko,YTOT} the anisotropic triangular lattice, 
\cite{Wataorg} 
the kagom\'e lattice, \cite{kagome} the checker-board lattice, 
\cite{CB} and a degenerate Hubbard model on the square lattice. 
\cite{twoband}
\par

\subsection{\label{sec:WFnow}Wave functions and VMC conditions}

In this paper, we continue to study the Mott transition induced by 
${\cal P}_Q$. 
Considering the lattice structure shown in Fig.~\ref{fig:model}(a), 
we introduce into ${\cal P}_Q$ the effect of doublon-holon binding 
between diagonal-neighbor sites $\mu'$ [${\bf r}=(\pm x,\pm y)$], 
in addition to that of the nearest neighbors, 
$\mu$ [${\bf r}=(\pm x,0)$ and $(0,\pm y)$]: 
\begin{eqnarray}
{\cal P}_Q=
\prod_i{\bigl(1-\mu Q_i^{\tau} \bigr)
\bigl(1-\mu'Q_i^{\tau'}\bigr)}, 
\label{eq:PQ}
\end{eqnarray}
\begin{eqnarray}
Q_i^{\tau(\tau')}=\prod_{\tau(\tau')} 
\left[d_i(1-e_{i+\tau(\tau')})+e_i(1-d_{i+\tau(\tau')})\right], 
\label{eq:Qi}
\end{eqnarray} 
in which $\tau$ ($\tau'$) varies over all the adjacent sites in the bond 
directions of $t$ ($t'$). 
The weight assignment of the correlation factors 
${\cal P}={\cal P}_Q{\cal P}_{\rm G}$ is explained in Fig.~\ref{fig:weight}. 
For the pure square lattice ($t'=0$), we use eq.~(\ref{eq:PQ1}) instead of 
eq.~(\ref{eq:PQ}) for simplicity, because we can confirm that 
the effect of $\mu'$ is negligible, even quantitatively, for $t'=0$. 
The wave function we deal with in this paper is 
\begin{equation}
\Psi_Q={\cal P}_Q{\cal P}_{\rm G}\Phi. 
\end{equation}
\par

As a one-body part $\Phi$, we primarily study a fixed-density BCS 
state: \cite{Lhuillier} 
\begin{equation}
\Phi_d =\left(\sum_{\bf k}a_{\bf k}
c_{{\bf k}\uparrow}^\dagger c_{{\bf -k}\downarrow}^\dagger
\right)^\frac{N_{\rm e}}{2}|0\rangle, 
\label{eq:singlet}
\end{equation}
\begin{equation}
a_{\bf k}=\frac{v_{\bf k}}{u_{\bf k}}=\frac{\Delta_{\bf k}}
{\varepsilon_{\bf k}-\zeta+
\sqrt{(\varepsilon_{\bf k}-\zeta)^2+\Delta_{\bf k}^2}}, 
\label{eq:BCSDelta}
\end{equation}
where $\zeta$ is a variational parameter that is reduced to the
chemical potential for $U/t\rightarrow 0$. 
Since we know that, at half filling, the simple $d_{x^2-y^2}$ wave is 
the most stable among the various gap shapes, \cite{YSSC,Gros,Hirashima} 
here we exclusively consider the $d$-wave gap: 
\begin{equation}
\Delta_{\bf k}=\Delta_d(\cos k_x-\cos k_y). 
\label{eq:gap}
\end{equation} 
Note that although the variational parameter $\Delta_d$ 
indicates the magnitude of the $d$-wave gap, a state 
($\Psi_Q^d={\cal P}\Phi_d$) with finite $\Delta_d$ does not necessarily 
mean a SC state. \cite{ZGRS} 
We fix the value of $t'$ in $\varepsilon_{\bf k}$ in the wave function 
[eq.~(\ref{eq:BCSDelta})] at the same value as that in the Hamiltonian, 
because the renormalization of $\varepsilon_{\bf k}$ \cite{Himeda-t'} 
is not significant when the system is conductive, \cite{Wataorg} 
and $\zeta$ compensates this effect to some extent in the insulating 
regime, as we will see later. 
For $\Delta_d=0$, $\Phi_d$ is reduced to $\Phi_{\rm F}$, which is 
explained next. 
\par 

As a reference state, we also study the Fermi sea, 
\begin{equation} 
\Phi_{\rm F}=\prod_{k<k_{\rm F},\sigma}
c_{{\bf k}\sigma}^\dagger|0\rangle. 
\label{eq:FS}
\end{equation} 
A complication is that 
$\Psi_Q^{\rm FS}$ ($={\cal P}_Q{\cal P}_{\rm G}\Phi_{\rm FS}$) 
does not merely represent a normal state; $\Psi_Q^{\rm FS}$ also undergoes 
a Mott transition, as we found for the attractive Hubbard model. 
\cite{Yoko} 
The results of the attractive model for the symmetric case ($t'=0$ and 
$n=1$) can be exactly mapped to those of the repulsive model through 
a canonical transformation. \cite{Shiba} 
As an extension in this paper, we study the properties of $\Psi_Q^{\rm FS}$ 
for asymmetric cases ($t'\ne 0$). 
\par

For the comparison made in \S\ref{sec:SC} and \S\ref{sec:AF}, we consider 
an ordinary mean-field solution $\Phi_{\rm AF}$ \cite{YS2} for a one-body 
AF state with a long-range order. 
In the trial AF state, 
\begin{equation}
\Psi_Q^{\rm AF}={\cal P}_Q{\cal P}_{\rm G}\Phi_{\rm AF}, 
\label{eq:AF} 
\end{equation} 
the AF gap $\Delta_{\rm AF}$ is optimized as a parameter, but $t'$ 
in $\varepsilon_{\bf k}$ is fixed at the model value, similarly to 
that in $\Psi_Q^d$. 
\par

Because our trial functions have at most five parameters to be 
optimized [$g$, $\mu$, $\mu'$, $\Delta_{d({\rm AF})}$, $\zeta$], 
we have used a simple version of optimization VMC methods, \cite{OVMC} 
namely a line minimization of one parameter with the others fixed. 
\cite{Numerical}
In one round of iteration, every parameter is optimized once. 
In most cases of optimization in this study, the parameters 
converge within 2 or 3 rounds, after which we continue the optimization 
process for another 15-20 rounds. 
The optimized values of the parameters and energy are determined 
by averaging the results of these rounds after convergence. 
Because each optimization procedure is carried out with typically 
$2.5\times10^5$ ($L=10$-14) samples, preserving the acceptance 
ratio of 0.5, our data are practically the averages of 3-5 million 
samples. 
Thereby, the accuracy in the total energy is markedly increased, 
typically to the order of $10^{-4}t$. 
Because the convergence of optimization becomes very slow near phase 
transitions, particularly continuous transitions, we carried out 
longer iterations ($\sim 50$ rounds) in such cases. 
With the optimized parameters thus determined, physical quantities are 
calculated in different VMC simulations with $2.5\times 10^5$ samples. 
We used lattices of $L\times L$ sites ($L=6$-18, mainly 10-14) with 
periodic-antiperiodic boundary conditions. 
Because the closed-shell condition cannot be satisfied in a wide 
range of asymmetric model parameters ($t'\ne 0$) at half filling, 
we are often obliged to calculate with open shells. 
\par

Finally, we mention finite-size analysis in this study. 
In the symmetric case ($t'=0$), the system-size dependence of various 
quantities is often monotonic, because the {\bf k}-point structure 
included in the Fermi surface becomes systematic with increasing $L$. 
In contrast, in asymmetric cases, the {\bf k}-point structure 
is unique to each system size $L$ and frustration strength $t'$; 
therefore, the system-size dependence becomes irregular, and the 
$t'/t$ dependence is not smooth. 
\par

%
\section{\label{sec:MottSC}Mott Transitions in $d$-Wave State}
In this section, we study Mott transitions arising in the $d$-wave 
singlet state $\Psi_Q^d$. 
In \S\ref{sec:energy}, we show that a first-order transition occurs 
by observing hysteresis in the $U/t$ dependence of total energy. 
In \S\ref{sec:para}, we identify this critical behavior as a Mott 
transition without relevance to magnetic orders by studying various 
quantities. 
In \S\ref{sec:mottprop}, we consider the properties of this transition 
with reference to other studies. 
\par

\begin{figure}[hob]
\vspace{-0.2cm}
\begin{center}
\includegraphics[width=8.7cm,clip]{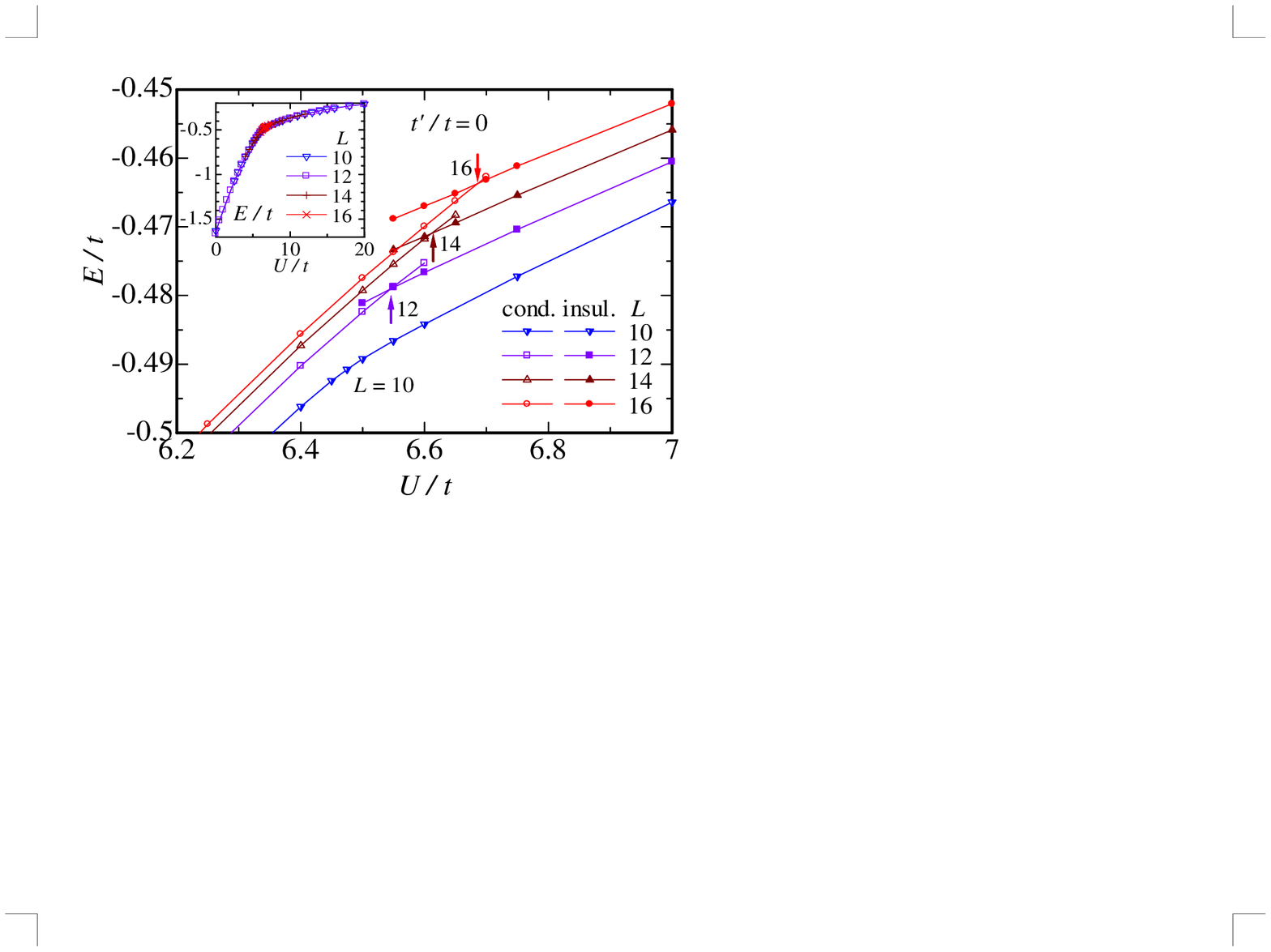}
\end{center}
\vskip -5mm
\caption{(Color online)
Total energies of $d$-wave state for pure square lattice ($t'=0$) 
for four system sizes near critical points ($U_{\rm c}$), indicated 
by arrows. 
Hysteresis is observed for $L=12$-16; the local minima 
for the conductive (insulating) sides are denoted by open (solid) 
symbols. 
For $L=10$, hysteresis is not observed using the 
present VMC calculations, and the parameters vary continuously. 
The critical values are $U_{\rm c}/t=6.54$, 6.61 and 6.69 for 
$L=12$, 14 and 16, respectively. 
Inset: The behavior of $E/t$ over a wider range of $U/t$. 
}
\label{fig:hyst0}
\end{figure}
%
\begin{figure*}[!t]
\vspace{-0.2cm}
\begin{center}
\includegraphics[width=17cm,clip]{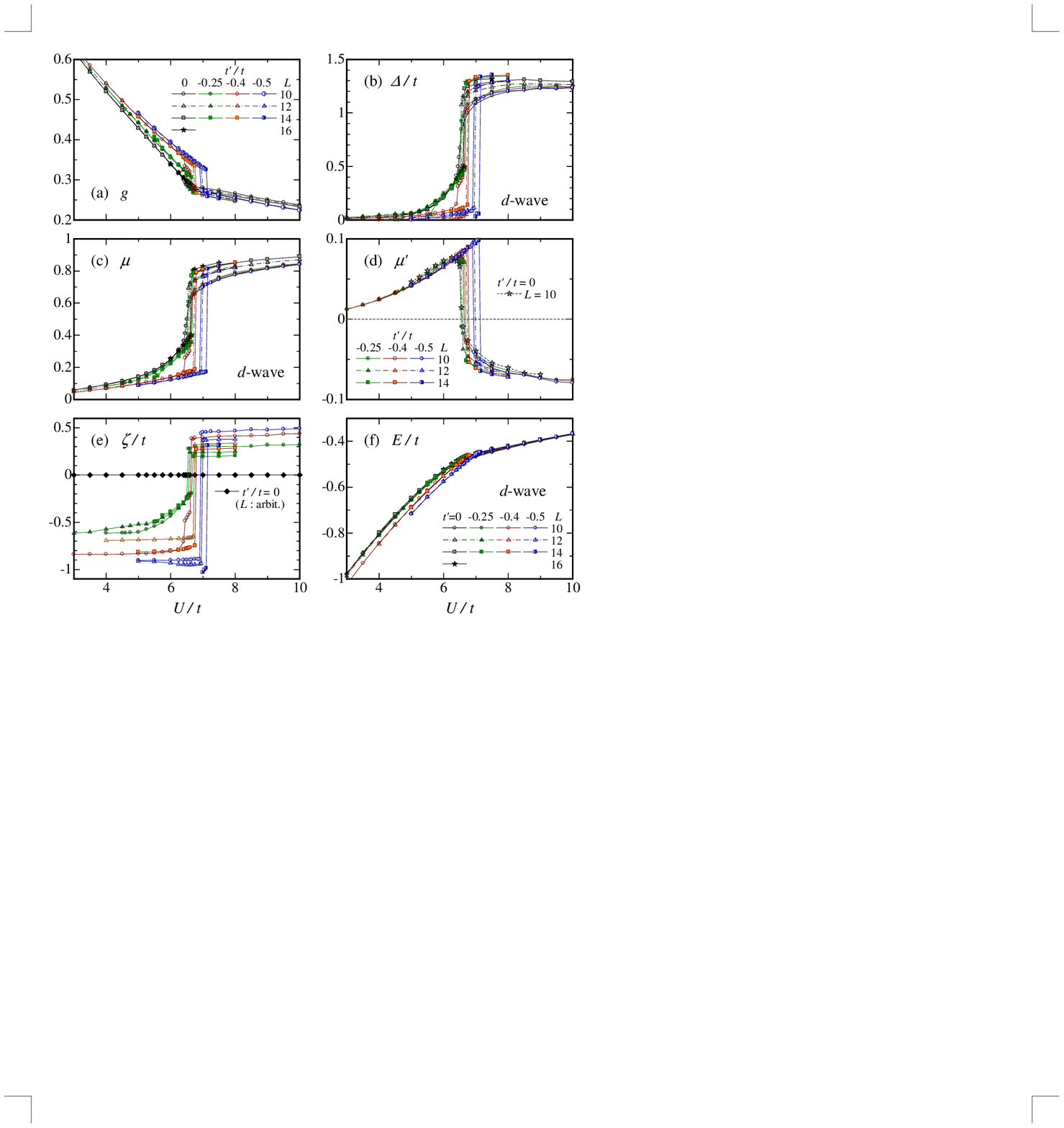}
\end{center}
\vskip -5mm
\caption{(Color online) 
(a)-(e) Optimized variational parameters (at the global minima of $E/t$) 
for $d$-wave state near critical values $U_{\rm c}$ as function 
of $U/t$: 
(a) Onsite correlation (Gutzwiller) factor.
(b) $d$-wave gap parameter. 
(c) Doublon-holon binding factor between nearest-neighbor sites. 
(d) The same factor between diagonal-neighbor sites. 
(e) Chemical potential. For $t'=0$, $\zeta$ is always zero owing to 
the band symmetry. 
(f) Total energies for four values of $t'/t$. 
The symbols and abscissa scales are common to all the panels. 
Data for four values of $t'/t$, and for three or four system sizes 
($L\times L$) are compared. 
}
\label{fig:para}
\end{figure*}
\subsection{Total energy and hysteresis behavior\label{sec:energy}}
First, let us consider the behavior of the optimized total energy 
per site, $E$. 
In the inset of Fig.~\ref{fig:hyst0}, $E/t$ for $t'/t=0$ is plotted 
for a wide range of $U/t$ and four system sizes ($L$). 
On this scale, $L$ dependence is imperceptible. 
One readily notices a cusp at $U/t\sim 6.5$, where the data 
points of various $L$ are concentrated. 
In the main panel of Fig.~\ref{fig:hyst0}, this cusp is magnified. 
We have optimized the wave function successively, from both weak- and 
strong-correlation sides toward this cusp. 
For the system with $L=10$, the optimized energies from the two sides 
coincide, and $E/t$ becomes a smooth function of $U/t$. 
On the other hand, for larger systems ($L\ge 12$ in Fig.~\ref{fig:hyst0}), 
the optimized values 
in the weak-correlation side are not smoothly connected to those in 
the strong-correlation side, and vice versa. \cite{notehyst} 
At the cusp, all optimized variational parameters have 
discontinuities, as shown in Figs.~\ref{fig:para}(a)-\ref{fig:para}(e). 
Such hysteresis and discontinuities indicate that a first-order phase 
transition occurs at the cusp point ($U=U_{\rm c}$). 
In the following subsection, we identify this transition as a Mott 
(conductor-to-nonmagnetic-insulator) transition. 
\par

\begin{figure}[hob]
\vspace{-0.2cm}
\begin{center}
\includegraphics[width=8.7cm,clip]{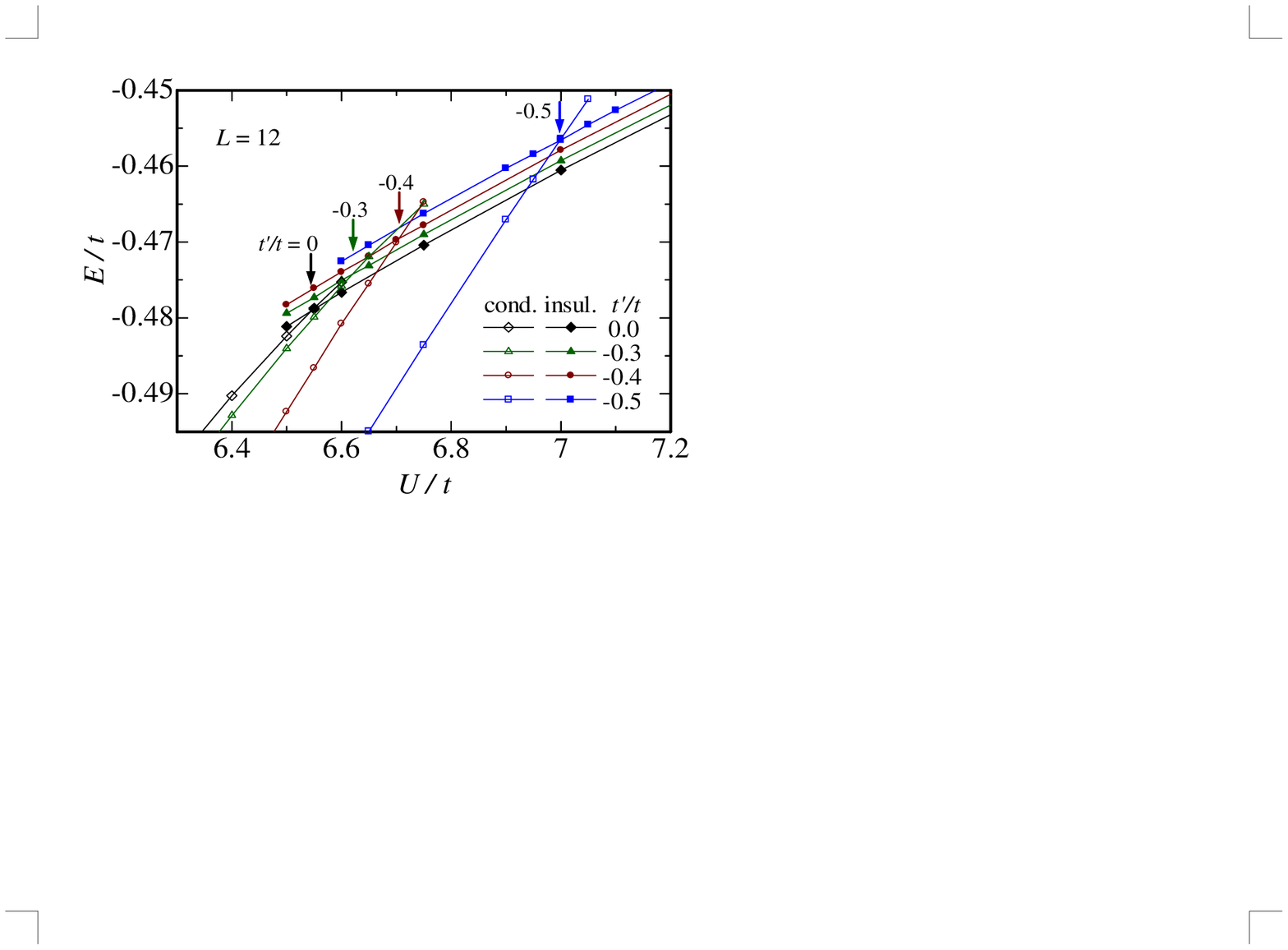}
\end{center}
\vskip -5mm
\caption{(Color online)
Total energies for four values of $t'/t$ near critical points 
indicated by arrows. 
The system size is fixed at $L=12$. 
Hysteresis is observed for each case. 
The local minima for the conductive (insulating) sides are shown by 
open (solid) symbols. 
The critical values are $U_{\rm c}/t=6.54$, 6.62, 6.70 and 7.00 
for $t'/t=0$, -0.3, -0.4 and -0.5, respectively. 
}
\label{fig:hysta}
\end{figure}
%
\begin{figure}[hob]
\vspace{-0.2cm}
\begin{center}
\includegraphics[width=8.7cm,clip]{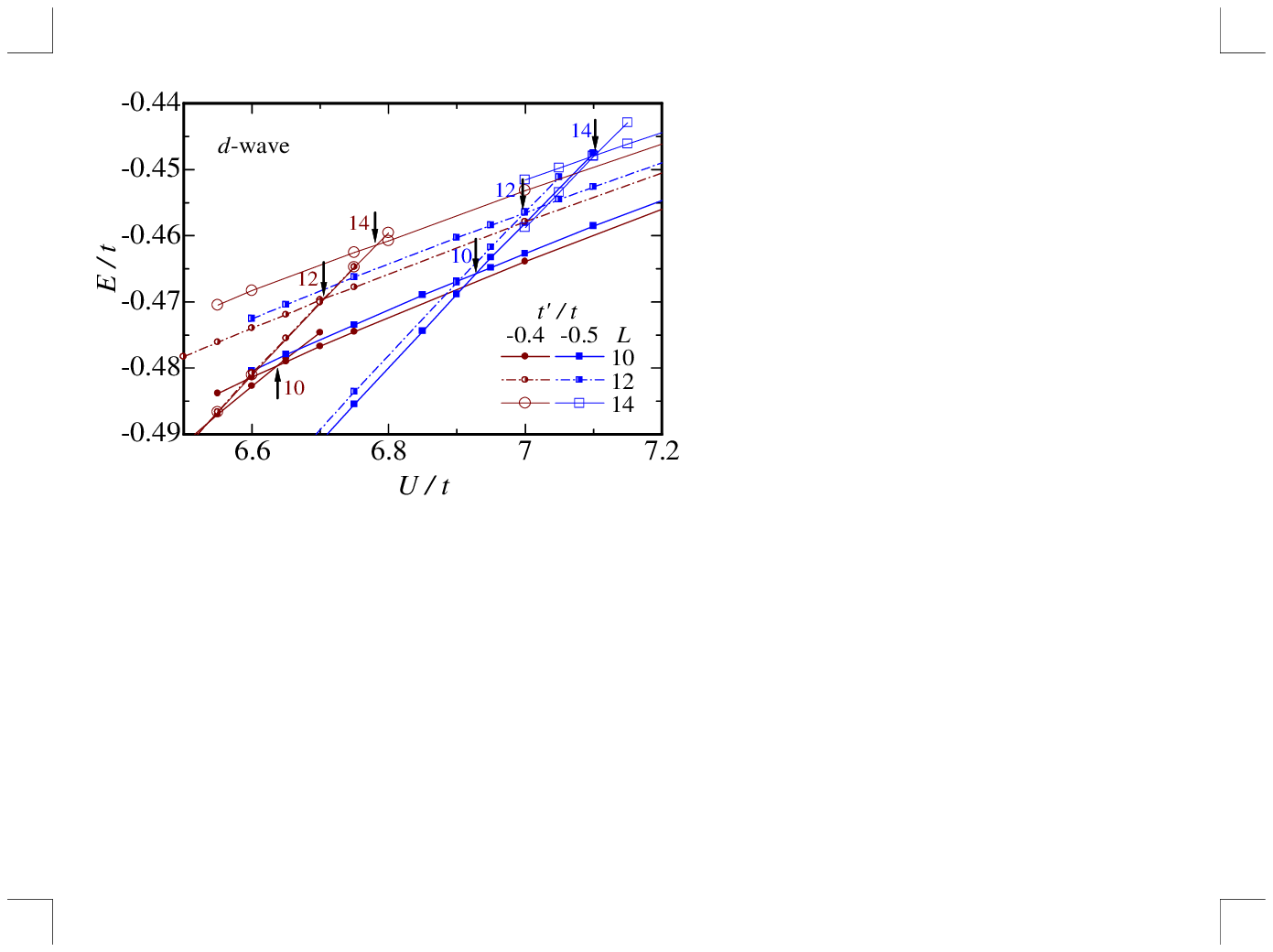}
\end{center}
\vskip -5mm
\caption{(Color online)
Total energies for strongly frustrated cases ($t'/t=-0.4$ and -0.5) 
near critical points, indicated by arrows, for three system sizes. 
Hysteresis is observed for every case; the conductive 
and insulating cases are denoted by the same symbols. 
The critical values are $U_{\rm c}/t=6.64$, 6.70 and 6.78 for 
$L=10$, 12 and 14, respectively, for $t'/t=-0.4$, and similarly, 
$U_{\rm c}/t=6.93$, 7.00 and 7.10 for $t'/t=-0.5$. 
}
\label{fig:hystl}
\end{figure}
%
Next, we study the $L$ dependence of the critical point. 
We first note that it is essential to check the system-size 
dependence when we consider critical phenomena, as we have learned from 
the distinct behavior between $L=10$ and $L\ge 12$ in Fig.~\ref{fig:hyst0}. 
As the system size increases ($L\ge 12$), the critical point shifts 
to a larger value of $U/t$ (Fig.~\ref{fig:hyst0}). 
This is partly because a large system size is more advantageous to 
conductive states, which have longer correlation lengths. 
For $t'/t=0$, because the system-size dependence of $E/t$ is fitted 
well with quadratic curves of $1/N_{\rm s}$, the critical value for 
$L=\infty$ can be estimated as $U_{\rm c}/t=7.0\pm 0.1$ using the 
method of least squares. 
\par

Here, we consider the $t'/t$ dependence. 
As shown in Fig.~\ref{fig:para}(f), the total energy $E/t$ also has a 
cusp for finite $t'/t$. 
The behavior of $E/t$ near the cusps is magnified in Fig.~\ref{fig:hysta}
for four values of $t'/t$ and for $L=12$. 
For $U<U_{\rm c}$, $E/t$ is significantly reduced as $|t'/t|$ increases; 
this behavior is common to the noninteracting case, in which the gain 
in $E_{t'}$ exceeds the loss in $E_t$ 
[$E_t$ ($E_{t'}$): the contribution from the $t$ ($t'$) term to 
$\langle{\cal H}_{\rm kin}\rangle$]. 
In contrast, for $U>U_{\rm c}$, $E/t$ changes only very slightly; 
we will return to this point in \S\ref{sec:mottprop}. 
Consequently, as $|t'/t|$ increases, the transition point shifts 
to a larger value of $U/t$, particularly rapidly for large $t'/t$, 
although the bandwidth remains constant, $8t$, for $0\le t'/t\le 0.5$. 
This $t'/t$ dependence of $U_{\rm c}/t$ can be verified from the behavior 
of the parameters [Figs.~\ref{fig:para}(a)-\ref{fig:para}(e)]. 
\par

In Fig.~\ref{fig:hystl}, we show the system-size dependence of $E/t$ 
near $U_{\rm c}/t$ for strongly frustrated cases ($t'/t=-0.4$ and -0.5). 
Here, even the system as small as $L=10$ exhibits clear hysteresis. 
The critical value $U_{\rm c}/t$ tends to increases monotonically 
as $L$ increases, although the extrapolation of $U_{\rm c}/t$ to 
$L=\infty$ is difficult using the present data, because of the nonmonotonic 
system-size dependence, as mentioned. 
However, we predict that $U_{\rm c}/t$ for $L=\infty$ is only 
slightly larger than those for finite $L$, because the rate of increase 
of $U_{\rm c}/t$ with respect to $L$ is similar to that for $t'/t=0$. 
\par

\subsection{Confirmation of Mott transition\label{sec:para}}
In this subsection, we confirm that the above transition is a Mott 
transition by studying various quantities. 
\par

\begin{figure}[hob]
\vspace{-0.2cm}
\begin{center}
\includegraphics[width=8.4cm,clip]{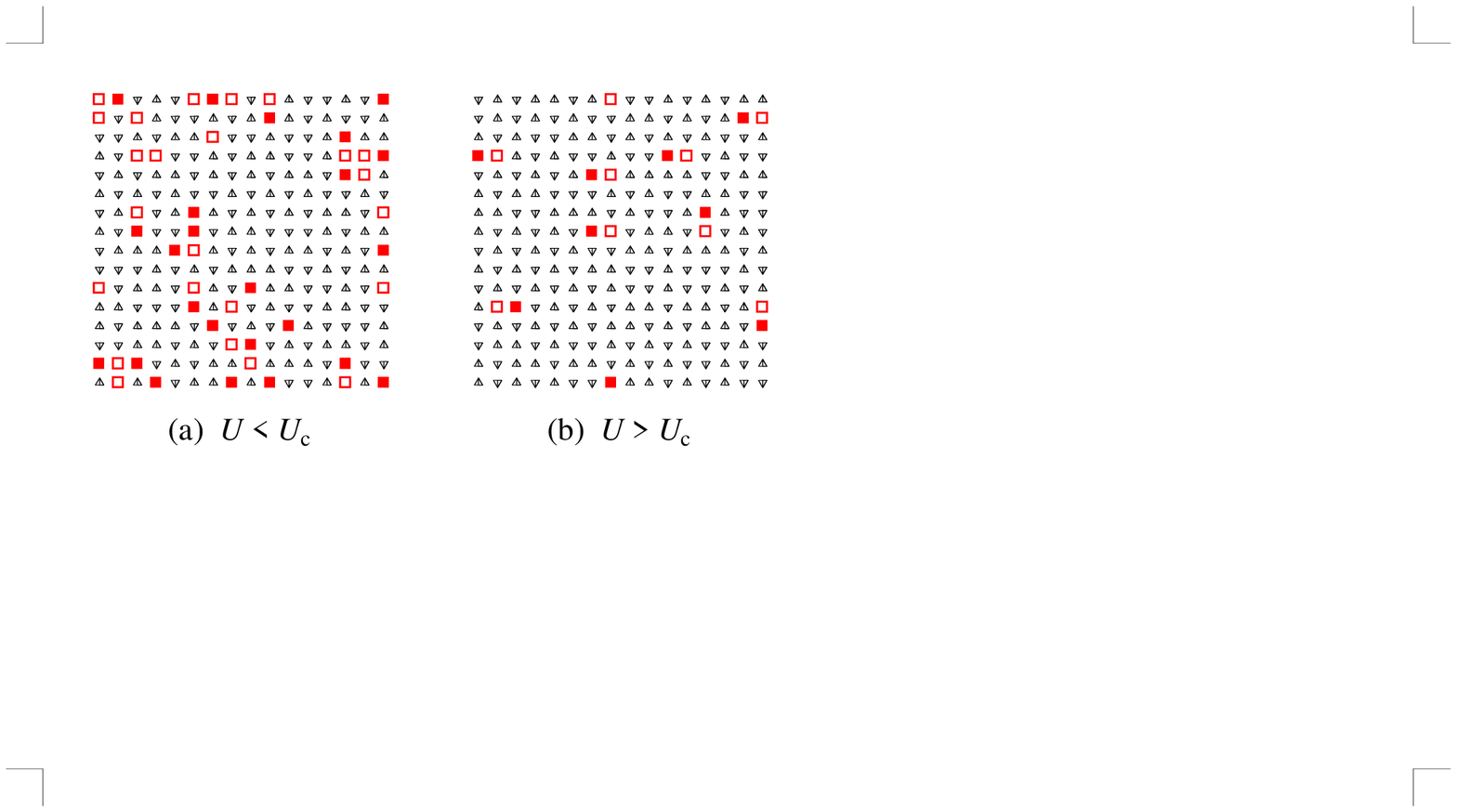}
\end{center}
\vskip -3mm
\caption{(Color online) 
Snapshots (typical electron configurations) taken in VMC 
sampling process of $d$-wave state for $t'/t=0$ and $L=16$
($U_{\rm c}/t=6.69$). 
(a) $U/t=6.25$ ($\mu=0.30$), slightly less than $U_{\rm c}$, and 
(b) $U/t=7.50$ ($\mu=0.85$), slightly greater than $U_{\rm c}$. 
Closed and open squares, upward and downward triangles denote 
doublons, holons, $\uparrow$- and $\downarrow$-spins, respectively. 
}
\label{fig:snap}
\end{figure}

First, we consider the doublon-holon binding factor $\mu$, which 
is a good indicator of the Mott transition. 
In Fig.~\ref{fig:para}(c), the optimized value of $\mu$ is plotted 
as a function of $U/t$. 
For each $t'/t$ and $L$, except for $t'=0$ and $L=10$, $\mu$ has a 
discontinuity and suddenly increases at $U=U_{\rm c}$. 
Note that for $U<U_{\rm c}$, $\mu$ primarily depends on $t'/t$ and 
slightly on $L$ in relation to SC, as mentioned later, whereas for 
$U>U_{\rm c}$, $\mu$ significantly increases as $L$ increases, but 
is almost independent of $t'/t$. 
Thus, in the strong-correlation side $U>U_{\rm c}$, $\mu$ has 
a value close to 1, \cite{notemuss} which means that almost all 
doublons and holons are bound within nearest-neighbor sites. 
This situation is shown in the snapshots taken in the VMC 
sampling process (Fig.~\ref{fig:snap}). 
On the weak-correlation side of $U_{\rm c}$ [Fig.~\ref{fig:snap}(a)], 
where $\mu$ has a relatively small value, doublons (negative charge 
carriers) are often isolated from holons (positive charge carriers). 
Thus, charge can move substantially, and the system is considered 
conductive. 
On the other hand, for $U>U_{\rm c}$ [Fig.~\ref{fig:snap}(b)], 
each doublon is, in most cases, paired with a holon, in addition 
to there being a decrease in the carrier number. 
It follows that free charged particles rarely exist. 
\par

\begin{figure}[hob]
\vspace{-0.2cm}
\begin{center}
\includegraphics[width=8.7cm,clip]{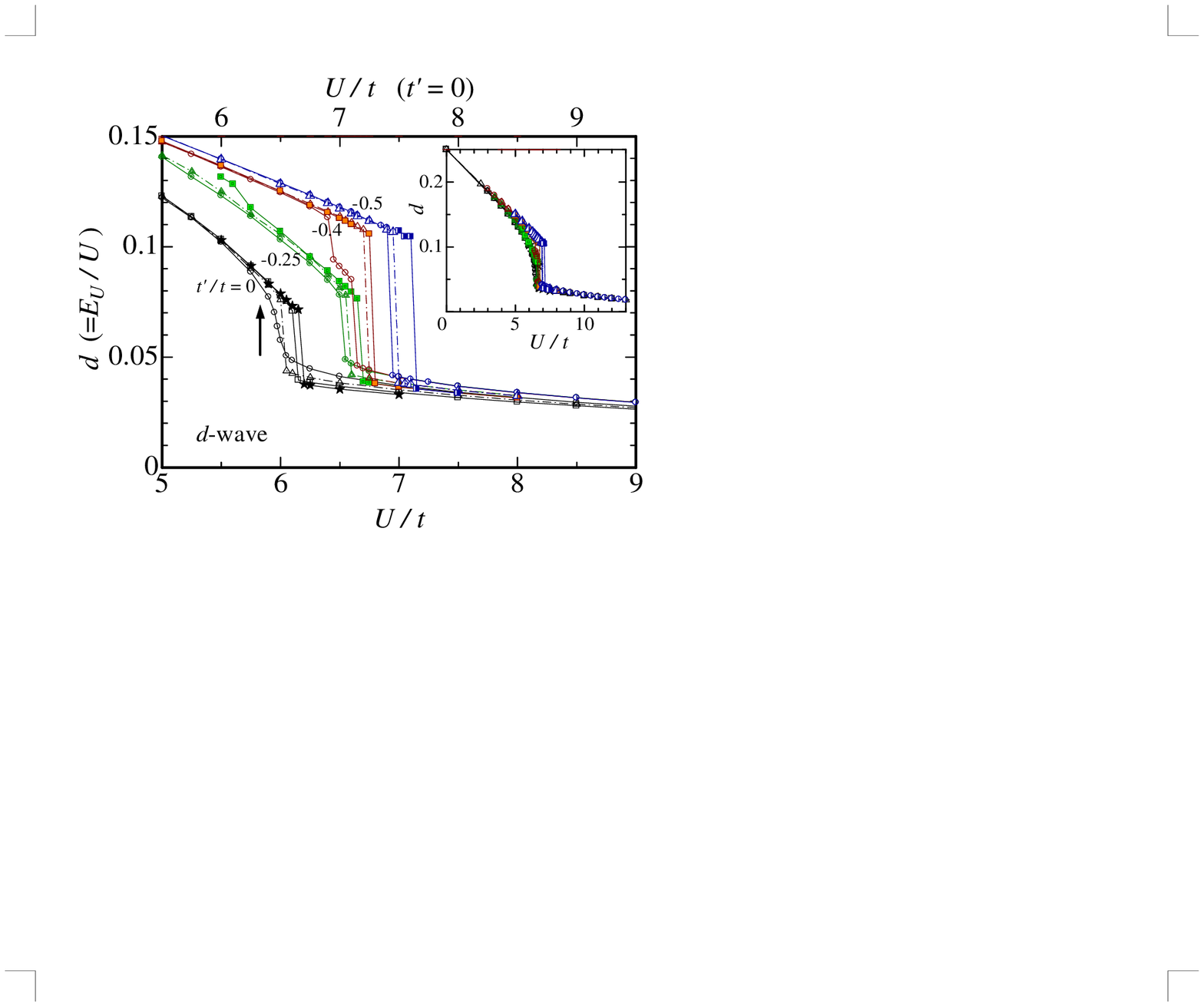}
\end{center}
\vskip -5mm
\caption{(Color online)
Density of doublons (doubly occupied sites) obtained with $\Psi_d$ 
as a function of $U/t$ for four values of $t'/t$. 
For each $t'/t$, various system sizes ($L=10$-16) are simultaneously 
plotted. 
The symbols are common to Fig.~\ref{fig:para}. 
For $t'/t=0$, the horizontal axis is shifted by 0.5 (see upper axis)
for clarity. 
In the inset, the same curves are shown over a wider range. 
}
\label{fig:d-dwave}
\end{figure}

Second, we consider the doublon density, 
\begin{equation}
d=\frac{1}{N_{\rm s}}\sum_i{n_{i\uparrow}n_{i\downarrow}}
=\frac{E_U}{U}, 
\label{eq:doublon}
\end{equation}
which is regarded as the order parameter of Mott transitions, 
\cite{BR,Castellani,Kotliar}
by analogy with the particle density in gas-liquid transitions. 
In eq.~(\ref{eq:doublon}), 
$E_U=\langle{\cal H}_{\rm int}\rangle/N_{\rm s}$. 
In the inset of Fig.~\ref{fig:d-dwave}, $d$ is plotted for four values 
of $t'/t$ and for various system sizes. 
As $U/t$ increases, $d$ decreases linearly from the noninteracting 
value 0.25, but at $U=U_{\rm c}$, it suddenly drops to a considerably 
smaller value, and then decreases slowly for $U>U_{\rm c}$. 
In the main panel of Fig.~\ref{fig:d-dwave}, the vicinity of 
the critical point is magnified; the discontinuity of $d$ at 
$U_{\rm c}$ is clear for each case. 
This abrupt decrease in doublon density can actually be verified in 
Fig.~\ref{fig:snap}, where the number of doublons is 24 ($d=0.094$) for 
$U/t=6.25$, and 9 ($d=0.035$) for $U/t=7.5$, which is consistent 
with the values in Fig.~\ref{fig:d-dwave}. 
Thus, the discontinuity of the doublon density at $U_{\rm c}$ is 
similar to that of the change in particle density in gas-liquid 
transitions. 
\par

\begin{figure}[hob]
\vspace{-0.2cm}
\begin{center}
\includegraphics[width=8.5cm,clip]{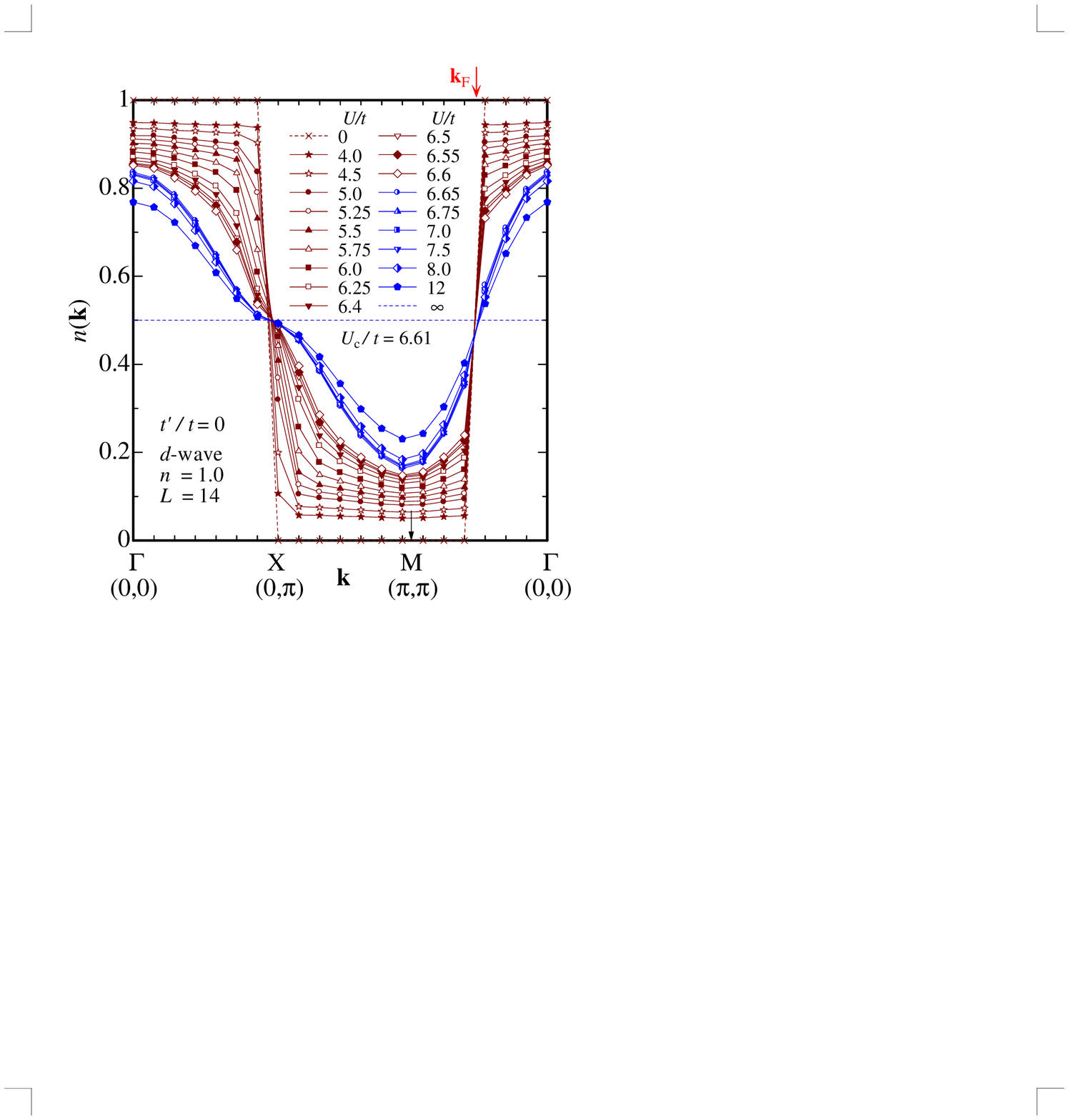}
\end{center}
\vskip -5mm
\caption{(Color online)
Momentum distribution function of the $d$-wave state for various 
values of $U/t$ for $t'/t=0$.
The half-closed symbols and star denote the data for $U>U_{\rm c}$. 
The arrow on the upper axis indicates the position of the Fermi 
surface in the node-of-gap ($\Gamma$-M) direction.
}
\label{fig:nk-d}
\end{figure}

Third, the behavior of the momentum distribution function, 
%
\begin{equation}
n({\bf k}) = \frac{1}{2N_{\rm s}} 
\sum_{k,\sigma}\langle c_{{\bf k}\sigma}^\dag c_{{\bf k}\sigma}\rangle, 
\label{eq:nk}
\end{equation} 
%
at the Fermi surface is another good indicator of a Mott transition. 
In Fig.~\ref{fig:nk-d}, we show the $U/t$ dependence of $n({\bf k})$ 
measured with the optimized parameters along the path 
$\Gamma$(0,0)-X$(\pi,0)$-M$(\pi,\pi)$-$\Gamma$ in the Brillouin zone 
for $t'/t=0$. 
Because the present trial state is a projected $d$-wave, 
there is a node in the gap function in the $\Gamma$-M direction, 
and $n({\bf k})$ has a discontinuity at the Fermi surface, 
${\bf k}_{\rm F}$, in this direction if the system is metallic or SC.
As shown in Fig.~\ref{fig:nk-d}, the discontinuity at ${\bf k}_{\rm F}$ 
is clear for $U<U_{\rm c}$, 
whereas, at $U=U_{\rm c}$, the behavior of $n({\bf k})$ abruptly 
changes, and becomes a smooth function for $U>U_{\rm c}$ also in the 
$\Gamma$-M direction; that is, the Fermi surface disappears. 
Thus, metallic properties are abruptly lost at $U=U_{\rm c}$, even 
in the nodal direction of the $d_{x^2-y^2}$ wave. 
\par

\begin{figure}[hob]
\vspace{-0.2cm}
\begin{center}
\includegraphics[width=8.0cm,clip]{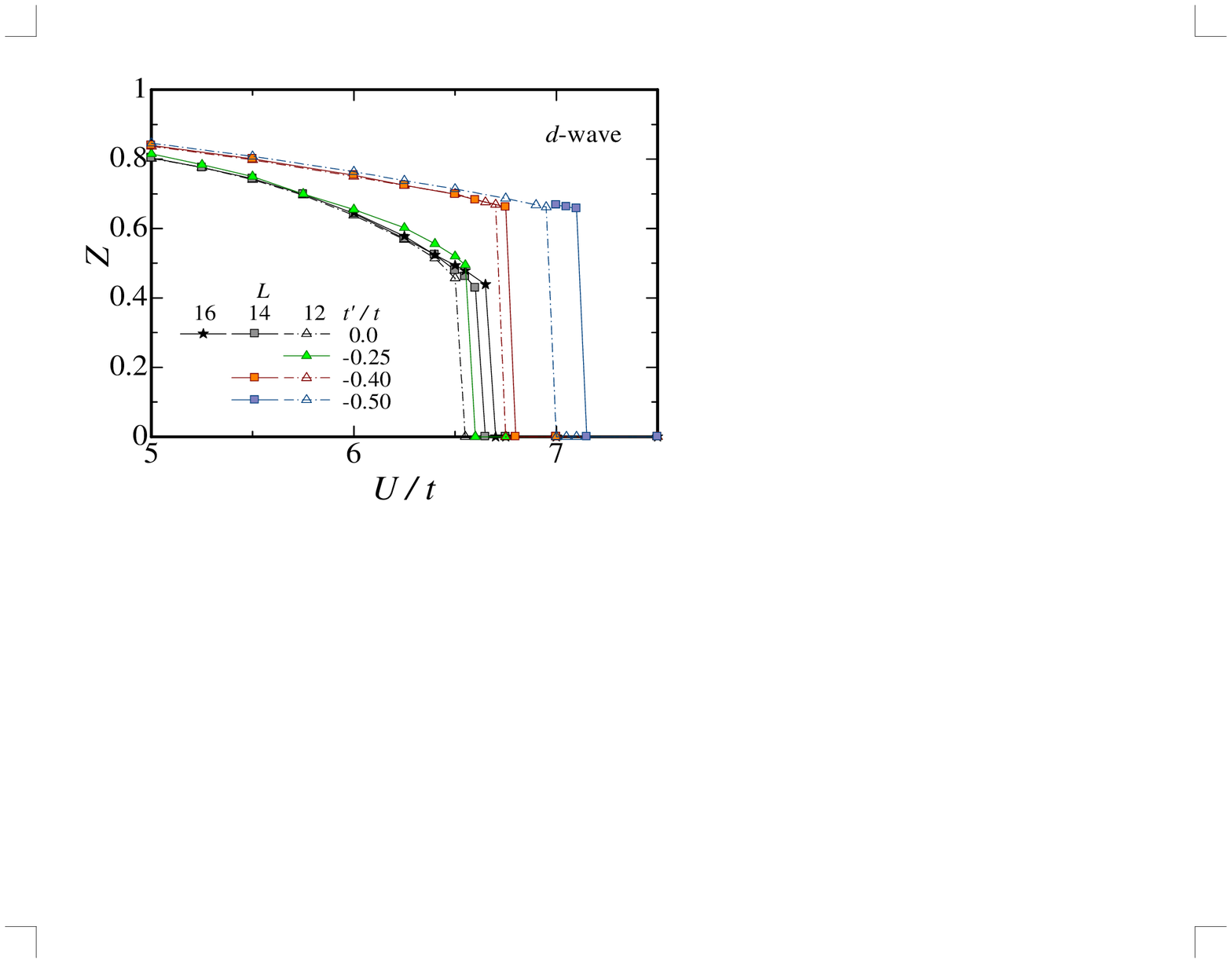}
\end{center}
\vskip -5mm
\caption{(Color online)
Quasi-particle renormalization factor $Z$ of $d$-wave singlet 
states, estimated from discontinuities of $n({\bf k})$ 
in node-of-gap direction. 
Data for four values of $t'/t$ are plotted as a function of $U/t$. 
}
\label{fig:scZ-U}
\end{figure}

To consider this behavior quantitatively, we employ the quasi-particle 
renormalization factor $Z$, which roughly corresponds to the inverse 
of effective mass, unless the ${\bf k}$-dependent renormalization of self 
energy is severe. 
We estimated $Z$ from the magnitude of the jump in $n({\bf k})$ at 
${\bf k}={\bf k}_{\rm F}$ in the nodal direction, and plotted it in 
Fig.~\ref{fig:scZ-U} for four values of $t'/t$. \cite{noteZ} 
For all the values of $t'/t$, $Z$ decreases slowly for $U<U_{\rm c}$, 
whereas at $U=U_{\rm c}$, $Z$ suddenly vanishes with a sizable 
discontinuity, reflecting the first-order character of the transition. 
The system-size dependence of $Z$ is very small, except for minor 
differences near $U_{\rm c}$. 
The behavior of $Z$ strongly suggests that the effective electron mass 
diverges for $U>U_{\rm c}$. 
\par

\begin{figure}[hob]
\vspace{-0.2cm}
\begin{center}
\includegraphics[width=8.5cm,clip]{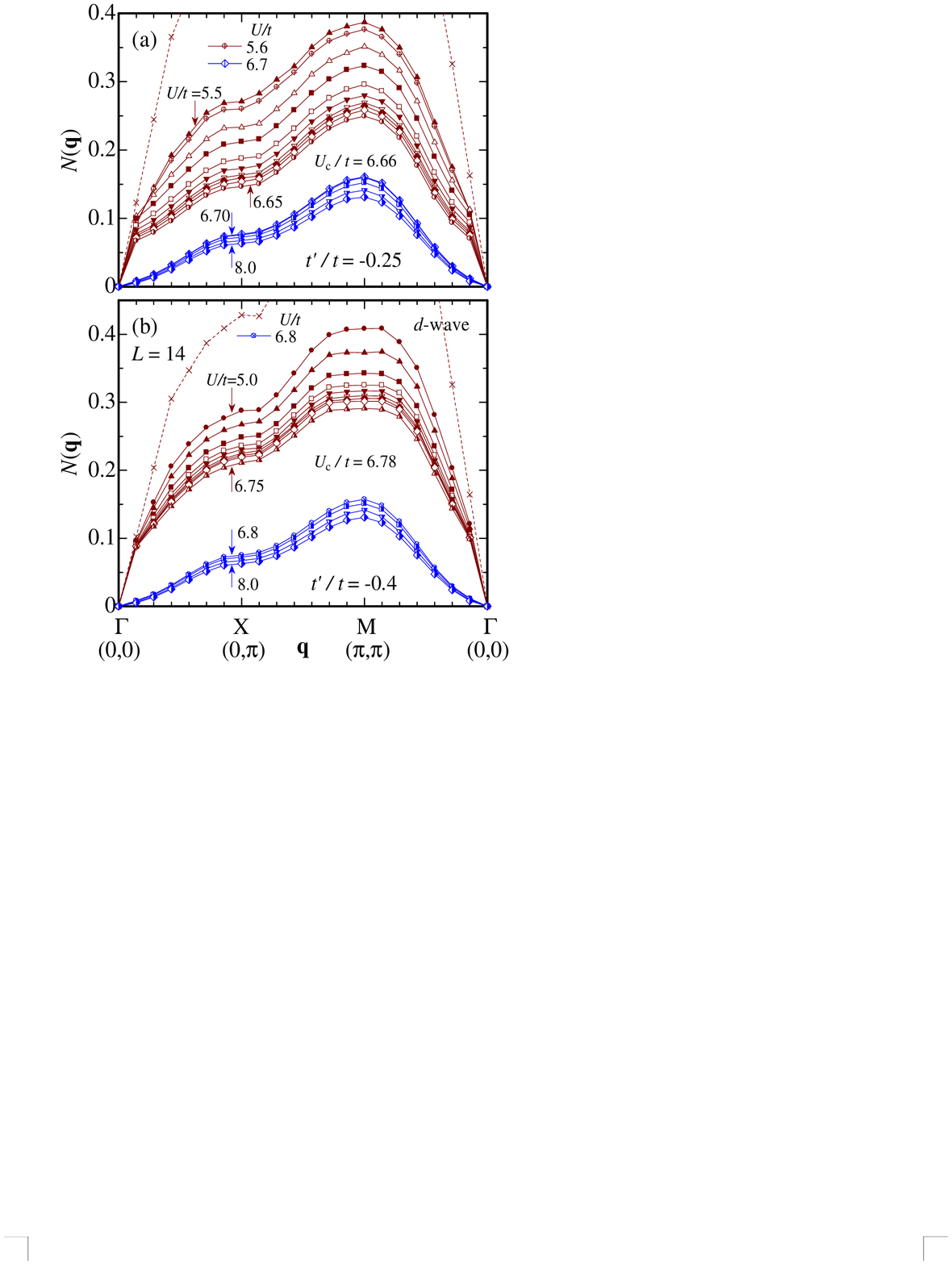}
\end{center}
\vskip -5mm
\caption{(Color online) 
$U/t$ dependence of charge structure factor for $d$-wave 
state ($L=14$) for (a) $t'/t=-0.25$ and (b) $t'/t=-0.4$. 
The two panels are in the same scale. 
The symbols in both panels are common to Fig.~\ref{fig:nk-d} 
except for the values of $U/t$ specified here. 
}
\label{fig:nq-d}
\end{figure}
\begin{figure}[hob]
\vspace{-0.2cm}
\begin{center}
\includegraphics[width=8.0cm,clip]{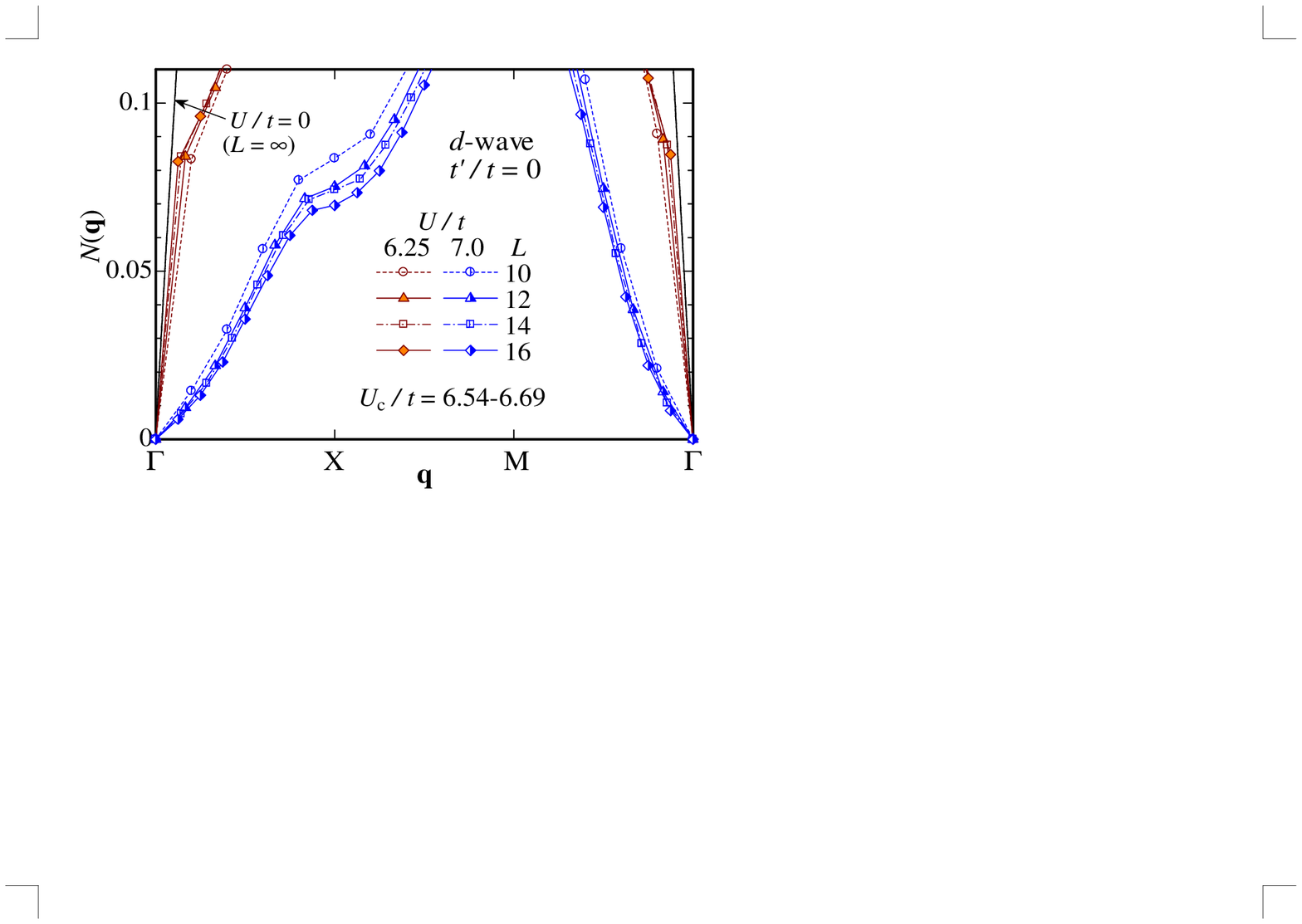}
\end{center}
\vskip -5mm
\caption{(Color online) 
System-size dependence of charge structure factor in $d$-wave 
state ($t'/t=0$) near $\Gamma$ point for $U=6.25t$ $(<U_{\rm c})$ 
and $U=7.0t$ $(>U_{\rm c})$. 
}
\label{fig:nqsize}
\end{figure}

Finally, let us consider the charge structure factor, 
\begin{equation}
N({\bf q})=\frac{1}{N_{\rm s}} 
\sum_{i,j}e^{i{\bf q}\cdot({\bf R}_i-{\bf R}_j)} 
\left\langle{n_{i} n_{j}}\right\rangle - n^2, 
\label{eq:nq}
\end{equation} 
with $n_{i} = n_{{i}\uparrow} + n_{{i}\downarrow}$. 
Within variation theory, it is known that 
$N({\bf q})\propto |{\bf q}|$ for $|{\bf q}|\rightarrow 0$ 
if the state does not have a gap in the charge degree of freedom, 
whereas $N({\bf q})\propto {\bf q}^2$ if a charge gap opens. 
In Fig.~\ref{fig:nq-d}, we show the $U/t$ dependence of $N({\bf q})$ 
for $t'/t=-0.25$ and $-0.4$. 
For $U<U_{\rm c}$, the behavior of $N({\bf q})$ near the $\Gamma$ 
point seems linear in $|{\bf q}|$ for both values of $t'/t$. 
The behavior of $N({\bf q})$ abruptly changes at the critical point, 
and seemingly becomes quadratic in $|{\bf q}|$ for $U>U_{\rm c}$. 
Although it is not easy to definitely determine the power of 
$N({\bf q})$ for the small systems used here, 
we find that the size dependence is different for $U<U_{\rm c}$ 
and for $U>U_{\rm c}$. 
As shown in Fig.~\ref{fig:nqsize}, the behavior of $N({\bf q})$ near 
the $\Gamma$ point for $U<U_{\rm c}$ approaches the analytic curve 
of $U/t=0$ ($L=\infty$) as $L$ increases. 
Conversely, for $U>U_{\rm c}$, the slope of $N({\bf q})$ for 
$|{\bf q}|\rightarrow 0$ becomes smaller as $L$ increases, 
suggesting the quadratic behavior of $N({\bf q})$. 
It follows that $\Psi_Q^d$ is gapless in the charge sector and 
conductive for $U<U_{\rm c}$, but a charge gap probably opens 
for $U>U_{\rm c}$ and $\Psi_Q^d$ becomes insulating. 
\par

Because of the behavior of all the quantities discussed above,
it is appropriate to judge that a first-order Mott transition occurs 
in the $d$-wave singlet state at $U=U_{\rm c}$ for the arbitrary 
values of $t'/t$ considered. 
\par

\subsection{Properties of Mott transitions\label{sec:mottprop}}
Most of the properties of the Mott transition in $\Psi_Q^d$ in the 
present model, eq.~(\ref{eq:model}), are shared with those studied in 
the preceding report for the anisotropic triangular lattice. \cite{Wataorg}
In such cases, we avoid repeating detailed explanations, and only 
give a brief summary. 
\par

\begin{figure}[hob]
\vspace{-0.2cm}
\begin{center}
\includegraphics[width=8.5cm,clip]{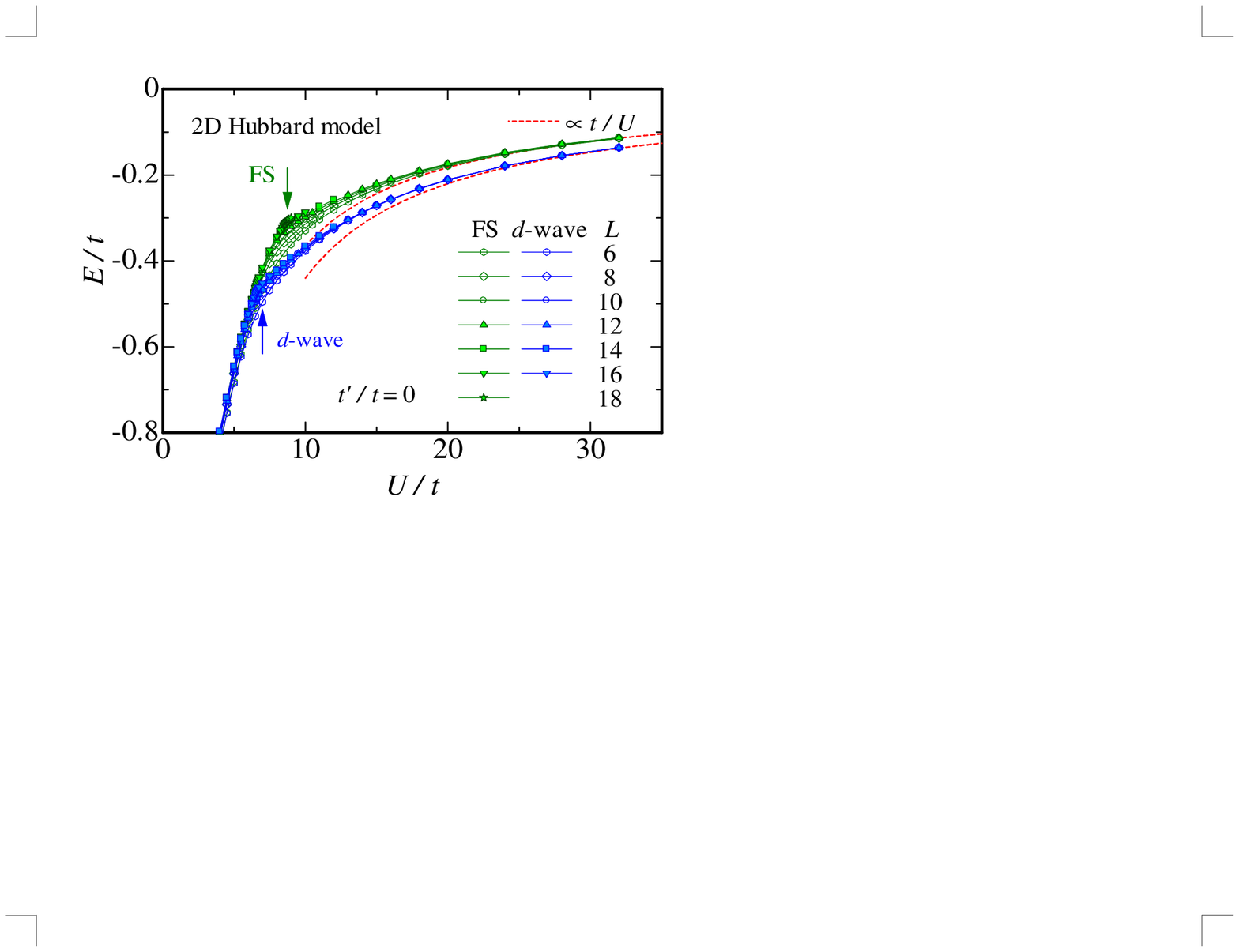}
\end{center}
\vskip -5mm
\caption{(Color online) 
Total energy of projected Fermi sea and $d$-wave states for 
$t'/t=0$ as a function of $U/t$ for several system sizes. 
The Mott critical points for both cases are indicated by arrows. 
Red dashed lines are fitted curves proportional to $t/U$. 
As expected from Fig.~\ref{fig:para}(f), the results for $|t'/t|>0$ 
almost coincide with the present curves for $t'/t=0$ in the insulating 
regime. 
}
\label{fig:etot}
\end{figure}
%
(1) In contrast to the behavior in Brinkman-Rice theory, \cite{BR} 
in which electrons cease moving and doublons completely vanish 
in the insulating regime, the present VMC result exhibits energy 
reduction broadly proportional to 
$-t^2/U$ $(=-J/4)$ for $U>U_{\rm c}$, 
as shown in Fig.~\ref{fig:etot}. 
Thus, as we argued in a previous letter, \cite{YTOT} the results of
strong-coupling theories ($t$-$J$-type models) are qualitatively useful 
for $U>U_{\rm c}$; the value of $U_{\rm c}$ is roughly equal to the 
bandwidth. \cite{Gros} 
\par

(2) As the frustration $t'/t$ increases, the character of the 
first-order phase transition becomes notable. 
For example, as $t'/t$ increases, the hysteresis in $E/t$ is observed 
in smaller systems, and the magnitude of discontinuity at $U=U_{\rm c}$ 
for the variational parameters and quantities such as $d$ and $Z$ 
increases. 
\par

\begin{figure}[hob]
\vspace{-0.2cm}
\begin{center}
\includegraphics[width=8.5cm,clip]{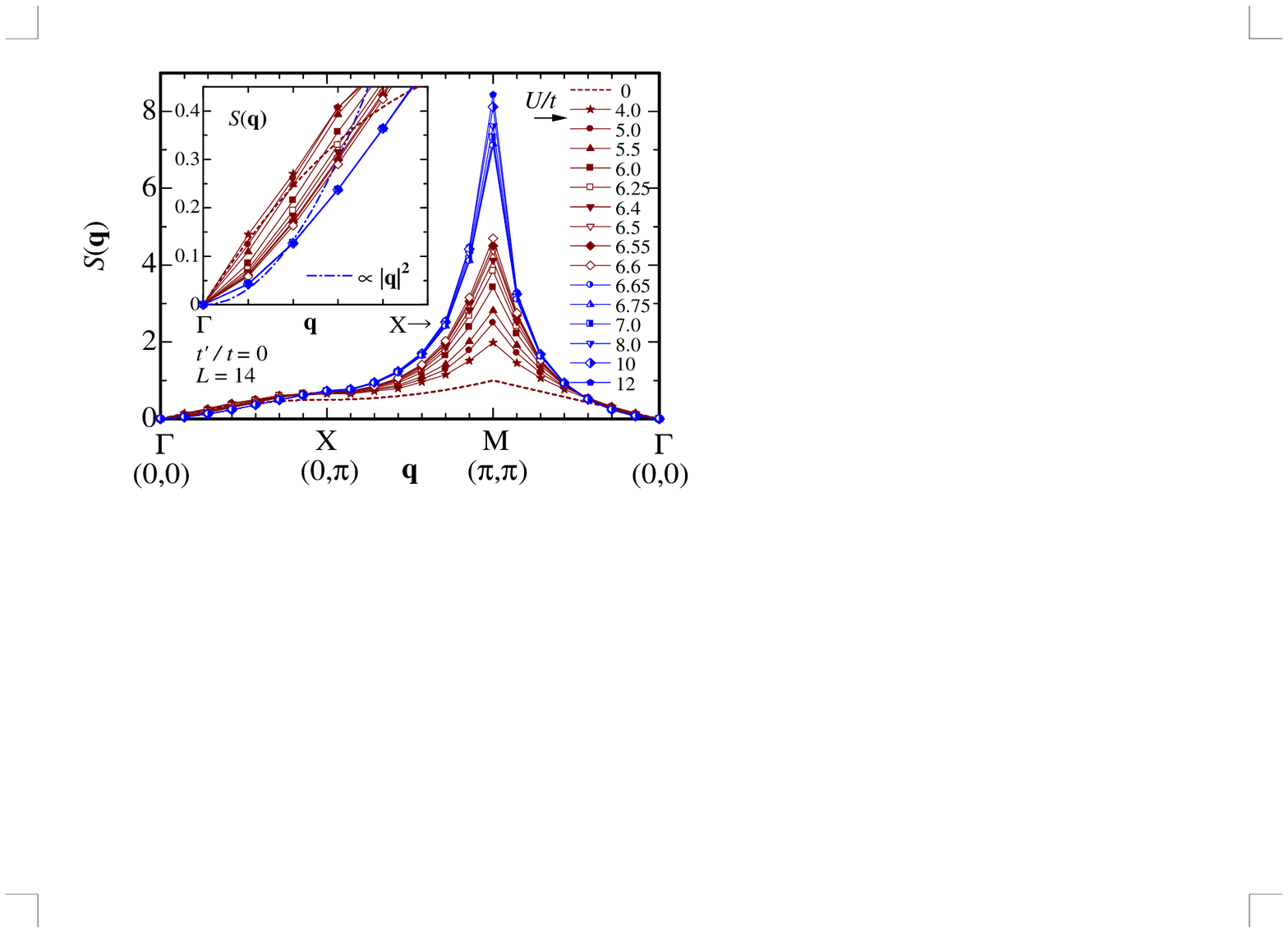}
\end{center}
\vskip -5mm
\caption{(Color online) 
$U/t$ dependence of spin structure factor in $d$-wave state 
for $t'/t=0$ and $L=14$ ($U_{\rm c}/t=6.61$). 
The inset shows the magnification of the region near the $\Gamma$ point 
on the $\Gamma$-X line, and the symbols are common to the main panel. 
$S({\bf q})$ on the $\Gamma$-M line (nodal direction) exhibits basically 
the same behavior for small $|{\bf q}|$. 
Data points of different $U$s for $U>U_{\rm c}$ almost overlap one 
another.
}
\label{fig:sq-d} 
\end{figure}

(3) In the insulating regime, $\Psi_Q^d$ tends to exhibit gaplike 
behavior in the spin structure factor, 
\begin{equation} 
S({\bf q})=\frac{1}{N_{\rm s}}\sum_{ij}{e^{i{\bf q}
\cdot({\bf R}_i-{\bf R}_j)} 
\left\langle{S_{i}^zS_{j}^z}\right\rangle},
\label{eq:sq}
\end{equation} 
for small $|{\bf q}|$. 
As seen in the inset of Fig.~\ref{fig:sq-d}, for $t'=0$, $S({\bf q})$ 
is a linear 
function of $|{\bf q}|$ at $U/t=0$; as $U/t$ increases, $S({\bf q})$ 
becomes a quadratic function of $|{\bf q}|$, suggesting that a SC gap 
opens and becomes large, as will be mentioned in \S\ref{sec:SC}. 
For $U>U_{\rm c}$, the quadratic behavior of $S({\bf q})$ becomes 
clearer; it is possible that a spin gap opens in the insulating 
regime. 
As discussed later, the frustration makes no difference at this point. 
This gaplike behavior is in contrast to the case of $\Psi_Q^{\rm FS}$, 
as will be argued in \S\ref{sec:MottFS}. 
Strictly, however, the insulating state represented by $\Psi_Q^d$ can be 
gapless in the spin sector, in the same manner that the $d$-wave SC is 
gapless in the node direction. 
To settle this point, further studies are necessary. 
\par

(4) In the preceding study for the anisotropic triangular lattice, 
\cite{Wataorg} a band renormalization effect \cite{Himeda-t'} is 
taken into account by optimizing $t'$ in $\Psi_Q^d$ as a variational 
parameter, independently of $t'$ given in the Hamiltonian. 
In the insulating regime, the effective $t'$ is significantly reduced 
to an almost nonfrustrated value, namely $t'/t\sim 0$. 
Thereby, the Fermi surface almost recovers the nesting condition 
for the square lattice, leading to highly developed short-range 
AF correlation. 
In the trial states studied here, the band renormalization effect 
is not included, but the various results are quantitatively similar 
to those of the previous study, and the AF correlation considerably 
increases for $U>U_{\rm c}$, as shown in Figs.~\ref{fig:sq-d} and 
\ref{fig:sq-dl}. 
This result is mainly caused by the behavior of the chemical potential 
$\zeta$ [Fig.~\ref{fig:para}(e)], \cite{notecp} which changes its sign 
for $U>U_{\rm c}$ so as to recover the nesting condition, instead of 
by the band renormalization. 
\par

\begin{table}
\caption{Optimized variational parameters of two wave functions, 
$\Psi_Q^d$ and $\Psi_Q^{\rm FS}$, studied in \S\ref{sec:MottFS}, 
in respective insulating regimes ($U>U_{\rm c}$) for various $t'/t$. 
The final digits may include some errors. 
$L=14$.
}
\vspace{1mm}
\label{table:para}
\begin{tabular}{c|c||c|c|c|c|c}
\hline 
$\Psi$ & $|t'/t|$ & $g$ & $\Delta_d/t$ & $\mu$ & $\mu'$ & $\zeta/t$ \\
\hline
$\Psi_Q^d$ & 0.0  & 0.266 & 1.291 & 0.831 &   ---   &  0.0   \\
($U=$      & 0.25 & 0.254 & 1.339 & 0.840 & -0.0677 &  0.200 \\
$\ 7.5t$)  & 0.3  & 0.254 & 1.335 & 0.837 & -0.0665 &  0.230 \\
           & 0.4  & 0.254 & 1.349 & 0.837 & -0.0690 &  0.279 \\
           & 0.5  & 0.253 & 1.355 & 0.832 & -0.0688 &  0.317 \\
\hline
$\Psi_Q^{\rm FS}$ & 0.0  & 0.140 & ---  & 0.923 &  ---    & ---  \\
($U=$             & 0.25 & 0.119 & ---  & 0.904 &  0.0776 & ---  \\
$\ 12t$)          & 0.4  & 0.121 & ---  & 0.829 &  0.1317 & ---  \\
\hline
\end{tabular}
\end{table}
\begin{table}
\caption{Three energy components, total energy and spin structure 
factor at AF wave number, calculated using $\Psi_Q^d$ and 
$\Psi_Q^{\rm FS}$ in respective insulating regimes ($U>U_{\rm c}$) 
for various $t'/t$. 
$L=14$. 
The final digits may include some errors. 
Because each energy component and $E/t$ are averaged independently, 
$E/t$ does not precisely coincide with the sum of its components. 
}
\vspace{1mm}
\label{table:ene}
\begin{tabular}{c|c||c|c|c|c|c} \hline
 $\Psi$      & $|t'/t|$ & $E_t/t$ & $E_{t'}/t$ & $E_U/t$ & $E/t$ & 
$S({\bf G})$ \\
\hline
 $\Psi_Q^d$  & 0.0  & -0.6922 &  0.0    & 0.2548 & -0.4384 &  7.54 \\
   ($U=$     & 0.25 & -0.6888 & -0.0008 & 0.2533 & -0.4375 &  7.28 \\
   $7.5t$)   & 0.3  & -0.6899 & -0.0012 & 0.2532 & -0.4370 &  7.27 \\
             & 0.4  & -0.6848 & -0.0019 & 0.2529 & -0.4357 &  7.06 \\
             & 0.5  & -0.6829 & -0.0033 & 0.2522 & -0.4340 &  6.90 \\
\hline                                                       
 $\Psi_Q^{\rm FS}$ & 0.0& -0.4383 & 0.0 & 0.1809 & -0.2575 & 15.93 \\
 ($U=$       & 0.25 & -0.3455 & -0.0018 & 0.1576 & -0.1899 &  3.81 \\
 $12t$)      & 0.4  & -0.3276 & -0.0053 & 0.1573 & -0.1752 &  2.09 \\
\hline                                                       
\end{tabular}                                                
\end{table}
%
(5) In Table \ref{table:para}, we show the optimized parameters 
in the insulating regime ($U>U_{\rm c}$) for various $t'/t$. 
Note that the parameters vary only very slightly, except for $\zeta/t$; 
the optimized $\Psi_Q^d$ is almost unchanged with varying $t'/t$. 
In Table \ref{table:ene}, we list the total energy and energy components
for $U>U_{\rm c}$, calculated using the optimized $\Psi_Q^d$. 
$E$ is again almost independent of $t'/t$, because $E_{t'}$ makes a very
slight contribution, even for large $t'/t$ (see also Table \ref{table:rho}). 
As shown in the final column of Table \ref{table:ene}, the AF spin 
correlation retains a considerably large value for large $t'/t$. 
This indicates that $\Psi_Q^d$ is stabilized by preserving the nesting 
condition for the square lattice; in other words, the quasi-Fermi 
surface is retained at the gap maxima $(\pi,0)$ and $(0,\pi)$, at the 
cost of the energy reduction due to the diagonal hopping or frustration. 
Thus, the AF correlation is a key factor for stabilizing $\Psi_Q^d$. 
\par

Regarding the AF correlation, it is known for the SC states with 
$d$-type symmetries that the gap maxima overlap with the hot spot, 
namely, the intersection of the Fermi surface and the magnetic Brillouin 
zone boundary. \cite{Hirashima}
Thus, it is possible that the shape of the gap function $\Delta_{\bf k}$ 
deviates from that of the simple $d$-wave, particularly for large $t'/t$. 
This is an interesting future problem. 
\par

\section{\label{sec:MottFS}Mott Transitions in Projected Fermi Sea}
In this section, we discuss the Mott transition in $\Psi_Q^{\rm FS}$ 
as a continuation of the previous studies for $t'/t=0$. \cite{Yoko,YTOT} 
This transition has features different from those of $\Psi_Q^d$, 
although $\Psi_Q^{\rm FS}$ always has a higher energy than $\Psi_Q^d$ 
within the present model. 
In \S\ref{sec:MottFS0}, we make a careful analysis for $t'=0$. 
In \S\ref{sec:MottFSa}, we consider the $t'/t$ dependence, and contrast 
the properties in $\Psi_Q^{\rm FS}$ with those in $\Psi_Q^d$.
\par

\subsection{\label{sec:MottFS0}Case for $t'=0$}
\begin{figure*}[!t]
\vspace{-0.2cm}
\begin{center}
\includegraphics[width=17cm,clip]{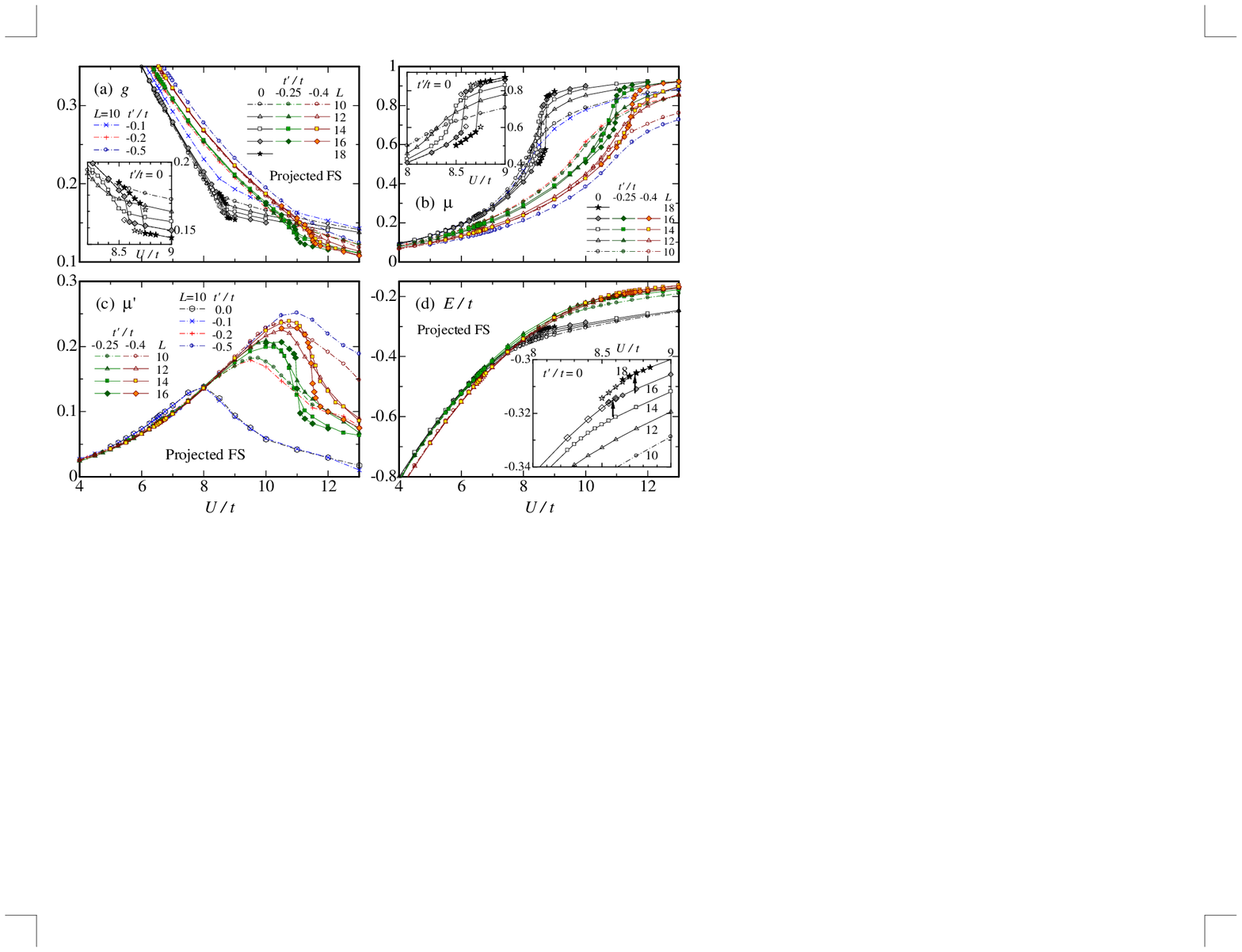}
\end{center}
\vskip -5mm
\caption{(Color online) 
(a)-(c) Optimized variational parameters (at global minima of $E/t$) 
for projected Fermi sea $\Psi_Q^{\rm FS}$ as a function of $U/t$: 
(a) Onsite correlation (Gutzwiller) factor.
(b) Doublon-holon binding factor between nearest-neighbor sites. 
The insets in (a) and (b) are magnifications near $U_{\rm c}$ 
for $t'/t=0$. 
The data of local (but not global) minima for $L=16$ and 18 are added 
using open symbols. 
(c) Same factor between diagonal-neighbor sites. 
(d) Optimized total energies for $t'/t=0$, $-0.25$ and $-0.4$. 
Data for several system sizes are compared. 
The inset represents a magnification near $U_{\rm c}$ for $t'/t=0$. 
For the data of $L=16$ and 18, open (closed) symbols are used for the 
metallic (insulating) regime to emphasize the hysteresis. 
The critical values indicated by arrows are $U_{\rm c}/t=8.59$ and 
8.73 for $L=16$ and 18, respectively. 
The other symbols and the abscissa scales are common to all the main 
panels. 
In (c), we add the data of $\Psi_Q^{\rm FS}$ including $\mu'$ for 
$t'/t=0$. 
Although $\mu'$ has a finite value in the metallic region, its effect 
is almost negligible. 
}
\label{fig:parafs}
\end{figure*}
%
The existence of a Mott transition in $\Psi_Q^{\rm FS}$ for $t'/t=0$ 
was first pointed out in ref.~\citen{Yoko}, in which the critical value 
was estimated as $U_{\rm c}/t\sim 8.8$ by analyzing the cusplike 
behavior in energy components and the discontinuity in $n({\bf k})$ 
for $L=10$ and 12. 
For these sizes, the transition appeared continuous. 
Certainly, we can find cusplike behavior in Fig.~\ref{fig:etot} 
[or in the main panel of Fig.~\ref{fig:parafs}(d)], where the $U/t$ 
dependence of total energy is shown. 
Because the system-size dependence is considerably large near the cusp, 
we show the magnification in the vicinity of the cusp in the inset 
of Fig.~\ref{fig:parafs}(d). 
For $L\le 14$, $E/t$ is a smooth function of $U/t$ and has a unique 
optimized value, whereas for $L=16$ and 18, $E/t$ exhibits hysteresis 
or double-minimum behavior near the critical point, $U_{\rm c}/t=8.59$ 
and $8.73$, respectively, as in the case of $\Psi_Q^d$ studied in the 
preceding section. 
Correspondingly, the two variational parameters, $g$ and $\mu$, exhibit 
discontinuities at $U_{\rm c}$ for $L\ge 16$, as shown in 
Figs.~\ref{fig:parafs}(a) and \ref{fig:parafs}(b), although their 
magnitudes are small 
compared with those of $\Psi_Q^d$ in Figs.~\ref{fig:para}(a) and 
\ref{fig:para}(c). 
Thus, this transition, at least for $t'/t=0$, is ascertained to be 
of the first order. 
\par 

\begin{figure}[hob]
\vspace{-0.2cm}
\begin{center}
\includegraphics[width=8.5cm,clip]{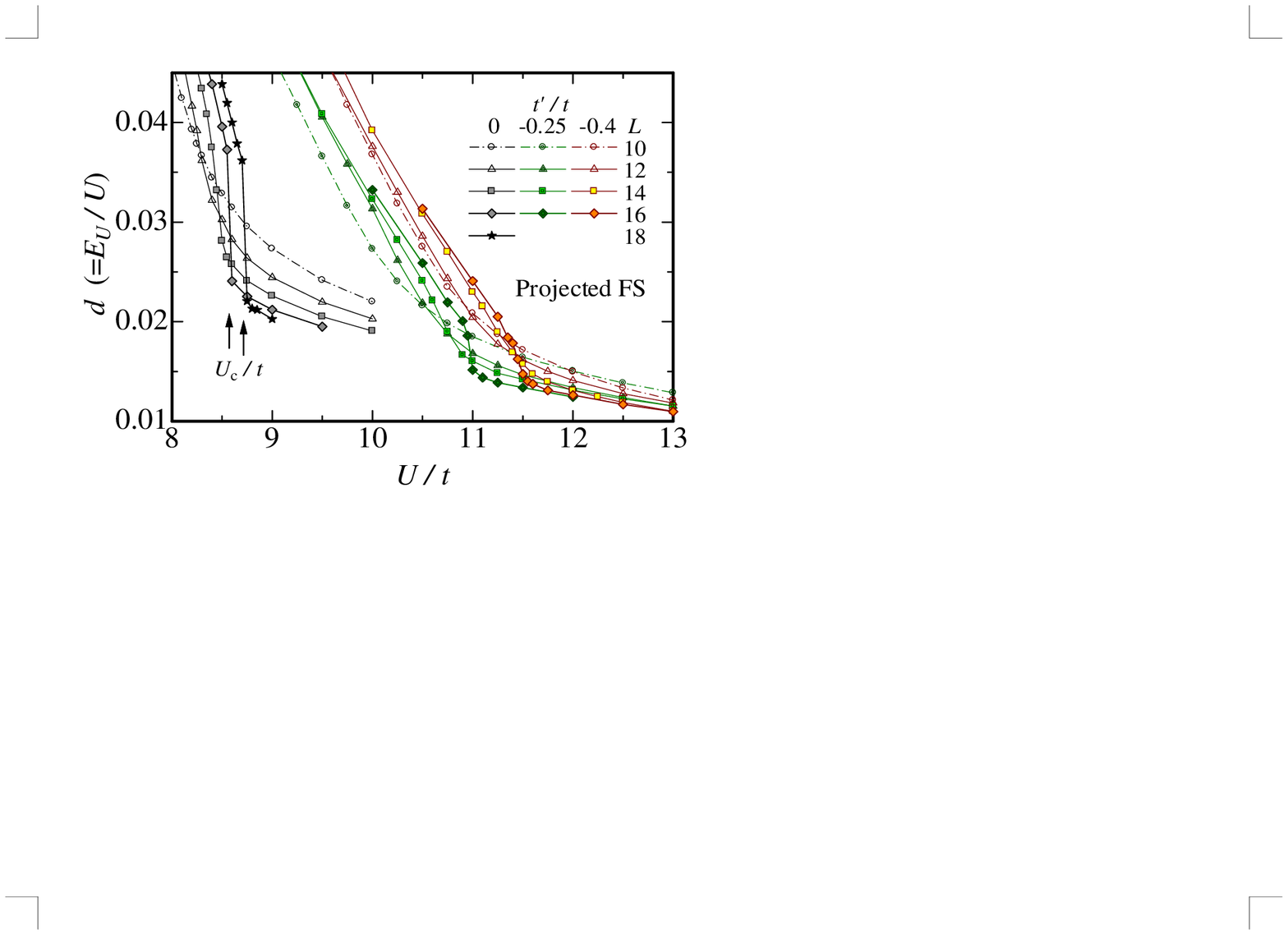}
\end{center}
\vskip -5mm
\caption{(Color online) 
Behavior of doublon density $d$ near the Mott critical points 
in the projected Fermi sea $\Psi_Q^{\rm FS}$ for three values of 
$t'/t$. 
Data of several system sizes are plotted for each $t'/t$. 
For $t'/t=0$ and $L=16$ and 18, the critical values of first-order 
transitions are indicated by arrows. 
}
\label{fig:euvsua}
\end{figure}

In Fig.~\ref{fig:euvsua}, the doublon density is plotted as a function 
of $U/t$. 
At the critical point $U_{\rm c}$, the order parameter $d$ of Mott 
transitions has a discontinuity for $L\ge 16$, and suddenly drops 
to a small value. 
Simultaneously, the doublon-holon binding parameter $\mu$ becomes 
large and approaches 1, as shown in Fig.~\ref{fig:parafs}(b). 
These results again suggest that this transition is a Mott transition with 
the doublon-holon binding mechanism similar to that of $\Psi_Q^d$. 
Actually, we have verified that the feature of electron configurations 
is considerably different for $U<U_{\rm c}$ and for $U>U_{\rm c}$ 
(not shown), as in Fig.~\ref {fig:snap}. 
\par 

\begin{figure}[hob]
\vspace{-0.2cm}
\begin{center}
\includegraphics[width=8.5cm,clip]{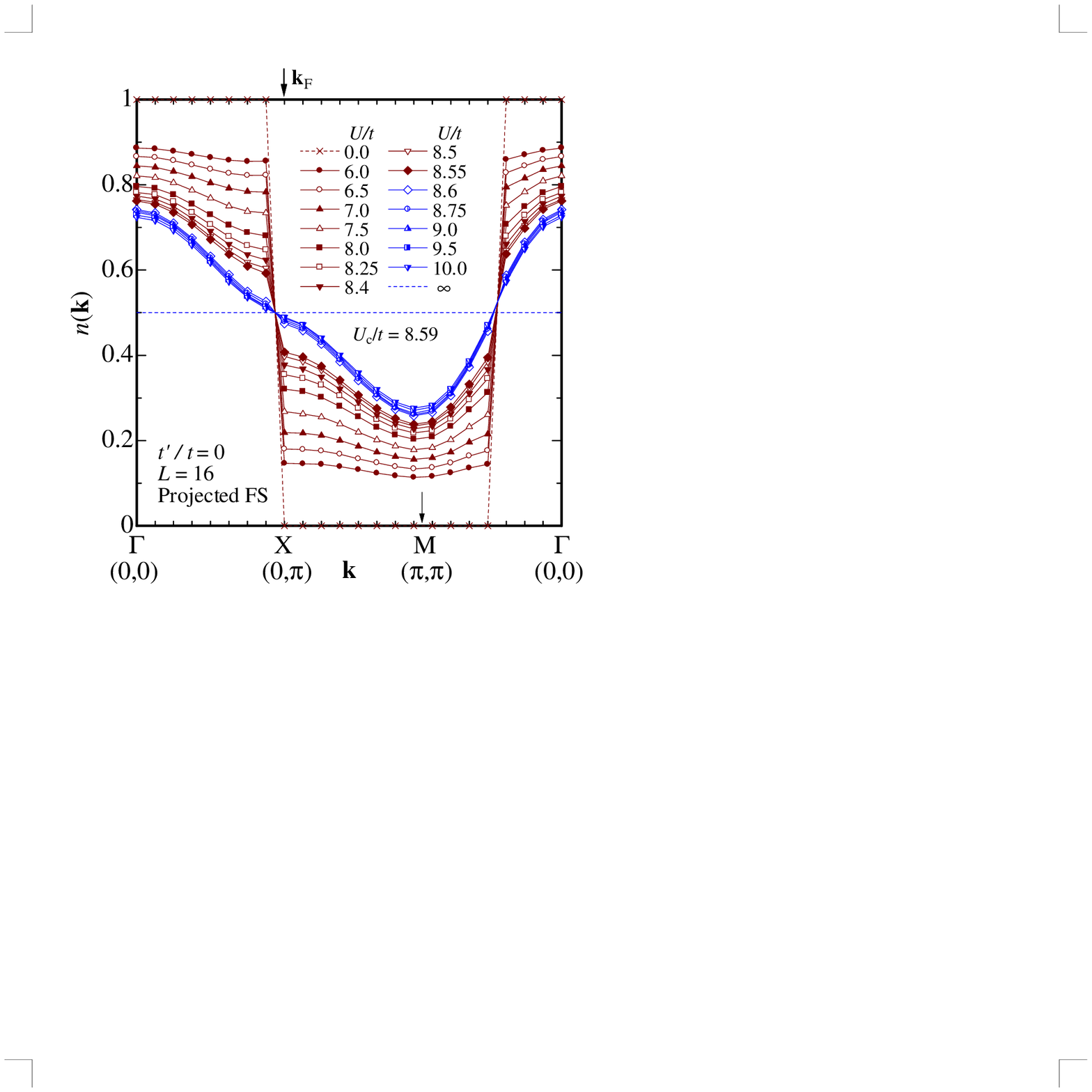}
\end{center}
\vskip -5mm
\caption{(Color online) 
$U/t$ dependence of momentum distribution function of $\Psi_Q^{\rm FS}$ 
for $t'/t=0$. 
In this case, $Z$ (Fig.~\ref{fig:sfZ-U}) is measured at ${\bf k}_{\rm F}$ 
indicated by an arrow on the upper axis (X point). 
}
\label{fig:nk-n}
\end{figure}
\begin{figure}[hob]
\vspace{-0.2cm}
\begin{center}
\includegraphics[width=8.7cm,clip]{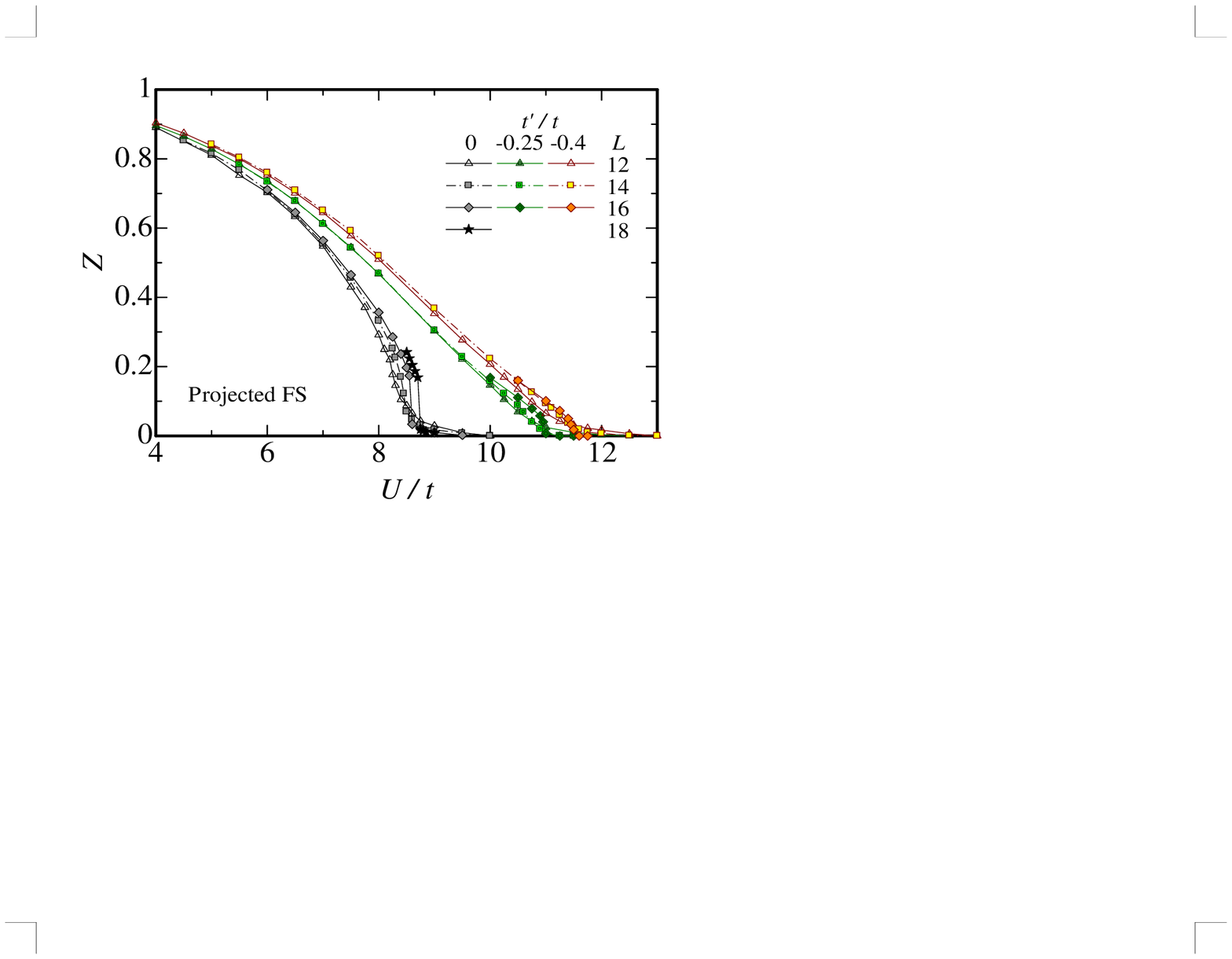}
\end{center}
\vskip -5mm
\caption{(Color online) 
Quasi-particle renormalization factor $Z$ of projected Fermi sea, 
estimated from discontinuities of $n({\bf k})$ on X-M line 
in Brillouin zone. 
Data for three values of $t'/t$ and various system sizes are plotted 
as a function of $U/t$. 
The tails for $U>U_{\rm c}$ are mainly caused by finite-sized effects. 
}
\label{fig:sfZ-U}
\end{figure}
%
The transition is corroborated by the behavior of $n({\bf k})$ and 
$N({\bf q})$, in the same way as for the $d$-wave state. 
In Fig.~\ref{fig:nk-n}, $n({\bf k})$ [eq.~(\ref{eq:nk})] is plotted 
along the path $\Gamma$-X-M-$\Gamma$ for various values of $U/t$. 
Because $\Psi_Q^{\rm FS}$ is metallic for $U<U_{\rm c}$, the Fermi 
surface can be defined in any direction; in this path, ${\bf k}_{\rm F}$ 
points are located at the X point and at the midpoint of the $\Gamma$-M 
segment. 
For $U<U_{\rm c}$, the discontinuities are apparent at both points of 
${\bf k}_{\rm F}$, whereas for $U>U_{\rm c}$, the discontinuities vanish, 
indicating a gap opening. 
In Fig.~\ref{fig:sfZ-U}, we show the magnitude of discontinuity, $Z$, 
measured at the X point, because the extrapolation error is small. 
As $U/t$ increases, the quasi-particle renormalization factor, $Z$, 
monotonically decreases, and vanishes at $U_{\rm c}$; 
the effective mass diverges, and $\Psi_Q^{\rm FS}$ becomes insulating. 
For $L\le 14$, $Z$ is continuous near $U_{\rm c}$, as reported 
in Fig.~17 in ref.~\citen{Yoko}, whereas $Z$ has an appreciable 
discontinuity at $U_{\rm c}$ for $L\ge 16$ and probably for $L=\infty$. 
Incidentally, the discontinuity in $Z$ in $\Psi_Q^{\rm FS}$ is smaller 
than that in $\Psi_Q^d$ (Fig.~\ref{fig:scZ-U}); the character of 
the first-order transition is less conspicuous in $\Psi_Q^{\rm FS}$. 
In contrast, in dynamical mean-field theory for the hypercubic 
lattice, \cite{Bulla} $Z$ continuously decreases and vanishes at 
$U=U_{\rm c}$ without a discontinuity. 
\par

\begin{figure}[hob]
\vspace{-0.2cm}
\begin{center}
\includegraphics[width=8.5cm,clip]{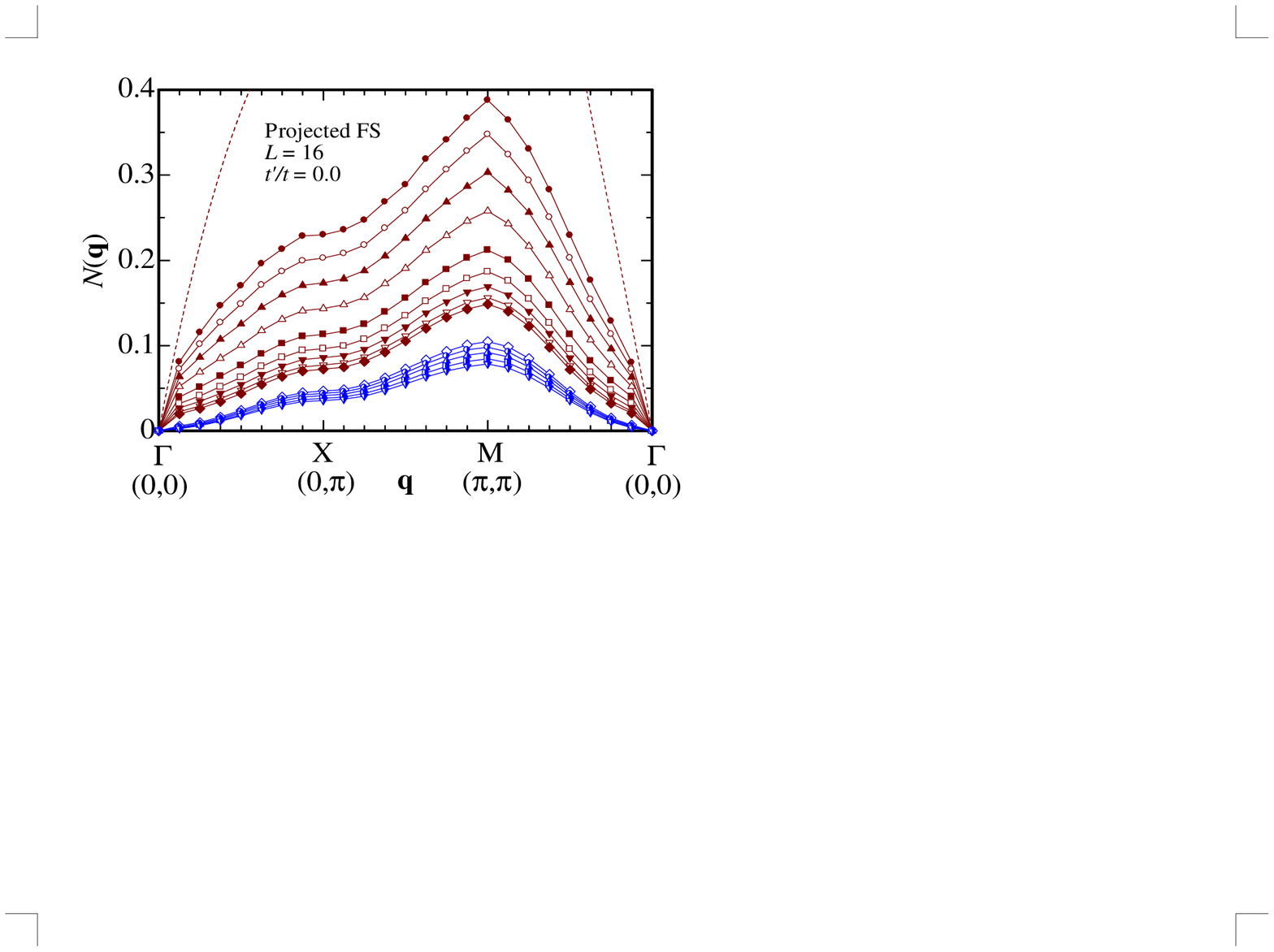}
\end{center}
\vskip -5mm
\caption{(Color online) 
$U/t$ dependence of charge density structure factor of 
$\Psi_Q^{\rm FS}$ for $t'/t=0$. 
The symbols denote the same values of $U/t$ as in Fig.~\ref{fig:sq-n}. 
$U_{\rm c}/t=8.59$. 
}
\label{fig:nq-n}
\end{figure}
%
In Fig.~\ref{fig:nq-n}, $N({\bf q})$ [eq.~(\ref{eq:nq})] is plotted 
for various values of $U/t$. 
The behavior for small $|{\bf q}|$ is basically the same as that for 
$\Psi_Q^d$ [Fig.~\ref{fig:nq-d}], namely, $N({\bf q})\propto |{\bf q}|$ 
for $U<U_{\rm c}$, whereas $N({\bf q})$ tends to be 
proportional to ${\bf q}^2$ for $U>U_{\rm c}$. 
This follows that a charge gap opens for $U>U_{\rm c}$. 
\par

From all the above results, we regard this transition as a Mott 
(metal-to-nonmagnetic-insulator) transition, which is caused by 
the doublon-holon binding mechanism, basically in the same way as 
in $\Psi_Q^d$. 
\par

\begin{figure}[hob]
\vspace{-0.2cm}
\begin{center}
\includegraphics[width=8.5cm,clip]{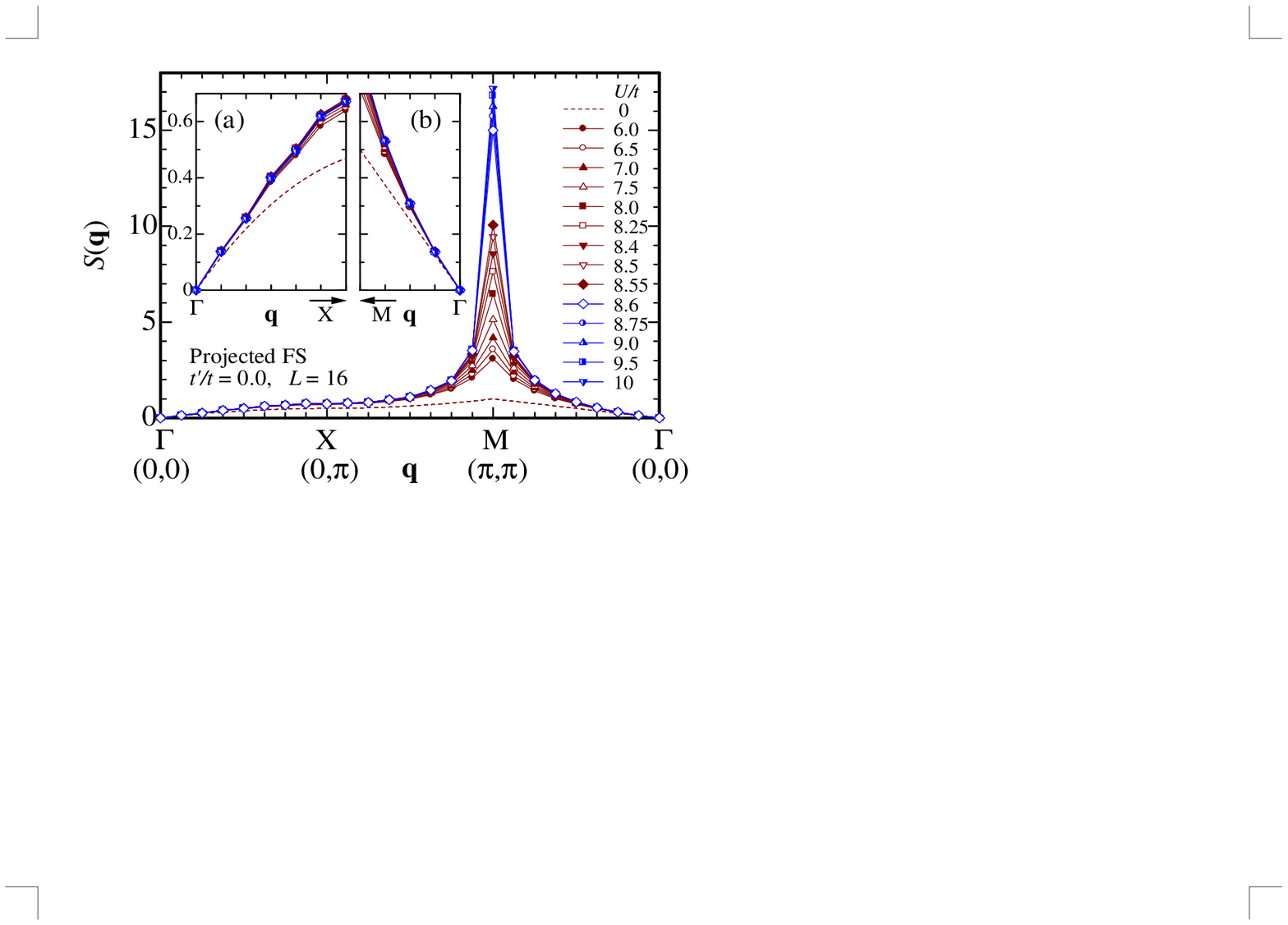}
\end{center}
\vskip -5mm
\caption{(Color online) 
$U/t$ dependence of spin structure factor of projected Fermi sea 
for $t'/t=0$. 
$U_{\rm c}/t=8.59$. 
The insets are the close-ups near the $\Gamma$ point on 
(a) the $\Gamma$-M line and (b) the $\Gamma$-X line. 
}
\label{fig:sq-n}
\end{figure}

Finally, we consider the spin degree of freedom. 
In Fig.~\ref{fig:sq-n}, $S({\bf q})$ [eq.~(\ref{eq:sq})] is plotted 
for various values of $U/t$. 
As $U/t$ increases, the AF correlation $S({\bf G})$ increases, and 
increases abruptly near the critical point. 
Although the behavior of $S({\bf q})$ is, as a whole, similar to that 
of $\Psi_Q^d$ (Fig.~\ref{fig:sq-d}), $S({\bf G})$ is twice 
as large for $\Psi_Q^{\rm FS}$. 
However, the sublattice magnetization, 
\begin{equation} 
m=\frac{1}{N_{\rm s}}
\left|\sum_ie^{i{\bf G}\cdot{\bf r}_i}\langle S_i^z\rangle\right|, 
\label{eq:m}
\end{equation}
remains zero within the statistical fluctuation. 
Thus, a long-range order is not realized, although the AF correlation 
is considerably enhanced in the insulating regime. 
Note that the behavior of $S({\bf q})$ for $|{\bf q}|\rightarrow 0$ 
is linear in $|{\bf q}|$ for arbitrary $U/t$, as shown in the insets 
of Fig.~\ref{fig:sq-n}, in contrast to the case of $\Psi_Q^d$ 
(inset of Fig.~\ref{fig:sq-d}). 
This strongly suggests that the spin gap is absent. 
Thus, $\Psi_Q^{\rm FS}$ for $U>U_{\rm c}$ represents a nonmagnetic 
insulator without a spin gap, which is considered to be realized 
in $\kappa$-(ET)$_2$Cu$_2$(CN)$_3$. \cite{Shimizu}
\par

\subsection{\label{sec:MottFSa}Effect of frustration}
First, we consider the effect of frustration on the properties of the 
transition. 
In Figs.~\ref{fig:parafs}(a)-\ref{fig:parafs}(c), we show the optimized 
variational 
parameters for $t'/t=-0.25$ and $-0.4$, as well as $t'=0$, for systems 
up to $L=16$. 
Let us consider the doublon-holon binding parameter $\mu$ 
[Fig.~\ref{fig:parafs}(b)] as a typical parameter. 
For small systems (\eg, $L=10$), $\mu$ is a smoother `S'-shaped 
function of $U/t$ than that for $t'=0$. 
As the system size increases, $\mu$ abruptly exhibits semicritical 
behavior at a slightly larger $U/t$ than that for $t'=0$ 
[$U_{\rm c}/t\sim 10.95$ (11.45) for $t'/t=-0.25$ ($-0.4$) for $L=16$]. 
In contrast to the case of $t'=0$, the cases of $t'/t=-0.25$ and $-0.4$ 
do not indicate a first-order-transition-like discontinuity or hysteresis, 
even for $L=16$. 
In comparing the results among the three values of $t'/t$, we notice 
that the critical behavior tends to be more continuous as the frustration 
becomes strong. 
We do not consider larger systems in this work, because the statistical 
fluctuation around the critical points increases rapidly 
as $L$ increases. 
However, we assume that this size-dependent critical behavior is a 
sign of a first-order transition. 
\par

These features can be seen in the other variational parameters 
[Figs.~\ref{fig:parafs}(a) and \ref{fig:parafs}(c)], total energy 
[Fig.~\ref{fig:parafs}(d)]
doublon density (Fig.~\ref{fig:euvsua}) and quasi-particle renormalization 
factor (Fig.~\ref{fig:sfZ-U}). 
The feature that the critical properties tend to become continuous 
as $t'/t$ increases is opposite to that of $\Psi_Q^d$ studied in 
\S\ref{sec:MottSC}, but similar to that of the 
path-integral-renormalization-group approach. \cite{Kashima}
\par 

\begin{table}
\caption{
Comparison of ratio $\rho=E_{t'}/E_t$ among the different phases 
indicated in the brackets for the $d$-wave state and the projected Fermi 
sea. 
The abbreviations `ins.' and `wSC' denote insulating and weak SC, 
respectively. 
The systems of $L=14$ is used. 
Values of $\rho$ are shown as percentages. 
}
\vspace{1mm}
\label{table:rho}
\begin{tabular}{c|c||c|c|c|c} \hline
             &          & \multicolumn{4}{c}{$U/t$}\\
\cline{3-6}
\multicolumn{1}{c|}{\raisebox{1.5ex}{$\Psi$}} & 
\multicolumn{1}{c||}{\raisebox{1.5ex}{$|t'/t|$}} & 
$0$ & 6.25 & $7.5$ & $13$ \\
\hline
             & 0.25 &  3.3  & 1.4 (SC) &  0.12 (ins.) & --- \\
\cline{2-6}
\multicolumn{1}{c|}{\raisebox{1.5ex}{$\Psi_Q^d$}}
             & 0.4  &  8.8  & 7.6 (wSC) &  0.28 (ins.) & --- \\
\hline                                                       
             & 0.25 &  3.3  & 3.3 (metal) & 3.2 (metal) & 0.4 (ins.) \\
\cline{2-6}
\multicolumn{1}{c|}{\raisebox{1.5ex}{$\Psi_Q^{\rm FS}$}} 
             & 0.4  &  8.8  & 8.8 (metal) & 8.5 (metal) & 1.1 (ins.) \\
\hline                                                       
\end{tabular}                                                
\end{table}
\begin{figure}[hob]
\vspace{-0.2cm}
\begin{center}
\includegraphics[width=8.7cm,clip]{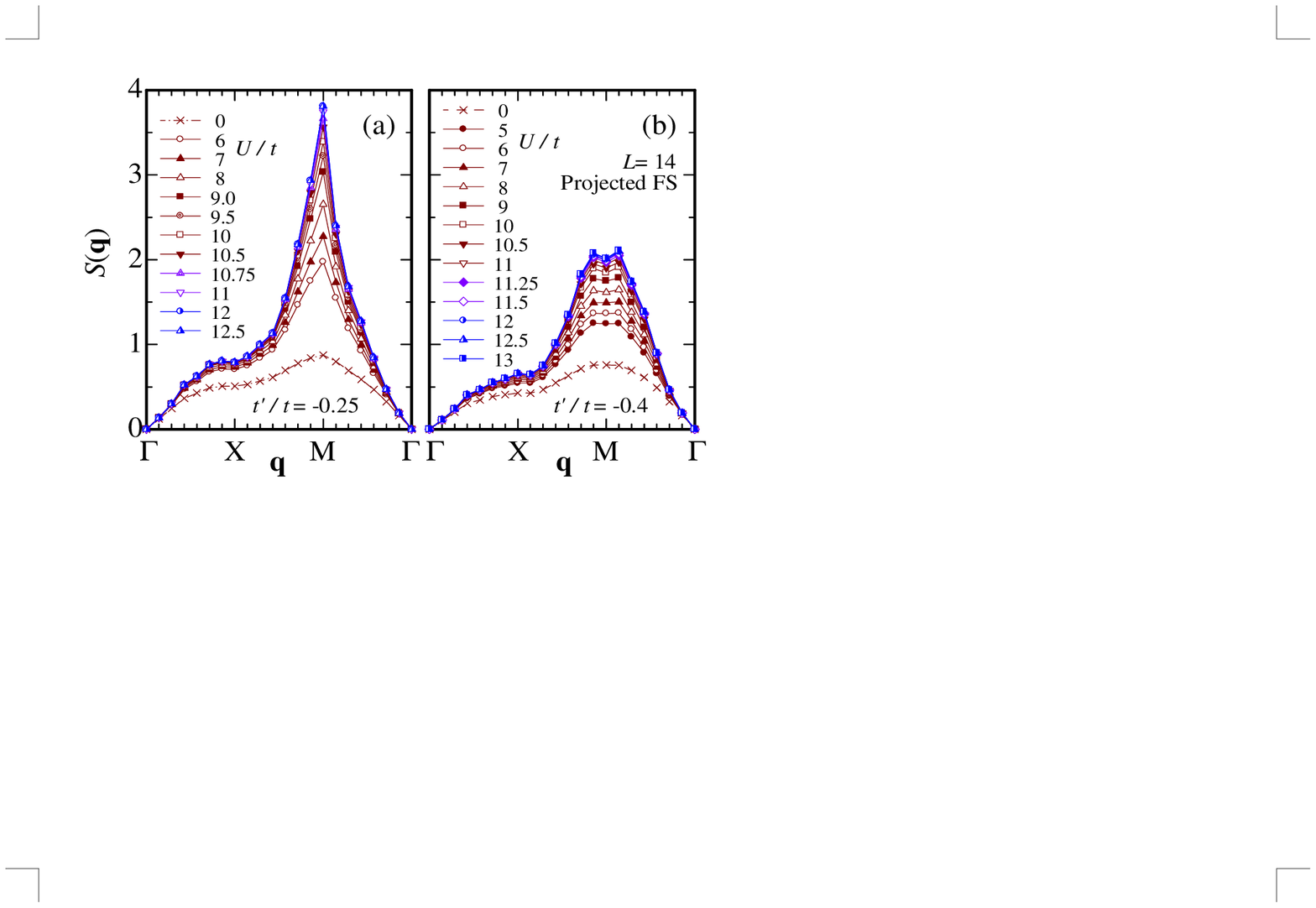}
\end{center}
\vskip -5mm
\caption{(Color online) 
$U/t$ dependence of the spin structure factor of the projected Fermi sea 
for (a) $t'/t=-0.25$ and (b) $t'/t=-0.4$. 
The scale of the vertical axis is common to both panels. 
}
\label{fig:sq-na}
\end{figure}

Next, we consider the effect of frustration on the insulating state. 
As shown in Table \ref{table:para}, when $t'/t$ varies, the optimized 
parameters, namely, the wave function, appreciably changes, in contrast to 
that of $\Psi_Q^d$. 
Accordingly, the physical quantities with respect to $\Psi_Q^{\rm FS}$ 
vary with $t'/t$, as shown in Table \ref{table:ene}. 
For energy, a notable point is that the contribution of $E_{t'}$ 
is still strongly suppressed in $\Psi_Q^{\rm FS}$, similarly to 
that for $\Psi_Q^d$. 
To consider quantitatively how $E_{t'}$ behaves when $U/t$ varies, 
we list the ratio $\rho\equiv E_{t'}/E_t$ in the different phases in 
Table \ref{table:rho}. 
$\rho$ almost maintains its value of the noninteracting case ($U=0$) 
in the metallic phase, but in the insulating phase, $\rho$ drops to 
a very small value, although not as small as in $\Psi_Q^d$. 
This feature is affected by $\mu'$. 
As seen in Fig.~\ref{fig:parafs}(c), $\mu'$ abruptly decreases at 
$U_{\rm c}$, indicating that the doublon-holon binding in the diagonal 
direction, 
which assists local diagonal hopping for large $U/t$, also becomes less 
advantageous for $U>U_{\rm c}$ in $\Psi_Q^{\rm FS}$. 
The decrease in $\mu'$ is anticipated to enhance AF correlation, 
mentioned below. 
\par

Returning to Table \ref{table:ene}, we find that the AF spin correlation 
$S({\bf G})$ markedly increases for $t'/t=0$, but abruptly decreases 
as $|t'/t|$ increases, in sharp contrast with that of $\Psi_Q^d$. 
To elucidate the situation, we plot $S({\bf q})$ for $t'/t=-0.25$ and 
$-0.4$ in Figs.~\ref{fig:sq-na}(a) and \ref{fig:sq-na}(b), respectively 
(cf. also Fig.~\ref{fig:sq-n} for $t'=0$). 
Although the magnitude of $S(\bf G)$ abruptly decreases as $|t'/t|$ 
increases from 0 to 0.25, the peak position of $S({\bf q})$ remains 
at ${\bf q}={\bf G}$. 
For $t'/t=-0.4$, however, in addition to the successive decrease in 
magnitude, the peak of $S({\bf q})$ moves to incommensurate wave 
numbers near the M point. 
\par

Thus, the effect of frustration is explicitly reflected in physical 
quantities estimated in the insulating state of $\Psi_Q^{\rm FS}$. 
\par

\section{\label{sec:SC}$d_{x^2-y^2}$-wave Superconductivity}
In this section, we study the properties of SC arising in the $d$-wave 
singlet state $\Psi_Q^d$. 
In \S\ref{sec:SCcond}, we deduce the area where SC appears in the $t'$-$U$ 
plane by distinguishing, in the condensation energy, between the 
contributions from a SC gap and from an insulating gap. 
In \S\ref{sec:pdr}, we confirm the appearance of SC by directly observing 
the $d$-wave pairing correlation function. 
In \S\ref{sec:SCprop}, we consider the origin of this SC. 
\par

\subsection{\label{sec:SCcond}Condensation energy}
\begin{figure}[hob]
\vspace{-0.2cm}
\begin{center}
\includegraphics[width=8.5cm,clip]{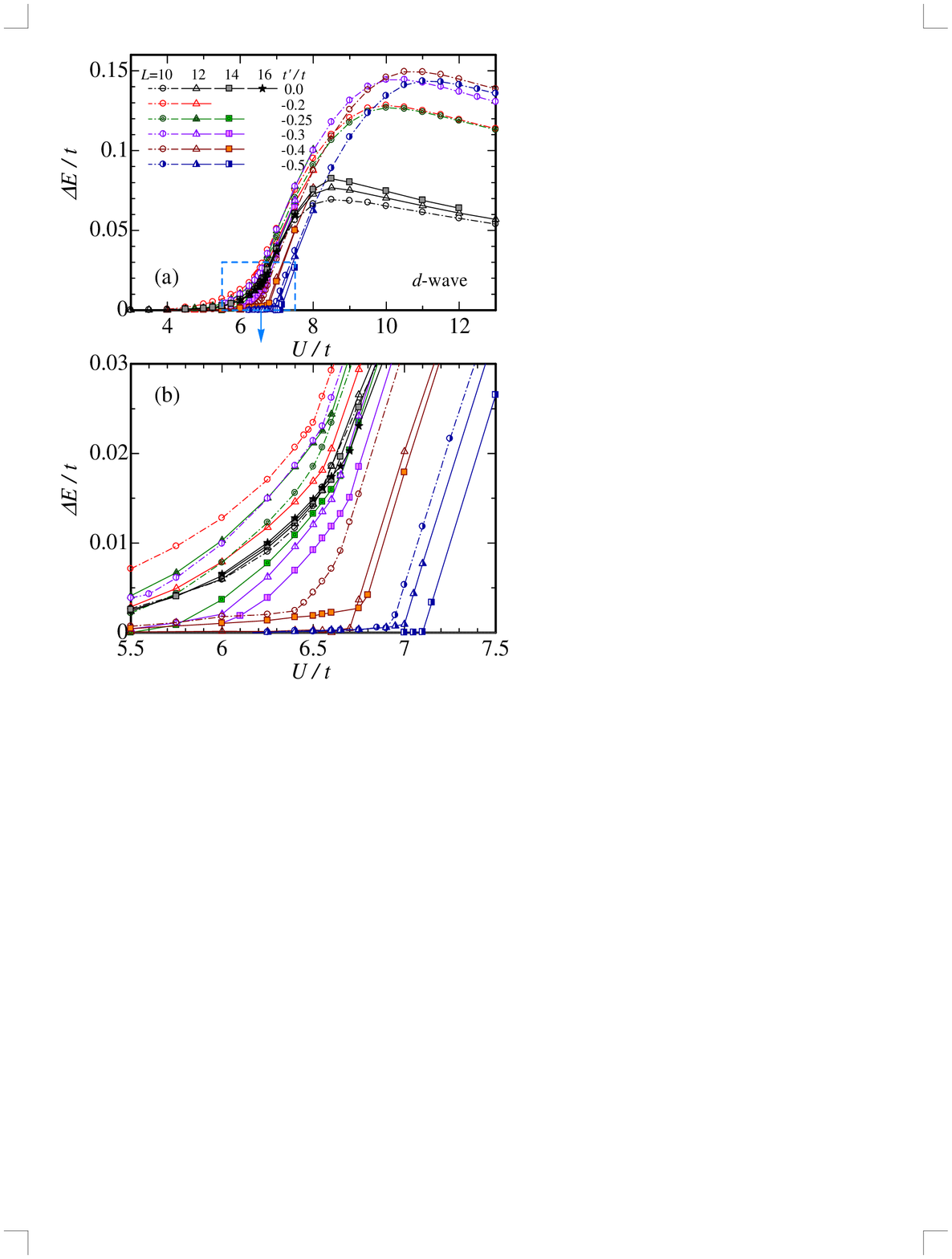}
\end{center}
\vskip -5mm
\caption{(Color online) 
Condensation energy, $\Delta E/t$, of $d$-wave singlet state as 
function of $U/t$ for various values of $t'/t$ and $L$. 
(a) $\Delta E/t$ over a wide range of $U/t$. 
(b) A close-up of $\Delta E/t$ near the Mott critical points; the range 
is indicated by a dashed-line box in (a). 
The symbols are common to (a) and (b).
}
\label{fig:conden}
\end{figure}

First, to determine the stability of the $d$-wave singlet state $\Psi_Q^d$, 
we consider its condensation energy given by 
\begin{equation}
\Delta E=E(\Psi_Q^d)-E(\Psi_Q^{\rm F}), 
\label{eq:cond}
\end{equation}
where $E(\Psi)$ denotes the optimized variational energy per site 
with respect to $\Psi$. 
In Fig.~\ref{fig:conden}(a), $\Delta E/t$ for various $t'/t$ is plotted 
as a function of $U/t$; in Fig.~\ref{fig:conden}(b), the region near 
the Mott critical points ($U_{\rm c}/t$) is magnified. 
\par

Because the behavior of $\Delta E$ depends on the value 
of $t'/t$, we first consider the weakly frustrated cases ($t'/t\lsim 0.3$). 
As pointed out in ref.~\citen{YTOT}, for small values of $U/t$ 
($\lsim 5$), $\Delta E/t$ is extremely small; at intermediate values 
of $U$ ($=U_{\rm onset}\sim 5t$-$6t$), $\Delta E/t$ starts to increase 
slowly at first, then abruptly at the Mott critical value $U_{\rm c}$ of 
$\Psi_Q^d$, where, in some cases, we can observe a mild cusp in 
Fig.~\ref{fig:conden}(b). 
As $U/t$ increases further, $\Delta E/t$ has a maximum and then slowly 
decreases. 
We are now aware (\S\ref{sec:MottSC}) that, for $U>U_{\rm c}$, $\Psi_Q^d$ 
becomes an insulating state. 
Hence, the marked increase in $\Delta E$ in this regime is considered 
to originate from the insulating $d$-wave gap. \cite{ZGRS} 
Consequently, the region where substantial energy reduction occurs 
owing to a SC gap is restricted to $U_{\rm onset}\lsim U<U_{\rm c}$. 
This idea is supported by the behavior of the $d$-wave gap parameter 
$\Delta/t$, which exhibits an appreciable increase for the corresponding 
values of $U/t$ and $t'/t$, as shown in Fig.~\ref{fig:para}(b). 
Incidentally, the doublon-holon binding parameter $\mu$ increases 
very similarly to $\Delta/t$ [Fig.~\ref{fig:para}(c)], suggesting that 
this binding plays an active role in the $d$-wave pairing in the 
nearest-neighbor sites. 
\par 

Next, we proceed to the strongly frustrated cases ($t'/t\gsim 0.4$). 
A major feature of $\Delta E$, different from that in the weakly 
frustrated case, is that there is no substantial increase for 
$U_{\rm onset}\lsim U<U_{\rm c}$. 
As seen in Fig.~\ref{fig:conden}(b), we cannot determine $U_{\rm onset}$ 
for $t'/t=-0.4$ and $-0.5$, except for a special case, $t'/t=-0.4$ 
and $L=10$. \cite{noteexcept} 
Correspondingly, the increase in the $d$-wave gap $\Delta/t$ 
[Fig.~\ref{fig:para}(b)] is firmly suppressed in the conductive region 
($U<U_{\rm c}$), compared with those in the weakly frustrated cases. 
The behavior of $\mu$ [Fig.~\ref{fig:para}(c)] again follows that 
of $\Delta/t$.  
Thus, SC, if there is any, is expected to be weak in this regime.
\par

Note that the value of $U/t$ at the maximum of $\Delta E/t$ in the 
insulating regime approximately corresponds to the Mott critical point 
of $\Psi_Q^{\rm FS}$, $U_{\rm c}^{\rm FS}/t$, considered 
in \S\ref{sec:MottFS}. 
$U_{\rm c}^{\rm FS}/t$ also corresponds to the crossover value, at 
which the character of SC changes from the interaction-energy origin 
to the kinetic-energy origin. \cite{YTOT} 
\par

\begin{figure}[hob]
\vspace{0.2cm}
\begin{center}
\includegraphics[width=8.7cm,clip]{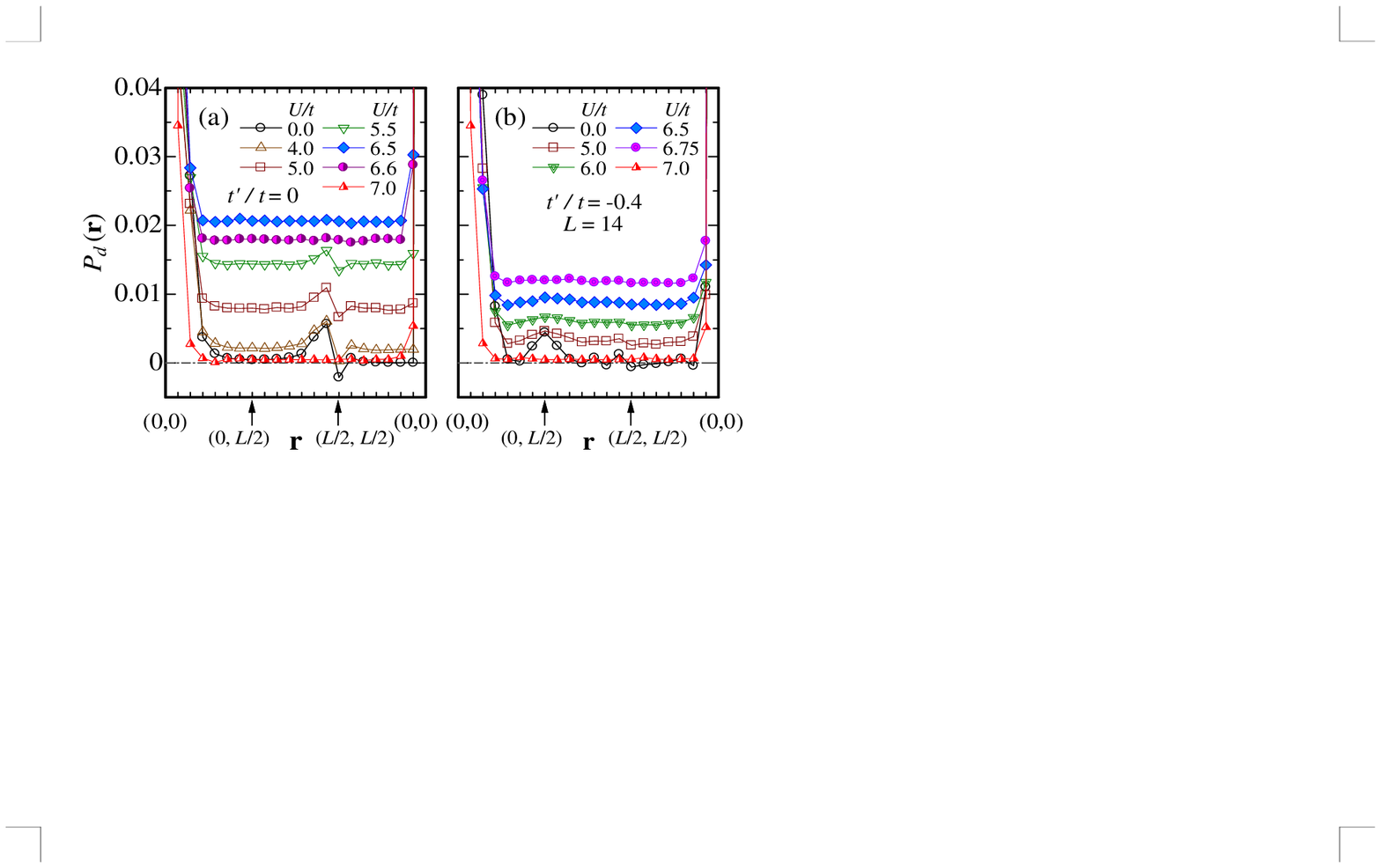}
\end{center}
\vskip -5mm
\caption{(Color online) 
Real-space pair correlation function of $d$-wave symmetry along 
path of {\bf r}, (0,0)-(0,$L/2$)-($L/2$,$L/2$)-(0,0), for various 
values of $U/t$. 
(a) $t'/t=0$ and (b) $t'/t=-0.4$, both for $L=14$. 
The scale of the ordinate axis is common to both panels. 
The data are obtained, using VMC calculations for $U\ne 0$, and from 
the analytic formula for $U=0$. 
}
\label{fig:pdr}
\end{figure}
\subsection{\label{sec:pdr}Pair correlation function}
%
To directly confirm the appearance of $d_{x^2-y^2}$-wave SC, 
the $d$-wave nearest-neighbor pair correlation function 
$P_d({\bf r})$ is convenient for the present approach: \cite{notePd} 
\begin{eqnarray}
P_d({\bf r})=&&\frac{1}{N_{\rm s}}
\sum_{i}\sum_{\tau,\tau'=\hat {\bf x},\hat {\bf y}}
(-1)^{1-\delta(\tau,\tau')}\times\qquad \nonumber\\ 
&& \left\langle{\Delta _\tau^\dag({\bf R}_i)\Delta_{\tau'}
({\bf R}_i+{\bf r})}\right\rangle, 
\label{eq:pd}
\end{eqnarray}
where $\hat{\bf x}$ and $\hat{\bf y}$ denote the lattice vectors 
in the $x$- and $y$-directions, respectively, and 
$\Delta_\tau^\dag({\bf R}_i)$ is the creation operator of a
nearest-neighbor singlet, 
\begin{equation}
\Delta_\tau^\dag({\bf R}_i)=
(c_{{i}\uparrow}^\dag c_{{i}+\tau\downarrow}^\dag+ 
 c_{{i}+\tau\uparrow}^\dag c_{{i}\downarrow}^\dag)
 /{\sqrt 2}. 
\end{equation}
%
If $P_d({\bf r})$ has a finite value for $|{\bf r}|\rightarrow\infty$, 
off-diagonal long-range order exists. 
For finite systems, however, we have to appropriately determine 
long-distance values of $P_d({\bf r})$, particularly, in the cases of 
small $U/t$, where the correlation length is long. 
In Fig.~\ref{fig:pdr}, $P_d({\bf r})$ is plotted for two values of 
$t'/t$. 
Although $P_d({\bf r})$ for large $|{\bf r}|$ should vanish for $U/t=0$,
spikes of sizable magnitude appear near ${\bf r}=(L/2,L/2)$ [$(0,L/2)$] 
for $t'/t=0$ [$-0.4$] for this system size. \cite{notepdrsize} 
Furthermore, a trace of this spike structure remains up to fairly 
large values of $U/t$. 
Thus, for small $U/t$, we should choose $P_d({\bf r})$ which does 
not have such peculiar finite-sized effects. 
Fortunately, we found that, in the noninteracting cases, the magnitude 
of $|P_d({\bf r})|$ for ${\bf r}=(L/2-1,L/2)$, which is almost the 
farthermost point, is very small (less than $10^{-4}$) for arbitrary 
values of $t'/t$ and $L$. 
Hence, we employ $P_d[(L/2-1,L/2)]$ as the large-$|{\bf r}|$ 
value, $P_d^\infty$, 
for small $U/t$, namely, $0\le U\le U_{\rm max}$, with $U_{\rm max}$ 
being the value at which $P_d({\bf r})$ becomes maximum. 
For the strong-correlation regime ($U>U_{\rm max}$), $P_d({\bf r})$ 
becomes almost constant for $|{\bf r}|\ge 3$, \cite{TKLee,metric} 
as shown in Fig.~\ref{fig:pdr}. 
Hence, in this regime, we adopt the average of $P_d({\bf r})$ for 
$|{\bf r}|\ge 3$ as $P_d^\infty$. 
\par

\begin{figure*}[!t]
\vspace{-0.2cm}
\begin{center}
\includegraphics[width=18.0cm,clip]{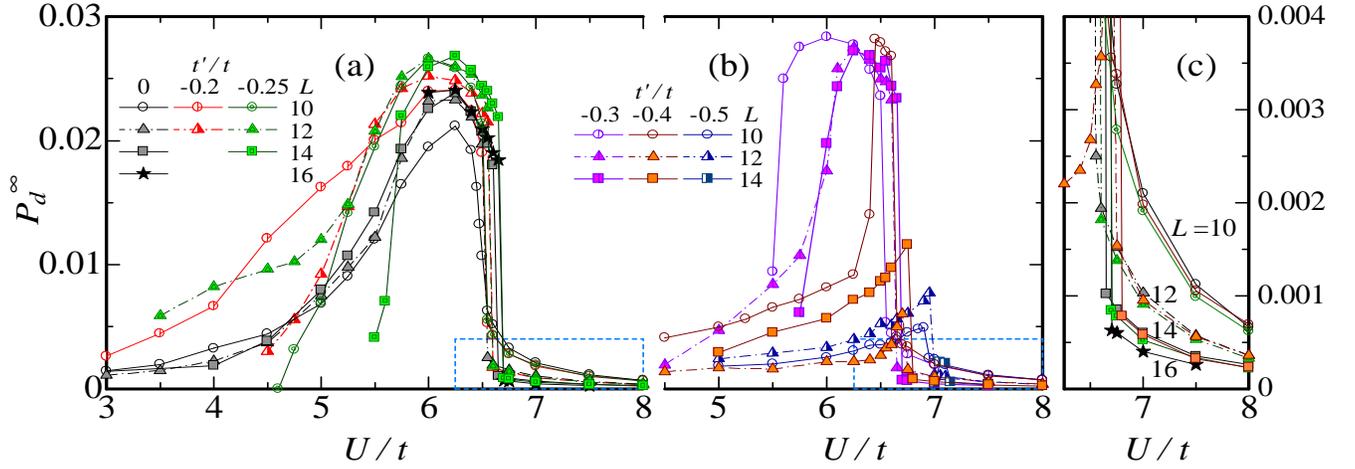}
\end{center}
\vskip -5mm
\caption{(Color online) 
Large-$|{\bf r}|$ value of $d$-wave pair correlation function 
$P_d({\bf r})$ for (a) small $|t'/t|$ ($\le 0.25$) and (b) large 
$|t'/t|$ ($\ge 0.3$). 
Data of various system sizes for each value of $t'/t$ are plotted. 
(c) Magnification of $P_d^\infty$ near the Mott critical points. 
The data for only $t'/t=0$, $-0.25$ and $-0.4$ are shown for clarity. 
The range of (c) is indicated by a dashed-line boxes in (a) and (b). 
}
\label{fig:pdrinf}
\end{figure*}
%
In Figs.~\ref{fig:pdrinf}(a) and \ref{fig:pdrinf}(b), 
$P_d^\infty$ thus obtained is plotted as a function of $U/t$. 
In the weakly correlated regime ($U/t\lsim 4$), the increase in 
$P_d^\infty$ is small, irrespective of the value of $t'/t$. 
For weakly frustrated cases ($|t'/t|\lsim 0.3$), $P_d^\infty$ starts 
to increases appreciably at $U\sim U_{\rm onset}$ as $U/t$ increases, 
has a peak at $U/t=6$-6.25, and then abruptly decreases at the Mott 
critical point $U=U_{\rm c}$. 
The system-size dependence of $P_d^\infty$ near the peak is weak for 
$L\ge 12$. 
Thus, in these cases, robust $d$-wave SC certainly occurs for 
$U_{\rm onset}\lsim U<U_{\rm c}$. 
On the other hand, for strongly frustrated cases [$|t'/t|\gsim 0.4$ in 
Fig.~\ref{fig:pdrinf}(b)], 
no sizable increase in $P_d^\infty$ is observed at the value corresponding 
to $U_{\rm onset}$. 
$P_d^\infty$ slowly and monotonically increases until it abruptly drops 
at $U=U_{\rm c}$. \cite{noteexcept} 
Moreover, the system-size dependence is very large. 
Eventually, robust $d$-wave SC occurs in a limited area, 
$U_{\rm onset}\lsim U<U_{\rm c}$ and $|t'/t|\lsim 0.3$, within $\Psi_Q^d$.
A similar result has been recently obtained using a fluctuation exchange 
approximation. \cite{Onari}
\par

In Fig.~\ref{fig:pdrinf}(c), we show the magnification of $P_d^\infty$ 
near the Mott critical points. 
In the insulating regime ($U>U_{\rm c}$), $P_d^\infty$ becomes almost 
independent of $t'/t$, as mentioned in item (5) in \S\ref{sec:mottprop}, 
decreases rapidly as the system size $L$ increases, and probably vanishes 
in the limit of $L\rightarrow\infty$. 
Because the statistical fluctuation in the VMC data is much smaller in 
the insulating regime than in the conductive regime, the data are more 
reliable. 
The disappearance of $P_d^\infty$ for $U>U_{\rm c}$ is expected in 
an insulating state. 
\par

\begin{figure}[hob]
\vspace{-0.2cm}
\begin{center}
\includegraphics[width=8.0cm,clip]{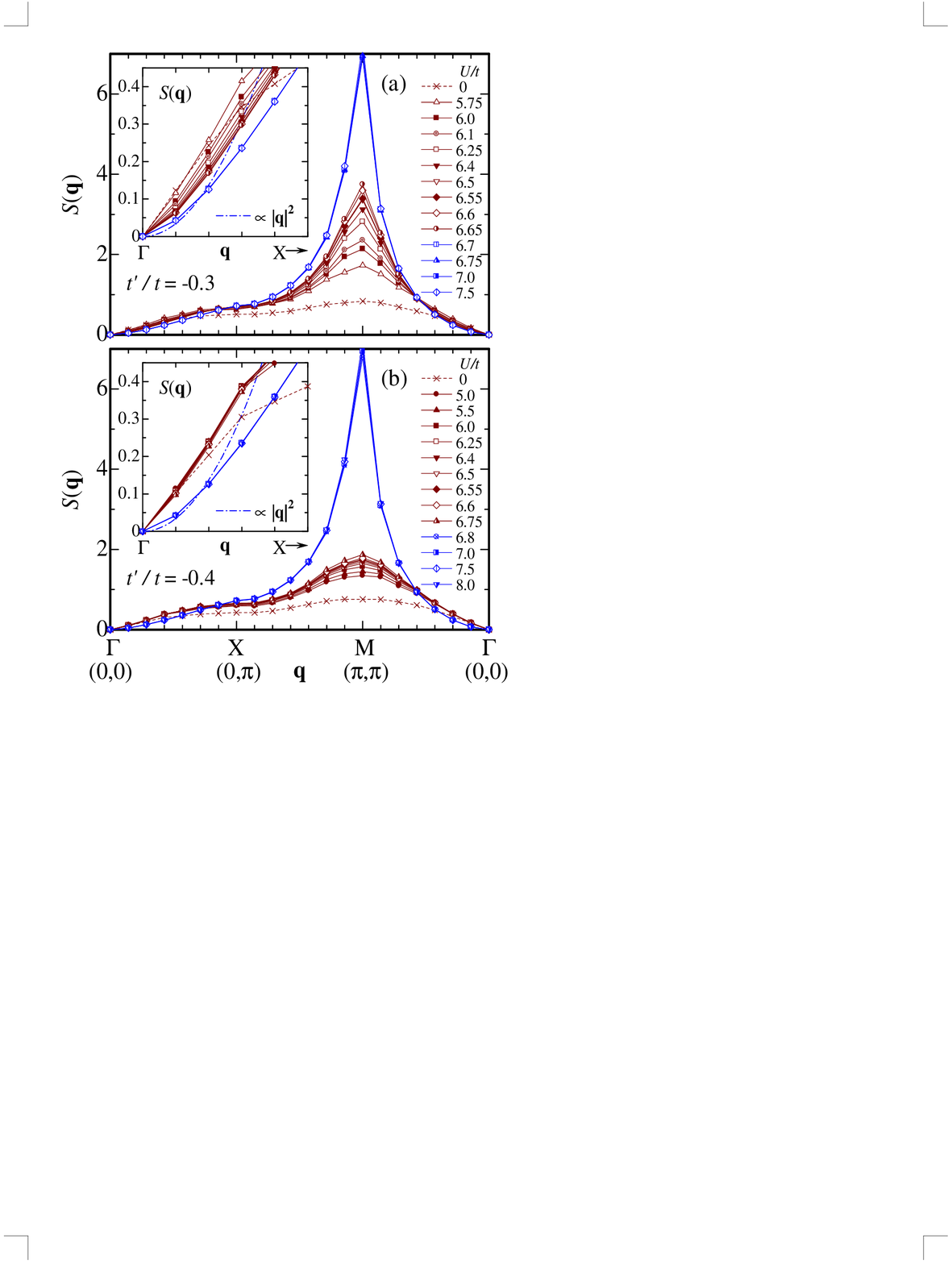}
\end{center}
\vskip -5mm
\caption{(Color online) 
Spin structure factor for (a) $t'/t=-0.3$ [$U_{\rm c}/t=6.69$] and 
(b) $t'/t=-0.4$ [$U_{\rm c}/t=6.78$] for various $U$ near $U_{\rm c}$
in $\Psi_Q^d$. 
The system size is $L=14$.
The insets show the magnification of the region near the $\Gamma$ point 
on the $\Gamma$-X line, and the symbols are common to the main panel. 
Data points of different $U$s for $U>U_{\rm c}$ almost overlap one 
another.
Similar data for $t'/t=0$ are given in Fig.~\ref{fig:sq-d}.
}
\label{fig:sq-dl}
\end{figure}
%
\subsection{\label{sec:SCprop}Properties of superconductivity}
%
First, we study the relation between $d$-wave SC and AF correlation. 
In Figs.~\ref{fig:sq-d}, \ref{fig:sq-dl}(a) and \ref{fig:sq-dl}(b), the 
$U/t$ dependence of $S({\bf q})$ in $\Psi_Q^d$ is shown for $t'/t=0$, 
$-0.3$ and $-0.4$, respectively. 
As mentioned, robust SC arises for $t'/t=0$ and $-0.3$, but does not 
for $t'/t=-0.4$ [Fig.~\ref{fig:pdrinf}]. 
This is supported by the small-$|{\bf q}|$ behavior of $S({\bf q})$, 
shown in the insets of Fig.~\ref{fig:sq-dl}. 
For $t'/t=-0.3$, $S({\bf q})$ for small $|{\bf q}|$ tends to be 
quadratic in $|{\bf q}|$ as $U/t$ increases, indicating that a SC gap 
develops, whereas for $t'/t=-0.4$, $S({\bf q})$ remains almost linear
in $|{\bf q}|$. 
\par

\begin{figure}[hob]
\vspace{-0.2cm}
\begin{center}
\includegraphics[width=8.7cm,clip]{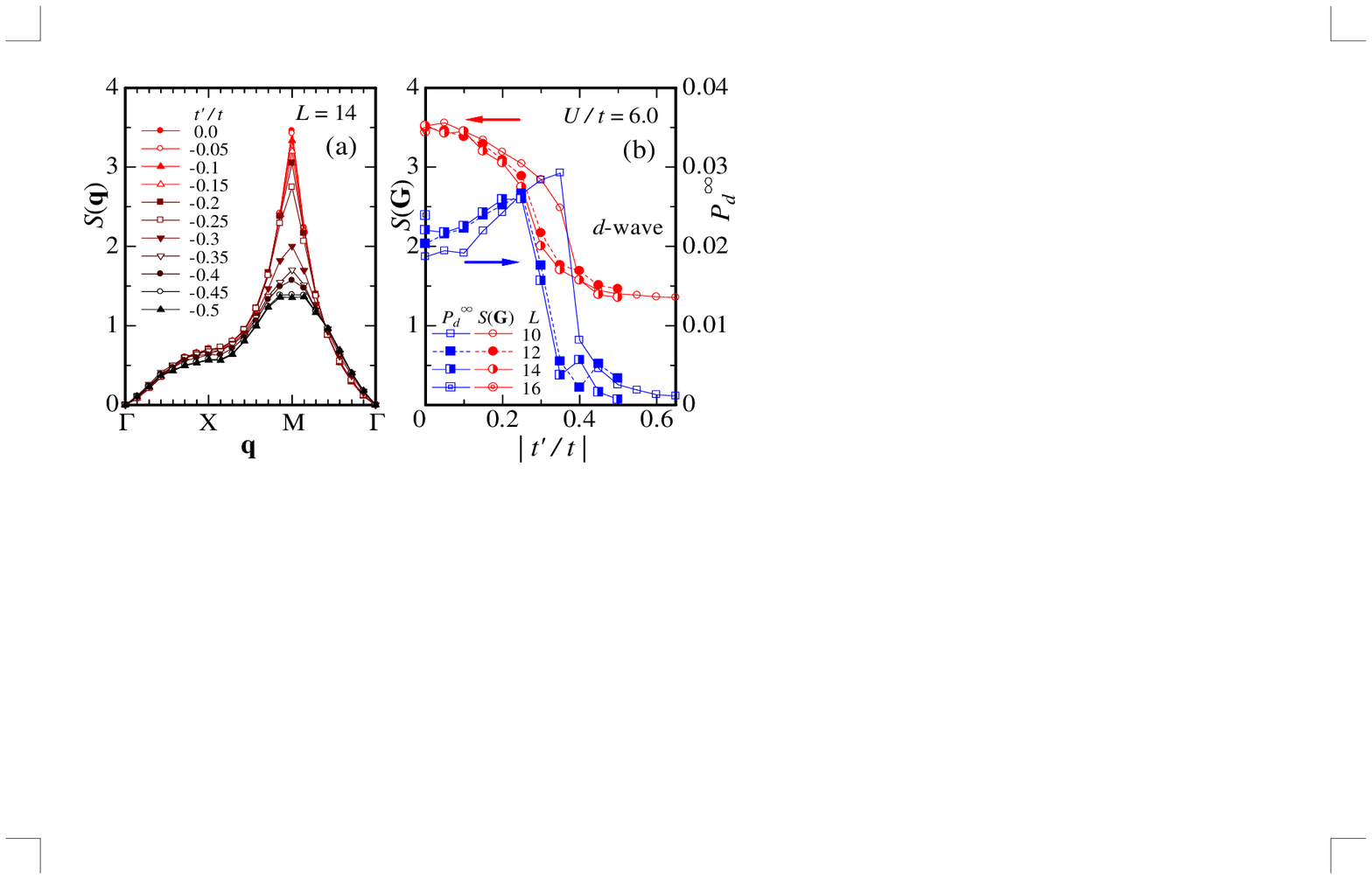}
\end{center}
\vskip -5mm
\caption{(Color online) 
(a) $t'/t$ dependence of spin structure factor for $U/t=6$, where 
$P_d^\infty$ is considerably enhanced particularly for small $t'/t$. 
(b) Spin structure factor at the AF wave number {\bf G} for $U/t=6$ 
as a function of $t'/t$ is denoted by circles (left axis). 
Simultaneously, the $d$-wave pair correlation function is plotted 
with squares (right axis). 
}
\label{fig:squ06a}
\end{figure}
%
\begin{figure}[hob]
\vspace{-0.2cm}
\begin{center}
\includegraphics[width=8.5cm,clip]{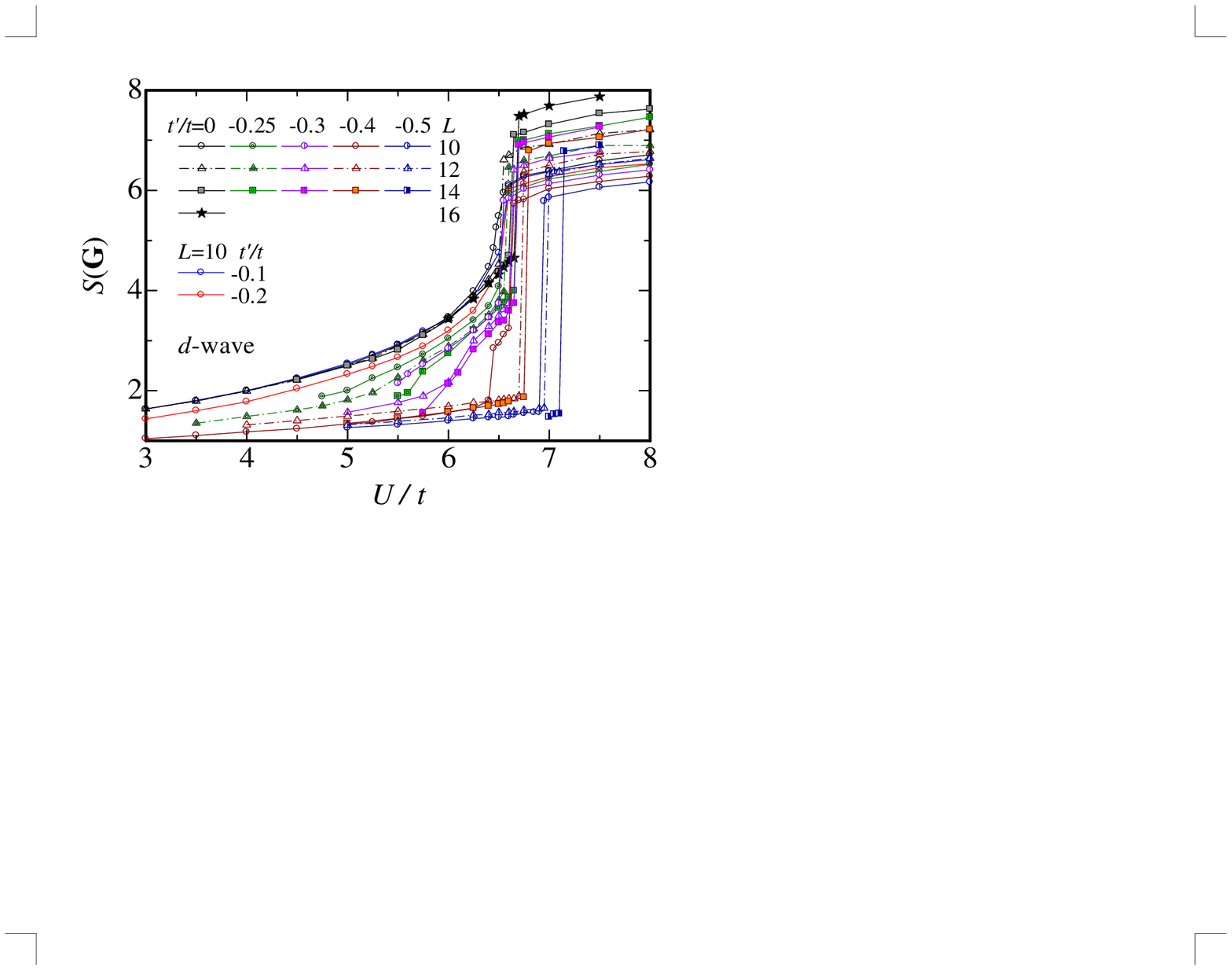}
\end{center}
\vskip -5mm
\caption{(Color online) 
$U/t$ dependence of spin structure factor at AF wave number {\bf G} 
for various values of $t'/t$ and $L$. 
}
\label{fig:sqvsu}
\end{figure}
%
We now focus on the AF wave number ${\bf G}$. 
For $U=0$, $S({\bf q})$ has a pointed peak for $t'=0$, a rounded peak 
for $t'/t=-0.3$ and a flat top for $-0.4$, due to the frustration. 
For the cases of $t'/t=0$ and $-0.3$, in which robust SC appears, 
$S({\bf G})$ steadily increases as $U/t$ increases, even in the conductive 
regime, $U<U_{\rm c}$ [Figs.~\ref{fig:sq-d} and \ref{fig:sq-dl}(a)]. 
In the strongly frustrated case ($t'/t=-0.4$), in which the SC 
correlation does not develop, the magnitude of $S({\bf G})$ reaches 
no more than half that for $t'/t=-0.3$, although $S({\bf q})$ 
increases slightly for $U<U_{\rm c}$ [Fig.~\ref{fig:sq-dl}(b)]. 
To consider the $t'/t$ dependence of $S({\bf q})$ explicitly, we plot 
$S({\bf q})$ for various values of $t'/t$ in Fig.~\ref{fig:squ06a}(a) 
at $U/t=6$, near which $P_d^\infty$ has a maximum (see 
Fig.~\ref{fig:pdrinf}). 
When $t'/t$ is varied, the change in $S({\bf q})$ is quantitatively 
insignificant, except near the M point. 
As $|t'/t|$ increases, $S({\bf G})$ sharply decreases, particularly 
at $t'/t\sim -0.3$, and ${\bf G}$ is no longer a characteristic 
wave number for $|t'/t|\gsim -0.45$. 
In Fig.~\ref{fig:squ06a}(b), we compare the $t'/t$ dependence of 
$S({\bf G})$ with that of $P_d^\infty$. 
In respective system sizes, when $S({\bf G})$ abruptly decreases 
for $|t'/t|>0.25$, $P_d^\infty$ similarly decreases. 
In Fig.~\ref{fig:sqvsu}, we show the $U/t$ dependence of $S({\bf G})$. 
Although in every case $S({\bf G})$ generally increases as $U/t$ 
increases, the range of significant increase [$S({\bf G})\gsim 2$] 
in the conductive regime roughly corresponds to 
$U_{\rm onset}\lsim U<U_{\rm c}$, and is accompanied by a marked 
increase in $P_d^\infty$ (Fig.~\ref{fig:pdrinf}). 
We have confirmed, for a wide range of model parameters, that 
whenever $P_d^\infty$ is appreciably enhanced, $S({\bf q})$ has an 
evident peak at ${\bf q}={\bf G}$. 
This result strongly supports the idea that the SC in this model is 
induced by AF spin correlation. 
Incidentally, this mechanism is reflected in the ratio of energy 
components. 
As shown in Table \ref{table:rho} ($U/t=6.25$), when SC is weak 
($|t'/t|=0.4$), $\rho$ is only slightly smaller than the noninteracting 
value, whereas for robust SC ($|t'/t|=0.25$), $\rho$ becomes less 
than half the value for $U=0$; the diagonal hopping is considerably 
suppressed. 
\par

\begin{figure}[hob]
\vspace{-0.2cm}
\begin{center}
\includegraphics[width=8.0cm,clip]{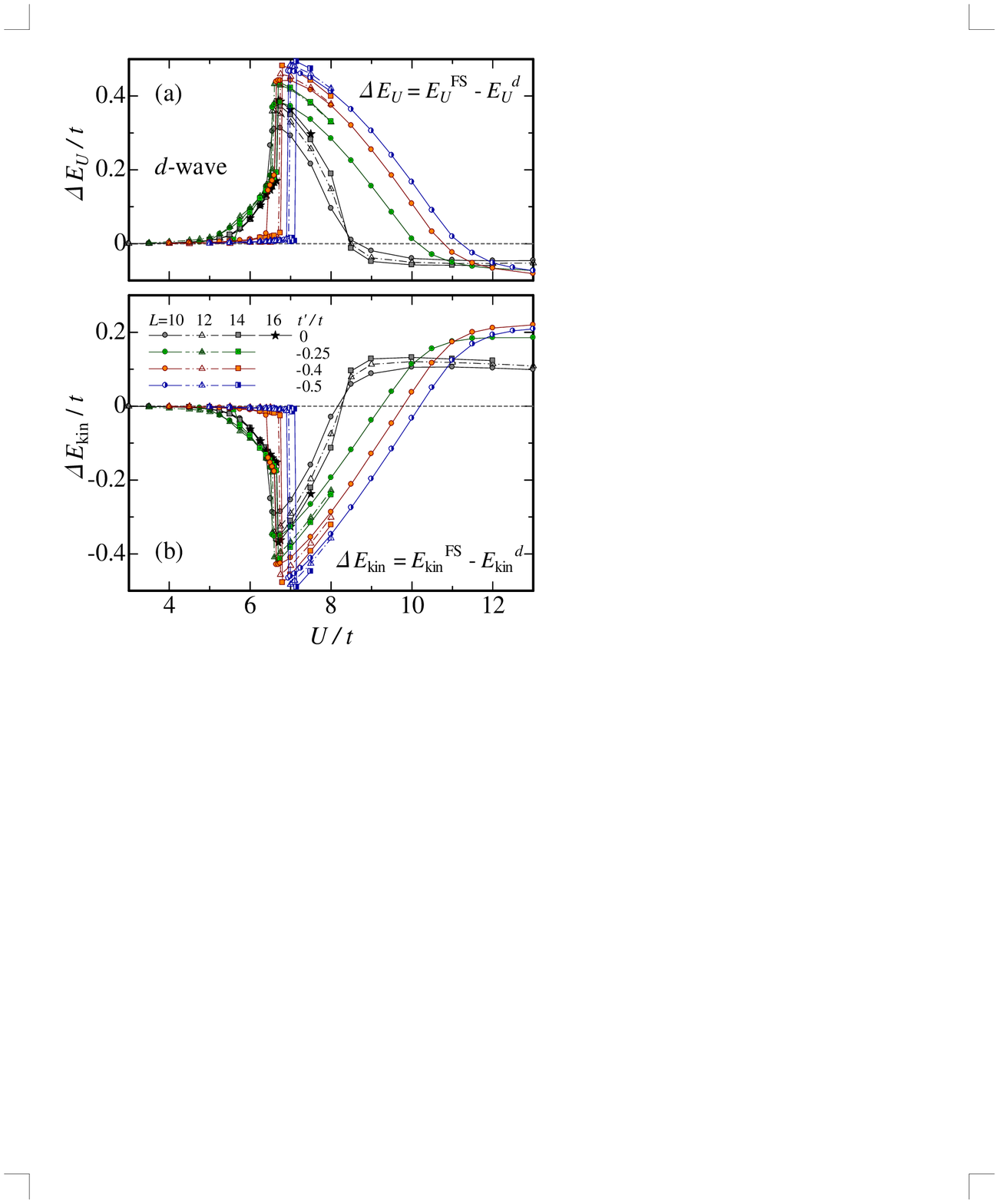}
\end{center}
\vskip -5mm
\caption{(Color online) 
(a) Interaction and (b) kinetic components of condensation energy 
due to $\Psi_Q^d$ as a function of interaction strength for four values 
of $t'/t$. 
Various system sizes are plotted. 
The symbols and scales are common to both panels. 
}
\label{fig:deteuvsa}
\end{figure}
%
Next, we consider the energy gain in the SC transition. 
The components of condensation energy are shown in Fig.~\ref{fig:deteuvsa}, 
that is, the differences in kinetic and interaction energies 
between $\Psi_Q^d$ and $\Psi_Q^{\rm FS}$; 
the actual expression is given in the figure. 
Here, $E_{\rm kin}=\langle{\cal H}_{\rm kin}\rangle=E_t+E_{t'}$, and 
$\Delta E_{\rm kin}+\Delta E_U=\Delta E\ (\ge 0)$ [eq.~(\ref{eq:cond})]. 
In the SC regime ($U<U_{\rm c}$), $\Delta E_{\rm int}$ 
($\Delta E_{\rm kin}$) is always positive (negative), regardless of 
the value of $t'/t$. 
This indicates that the SC transition is induced by the gain in the 
interaction energy at the expense of kinetic energy. 
This feature smoothly continues to the weak-coupling limit 
($U/t\rightarrow 0$), and is common to conventional BCS superconductors. 
Although the component of energy gain switches to the kinetic energy 
at $U=8t$-$11t$, which broadly corresponds to $U_{\rm c}^{\rm FS}$, 
SC is excluded for $U>U_{\rm c}$. 
The kinetic-energy-driven SC is not realized at half filling, in 
contrast to that in the doped cases. \cite{YTOT}
\par 

\begin{figure}[hob]
\vspace{-0.2cm}
\begin{center}
\includegraphics[width=8.7cm,clip]{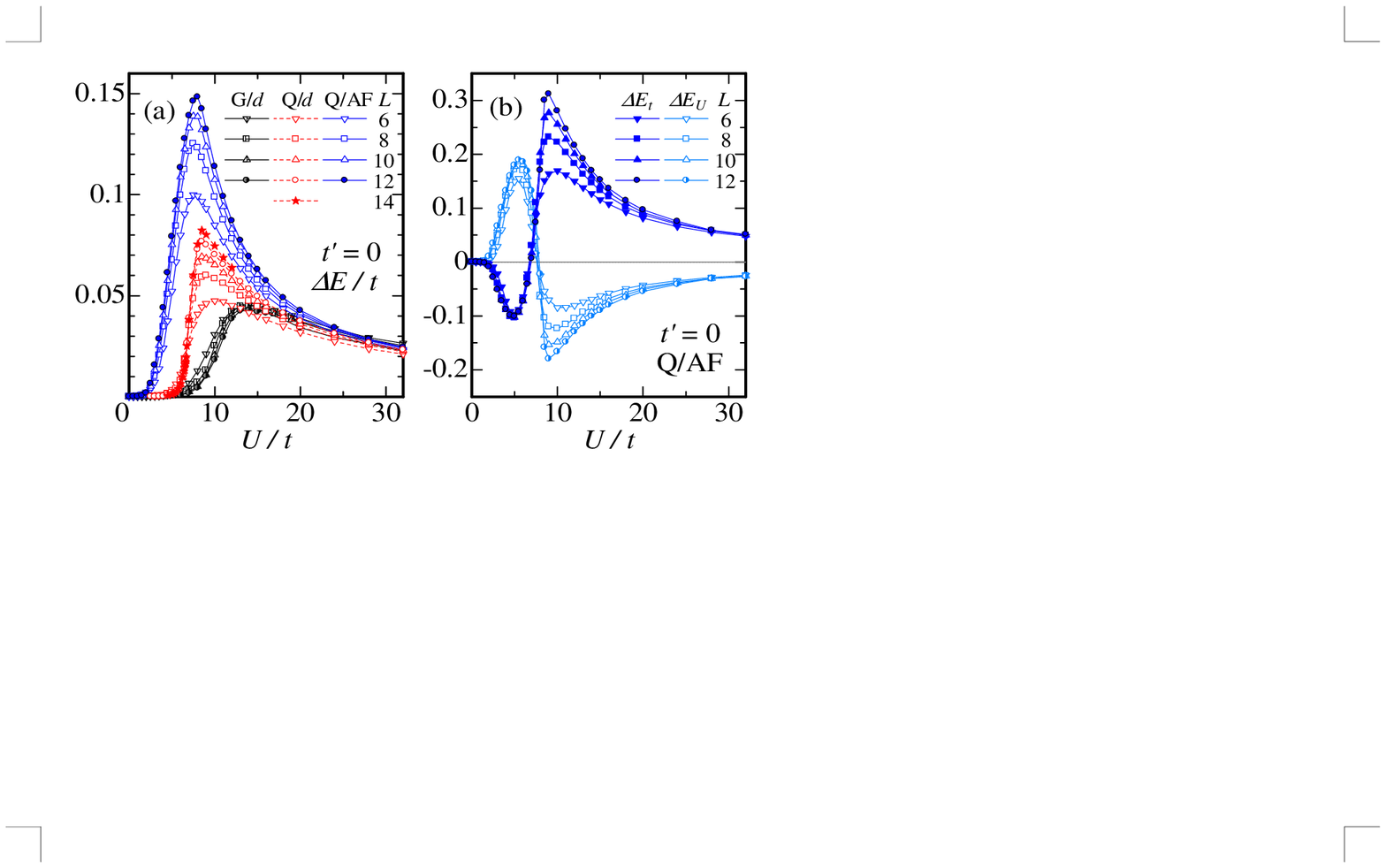}
\end{center}
\vskip -5mm
\caption{(Color online) 
(a) Comparison of the condensation energy for $t'/t=0$ among 
$\Psi_Q^d$ (indicated by Q/$d$), $\Psi_Q^{\rm AF}$ (Q/AF) and 
${\cal P}_{\rm G}\Phi_d$ (G/$d$), for which 
$\Delta E=E({\cal P}_{\rm G}\Phi_{\rm FS})-E({\cal P}_{\rm G}\Phi_d)$. 
Several system sizes are plotted to consider the $L$ dependence. 
(b) Kinetic (closed symbols) and interaction (open symbols) parts of 
the condensation energy due to $\Psi_Q^{\rm AF}$ for $t'/t=0$. 
Four system sizes are used. 
}
\label{fig:condfs}
\end{figure}
%
Incidentally, this behavior of $\Delta E_{\rm kin}$ and $\Delta E_U$ 
is not restricted to the $d$-wave state, but is also found in some 
order-disorder transitions. 
As an example, we plot, in Fig.~\ref{fig:condfs}(b), $\Delta E_{\rm kin}$ 
($=\Delta E_t$) and $\Delta E_U$ for the AF state $\Psi_Q^{\rm AF}$ 
[eq.~(\ref{eq:AF})], which will be discussed in the next section, 
for $t'/t=0$. 
The interaction part $\Delta E_U$ makes a positive contribution to 
$\Delta E$ for $U/t\lsim 8$, whereas the kinetic part $\Delta E_t$ 
contributes for $U/t\gsim 7$. 
Hence, the behavior is qualitatively identical to that of $\Psi_Q^d$. 
Such behavior is also observed in SC and CDW states for the two-dimensional 
attractive Hubbard model. \cite{Attractive} 
\par

\section{\label{sec:AF}Phase Diagram and Antiferromagnetism}
Up to this point, we have not considered the competition with the AF state, 
but it is a crucial problem when drawing a phase diagram. 
We thus compare the stability between $\Psi_Q^d$ and the AF-ordered state 
$\Psi_Q^{\rm AF}$ [eq.~(\ref{eq:AF})] into which we do not introduce band 
renormalization parameters to equalize the condition with $\Psi_Q^d$. 
Whenever $\Delta_{\rm AF}$ is finite at half filling, $\Psi_Q^{\rm AF}$ 
is insulating and has an AF long-range order; the sublattice magnetization 
$m$ [eq.~(\ref{eq:m})] behaves similarly to $\Delta_{\rm AF}$. 
\par

\begin{figure}[hob]
\vspace{-0.2cm}
\begin{center}
\includegraphics[width=7.5cm,clip]{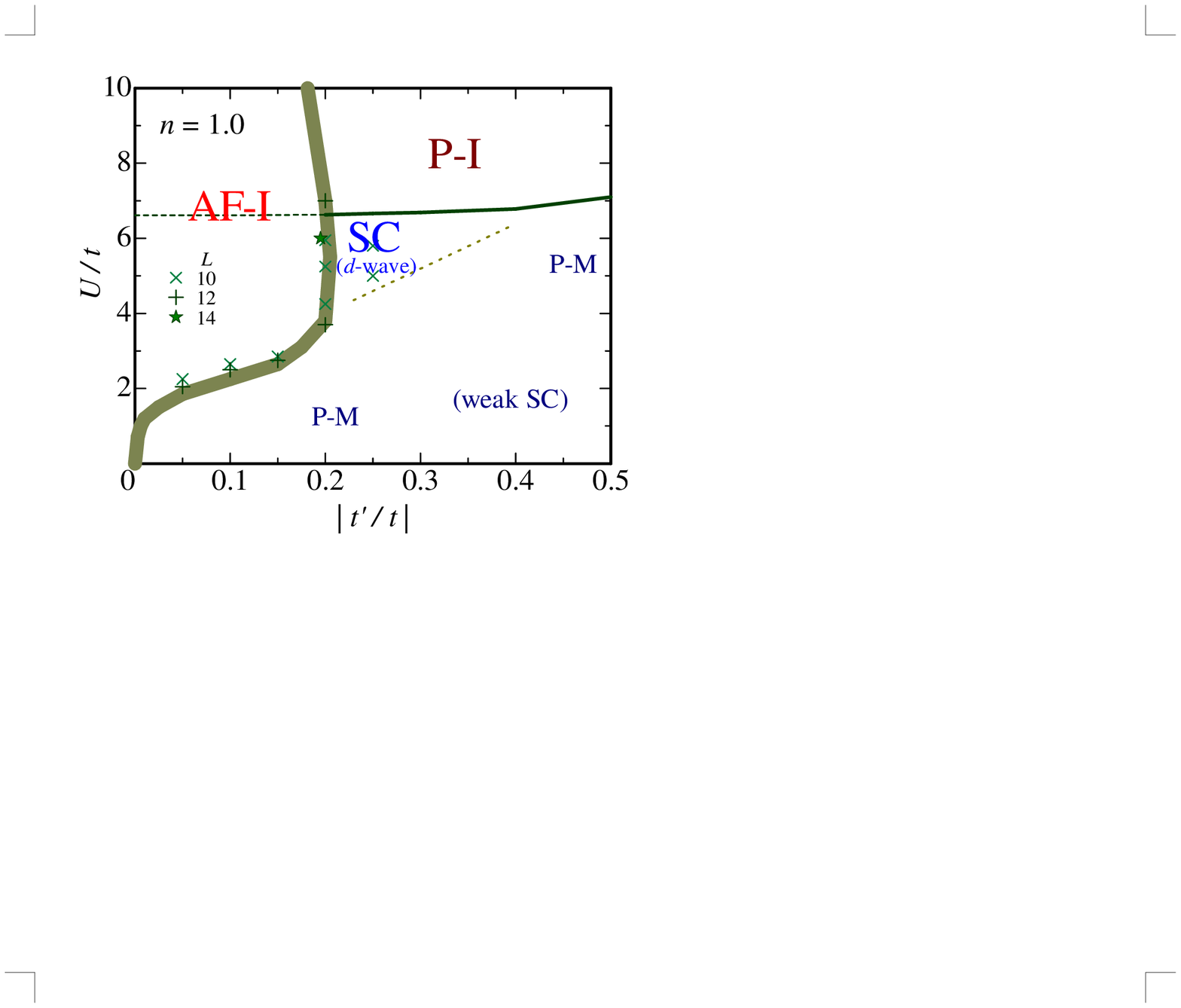}
\end{center}
\vskip -5mm
\caption{(Color online) 
Ground-state phase diagram in $t'$-$U$ space constructed from VMC 
calculations performed in this study. 
The abbreviations AF-I, P-I (P-M) and SC denote AF insulator, 
paramagnetic insulator (metal) and superconductor, respectively. 
}
\label{fig:phased}
\end{figure}
%
In Fig.~\ref{fig:phased}, we show a phase diagram constructed as follows. 
The boundary between AF-I and P-M ($|t'/t|\le 0.2$) is determined 
by the points where the extrapolated values of $\Delta_{\rm AF}$ vanish. 
The boundary between AF-I and SC (or P-I) is determined by the comparison 
of the total energies. 
The boundary between P-I and P-M is determined by $U_{\rm c}/t$ for the 
largest systems for each $t'/t$ obtained in \S\ref{sec:MottSC}. 
The last boundary is extended to $t'/t=0$ if $\Psi_Q^{\rm AF}$ is not 
allowed, as indicated by a dashed line. 
It is unnecessary to fix the boundary between SC and P-M, because the 
small magnitude of $\Delta$ often survives for extremely small $U/t$; 
instead, we show $U_{\rm onset}/t$ with a dotted line. 
\par

For $t'/t=0$, a continuous metal-to-AF-insulator transition occurs 
at $U=U_{\rm c}^{\rm AF}=0$, \cite{Hirsch} owing to the complete 
nesting condition. 
This AF state is very stable and continues to the Heisenberg limit 
($U/t\rightarrow\infty$). 
Our result is consistent with it, as shown in Fig.~\ref{fig:condfs}(a),
where the condensation energy for $\Psi_Q^{\rm AF}$ is always larger 
than that for $\Psi_Q^d$. 
The nature of the continuous transition seems to be preserved in 
the region of small $|t'/t|$, although $U_{\rm c}^{\rm AF}$ becomes 
finite, and the tendency toward a first-order transition gradually 
develops as $|t'/t|$ increases. 
Note that as the frustration becomes strong, $\Psi_Q^{\rm AF}$ is 
rapidly destabilized, and surrenders to $\Psi_Q^d$ at 
$t'=t'_{\rm c}\sim -0.2t$. 
Moreover, $|t'_{\rm c}/t|$ tends to decrease as $U/t$ increases; 
this feature is consistent with the result that the AF state has 
the largest range of $n$ at $U\sim W$ in a phase diagram in the 
$n$-$U$ plane. \cite{YTOT} 
However, this tendency is not in accord with the arguments of the 
$J$-$J'$ spin model, which is an effective model of eq.~(\ref{eq:model}) 
for large $U/t$. 
Various studies of the $J$-$J'$ model \cite{J-J'} concluded that 
the AF order vanishes at much larger values: $J'/J\sim 0.4$ 
($|t'/t|\sim 0.63$). 
\par

We consider that this discrepancy is primarily due to the choice of 
variational states: 
(1) $\Psi_Q^d$ does not have a seed of AF long-range order, although 
the AF short-range correlation appreciably develops, and 
(2) in $\Psi_Q^{\rm AF}$, we have not allowed for the band renormalization 
effect. 
These two requirements are satisfied simultaneously by adopting 
a coexisting state of AF and SC gaps, \cite{Giamarchi,Himeda} 
with a band renormalization effect. \cite{Himeda-t'}
In fact, we have performed VMC calculations using such a wave function 
for the anisotropic triangular lattice and found that the area 
of the AF phase considerably expands. \cite{Wataorg2} 
A similar result has been obtained independently by Chen, \cite{Chen}
and also for the checkerboard lattice by Koga \etal\ \cite{CB}
Thus, it is urgent that the wave function is improved in this line 
to refine the phase diagram. 
\par

\section{\label{sec:summary}Conclusions}
\subsection{Summary}
Using an optimization variational Monte Carlo method, we have studied 
the half-filled-band Hubbard model on frustrated square lattices, given 
by eq.~(\ref{eq:model}). 
Our primary aim is to understand the mechanisms of the Mott transition 
and of the $d_{x^2-y^2}$-wave SC arising in the Hubbard model. 
To this end, we introduce an intersite correlation factor that controls 
the binding between a doublon and a holon into the trial 
functions: normal (Fermi sea), $d$-wave singlet and AF states. 
We have succeeded in describing the $d$-wave SC and a Mott transition 
simultaneously in a single approach. 
We itemize our main findings: 
\par

(1) Within the $d$-wave singlet state, a first-order Mott 
(conductor-to-nonmagnetic-insulator) transition occurs at $U_{\rm c}$, 
which is approximately the bandwidth, for arbitrary $t'/t$. 
In the insulating regime, most doublon-holon pairs are actually 
confined within the nearest-neighbor sites, in contrast to the case 
in the conductive regime. 
The critical $U_{\rm c}/t$ gradually increases as the frustration 
becomes strong. 
This transition is not directly related to a magnetic order. 
\par

(2) We have confirmed that in the projected Fermi sea, a first-order 
Mott transition without relevance to magnetism also arises 
at a larger $U/t$ than the bandwidth for arbitrary $t'/t$, 
although the state does not have the lowest energy. 

(3) The nonmagnetic insulating state ($d$-wave singlet state for 
$U>U_{\rm c}$) has a considerably low energy and a strong short-range 
AF correlation. 
According to the small-$|{\bf q}|$ behavior of $S({\bf q})$, 
the $d$-wave state tends to have short singlet bonds owing to the 
nearest-neighbor pairing [eq.~(\ref{eq:gap})], in contrast to the 
projected Fermi sea, which clearly does not have a spin gap. 
\par

(4) Robust SC with $d_{x^2-y^2}$-wave symmetry appears for moderate 
values of $U/t$ ($\sim 6$) and $t'/t$ ($0.2\lsim|t'/t|\lsim 0.35$). 
This area is adjacent to both domains of a Mott insulator and an AF 
long-range order. 
The phase diagram obtained in this study is shown in 
Fig.~\ref{fig:phased}. 
\par

(5) By comparing the pair correlation function with $S({\bf q})$, 
it is found that robust SC is always accompanied by appreciably enhanced 
short-range AF spin correlation, which is weakened by the frustration 
and almost vanishes for $|t'/t|\gsim 0.35$. 
The SC transition is induced by the gain in interaction energy; 
this mechanism is identical to that in the weak-correlation limit 
as well as that of conventional BCS superconductors. 
\par

(6) The AF long-range order prevailing in the weakly frustrated 
cases ($|t'/t|\lsim 0.2$) is rapidly destabilized as $|t'/t|$ increases 
if a band renormalization effect is not introduced. 
\par

\subsection{Further discussions}
(1) In comparing the present study for the frustrated square lattice 
[Fig.~\ref{fig:model}(a)] with the preceding study, \cite{Wataorg} 
in which almost the same wave functions are applied to the anisotropic 
triangular lattice [Fig.~\ref{fig:model}(b)], the results for the 
two lattices are qualitatively identical, indicating that the two types 
of frustration work similarly unless $|t'/t|$ is too large. 
However, the critical values with respect to $t'/t$ are approximately 
doubled for the latter lattice; namely, the AF state becomes unstable 
at approximately $|t'/t|=0.2$ for the former and 0.4 for the latter, 
and the robust SC disappears at approximately $|t'/t|=0.35$ for the 
former, and 0.8 for the latter. 
This can be explained by the difference in the number of frustrated 
bonds. 
\par

(2) Recently, using a VMC method with a two-body long-range Jastrow 
factor for the square lattice ($t'=0$), Capello \etal\ \cite{Sorella} 
found that a metal-to-insulator transition arises, similar to that 
in $\Psi_Q^{\rm FS}$, for example, in the behavior of $Z$ and $d$. 
The critical value of their function is $U_{\rm c}/t\sim 8.5$, 
which is close to $8.73$ ($L=18$) in $\Psi_Q^{\rm FS}$. 
Although their transition is regarded as continuous, evidence of 
the first order is possibly be found by a detailed analysis of larger 
systems.
In their wave function, no explicit (four-body) doublon-holon binding 
factor is introduced, but a short-range part of the two-body Jastrow 
factor may substantially work as a binding factor under the condition 
of strong electron repulsion at half filling. 
It will be interesting to reveal the relation between the two wave 
functions. 
\par

(3) In this paper, we have restricted the electron density to half 
filling ($n=1$). 
When carriers are doped, unless the doping rate $|1-n|$ is too large, 
the doublon-holon-binding effect remains significant, as we showed 
for $t'=0$ in the previous letter. \cite{YTOT} 
The Mott transition at half filling changes to a crossover from 
weakly to strongly correlated regimes. 
As $|1-n|$ increases, the AF order is rapidly destabilized, and the SC 
phase expands to the region of large $U/t$, consistent with 
the behavior of high-$T_{\rm c}$ cuprates. 
We will report a detailed description of doped cases with the effect 
of frustration elsewhere.
\par

\begin{acknowledgments}
We would like to thank Kenji Kobayashi and Tsutomu Watanabe for useful 
discussions. 
We appreciate the beneficial communication with Yung-Chung Chen on the 
stability of the AF state. 
This study is partly supported by Grants-in-Aid from 
the Ministry of Education, Culture, Sports, Science and Technology, and 
by the Supercomputer Center, ISSP, University of Tokyo.
\end{acknowledgments}




\begin{thebibliography}{99}
\bibitem{note1D} In the results of ref.~\citen{YS3}, $\Psi_q^{\rm FS}$, 
even in one dimension, seems to induce a Mott transition between 
$U/t=4$ and 16. 
We speculate that a longer-range doublon-holon binding factor is 
crucial in one dimension to reduce $U_{\rm c}$ down to zero.
\bibitem{notehyst} In the present VMC procedure, if the energy barrier 
between the two local energy minima is sufficiently high, the optimized 
variational parameters are trapped in the local minima to which the 
initial parameter values tend. 
Thus, as seen in Figs.~\ref{fig:hyst0}, \ref{fig:hysta} and 
\ref{fig:hystl}, we can obtain the optimized values for both local 
minima by giving the respective initial parameters anticipated from 
those obtained for nearby data points. 
On the other hand, if the barrier becomes as low as the statistical 
fluctuation in $E/t$ (we assume there are still two energy minima), 
the parameters readily arrive at the global minima, jumping over 
the barrier. 
Therefore, it is not easy to determine the values of $U/t$ at which 
the double-minimum structure of $E/t$ vanishes, namely, to draw 
the accurate hysteresis curves.
\bibitem{notemuss} The data of $\mu$ for $L=8$-16 seems fitted well 
with a linear function of $N_{\rm s}^{-1}$ ($=L^{-2}$). 
Thereby, we estimated approximate values of $\mu$ for $L=\infty$ 
and $t'=0$ as $\mu=0.912$-0.945 for $U/t=6.75$-10; as $U/t$ increases, 
$\mu$ tends to increase.
\bibitem{notecp} In ref.~\citen{Wataorg}, the optimized chemical 
potential $\zeta$ is positive, because the sign of $t'/t$ is assumed 
positive, contrary to the present case. 
The behavior of $|\zeta|$ for $U<U_{\rm c}$ in ref.~\citen{Wataorg} 
is broadly similar to that in Fig.~\ref{fig:para}(e), whereas 
for $U>U_{\rm c}$, $\zeta$ decreases to nearly zero, which is 
the value for $t'/t=0$, but never becomes negative.
\bibitem{notepdrsize} As $L$ is increased, the magnitude of spikes 
gradually decreases. 
We need to use the systems of $L=30$-50 in which the spikes for $U/t=0$ 
appear negligible on the scale of Fig.~\ref{fig:pdr}.
\end{thebibliography}
\end{document}